\documentclass[journal]{IEEEtran}
\usepackage[pdftex]{graphicx}
\usepackage[cmex10]{amsmath}
\usepackage{amssymb}
\usepackage{caption}
\usepackage{subcaption}
\usepackage{cite}
\usepackage{psfrag}
\usepackage{epstopdf}
\usepackage{algorithm}
\usepackage{algorithmic}
\usepackage{widetext}
\usepackage{float}
\usepackage{color}
\usepackage{soul}
\usepackage{wrapfig}
\usepackage[normalem]{ulem}
\usepackage{upgreek}
\usepackage{multirow}
\usepackage[table,xcdraw]{xcolor}
\usepackage[justification=centerlast]{caption}
\usepackage{tikz,pgfplots}
\usepackage{dblfloatfix}
\usepackage{adjustbox}
\DeclareUnicodeCharacter{2005}{\hspace{0.01em}}
\usetikzlibrary{calc}
\usepackage{mathtools}
\usetikzlibrary{arrows}
\pgfplotsset{compat=newest}
\usepgfplotslibrary{fillbetween}
\usepackage{xcolor}
\usepackage{pgfplots}
\definecolor{bblue}{HTML}{4F81BD}
\definecolor{rred}{HTML}{C0504D}
\definecolor{ggreen}{HTML}{9BBB59}
\definecolor{ppurple}{HTML}{9F4C7C}

\pgfplotsset{
/pgfplots/my legend/.style={
legend image code/.code={
\draw[thick,black](-0.05cm,0cm) -- (0.3cm,0cm);%
   }
  }
}

\definecolor{myBlue}{RGB}{72,125,215}
\definecolor{myOrange}{RGB}{118,54,45}
\definecolor{InfinBlue}{RGB}{72,72,51}
\definecolor{bblue}{HTML}{4F81BD}
\definecolor{rred}{HTML}{C0504D}
\definecolor{ggreen}{HTML}{9BBB59}
\definecolor{ppurple}{HTML}{9F4C7C}

\usetikzlibrary{patterns}
\DeclareMathOperator{\EX}{\mathbb{E}}

\ifCLASSINFOpdf
\else
\fi
\usepackage{caption}
\usepackage{subcaption}
\usepackage{graphicx}
\usepackage[usestackEOL]{stackengine}
\begin{document}
%
\title{Neural Networks-based Equalizers for Coherent Optical Transmission: Caveats and  Pitfalls}

%
%
     \pgfplotsset{
        compat=1.3, 
        my axis style/.style={
            every axis plot post/.style={/pgf/number format/fixed},
            ybar=5pt,
            bar width=8pt,
            x=1.7cm,
            axis on top,
            enlarge x limits=0.1,
            symbolic x coords={MLP, biLSTM, ESN, CNN+MLP, CNN+biLSTM},
            visualization depends on=rawy\as\rawy, 
            nodes near coords={%
                \pgfmathprintnumber[precision=2]{\rawy}
            },
            every node near coord/.append style={rotate=90, anchor=west},
            tick label style={font=\footnotesize},
            xtick distance=1,
        },
    }
%

\author{Pedro J. Freire, Antonio Napoli, Bernhard Spinnler, Nelson Costa, Sergei K. Turitsyn, Jaroslaw E. Prilepsky
\thanks{This paper was supported by the EU Horizon 2020 program under the Marie Sklodowska-Curie grant agreement 813144 (REAL-NET). JEP is supported by Leverhulme Trust, Grant No. RP-2018-063. SKT acknowledges support of the EPSRC project TRANSNET. (Corresponding Author: Pedro J. Freire)}
\thanks{Pedro J. Freire, Jaroslaw E. Prilepsky and Sergei K. Turitsyn are with Aston Institute of Photonic Technologies, Aston University, United Kingdom, p.freiredecarvalhosouza@aston.ac.uk.}
\thanks{Antonio Napoli and Bernhard Spinnler are with Infinera R\&D, Sankt-Martin-Str. 76, 81541, Munich, Germany.}
\thanks{Nelson Costa is with Infinera Unipessoal, Lda, Rua da Garagem nº1, 2790-078 Carnaxide, Portugal, ncosta@infinera.com.}

\thanks{Manuscript received xxx 19, zzz; revised January 11, yyy.}}

\maketitle
\begin{abstract}
This paper performs a detailed, multi-faceted analysis of key challenges and common design caveats related to the development of efficient neural networks (NN) based nonlinear channel equalizers in coherent optical communication systems. The goal of this study is to guide researchers and engineers working in this field.
We start by clarifying the metrics used to evaluate the equalizers' performance, relating them to the loss functions employed in the training of the NN equalizers. The relationships between the channel propagation model's accuracy and the performance of the equalizers are addressed and quantified. Next, we assess the impact of the order of the pseudo-random bit sequence used to generate the -- numerical and experimental -- data as well as of the DAC memory limitations on the operation of the NN equalizers both during the training and validation phases. Finally, we examine the critical issues of overfitting limitations, the difference between using classification instead of regression, and batch-size-related peculiarities. We conclude by providing analytical expressions for the equalizers' complexity evaluation in the digital signal processing (DSP) terms and relate the metrics to the processing latency.  
\end{abstract}



%
\IEEEpeerreviewmaketitle

\section{Introduction} 
\IEEEPARstart{M}{achine} learning techniques and, more specifically, deep artificial neural networks (NN), are rapidly finding their way into the telecommunication sector, in particular due to their ability to efficiently mitigate transmission and device impairments (see, e.g., \cite{jarajreh2014artificial,Darko2015,Hoydis,Hunt_2009,eriksson2017applying,WANG20171,hager2018nonlinear,zhang2019field,Zibar19,Khan19,freire2020complex,schadler2021soft}). Indeed, a large number of NN-based techniques have already been proposed and tested for the channel equalization in coherent optical communication systems \cite{Zibar19,Khan19,9645206, freire2020complex,freire2021performance,freire2021transfer}. 

The NNs are known to be universal approximators that can mimic almost any complex function, provided that the NN structure possesses enough computational capacity and is trained on a sufficient dataset. NNs are especially useful in solving complex nonlinear problems, where the results rendered by more conventional approaches often have limited applicability. The capability of NNs to learn from data, and their inherent adaptability to diverse operating conditions, make them a natural tool for the equalization of fiber-optic channels where the data-carrying signal experiences nonlinear interactions, signal-noise interference, and memory effects. The considerable speed of optical data transmission results in large datasets obtained in a short time, making the optical channel a suitable playground for machine learning techniques. Despite various acknowledged advantages and benefits, there are still challenges and pitfalls that hinder the use of machine learning and NNs, particularly in optical transmission-related tasks. In this paper, we discuss common misunderstandings and misinterpretations that occur when using ML approaches for channel equalization in coherent optical communications.

When applying known NN techniques to optical fiber transmission, the customization and adaptation of these algorithms can be necessary for reaping the full benefits of machine learning\footnote{Similarly, when the telecommunication industry began to use DSP techniques -- already employed in wireless -- in optics, adaptions were also required.}. For example, when using NN approaches in image recognition tasks, an accuracy of 99\% is typically regarded as ``more than superb''. Alternatively, this outcome may be viewed with skepticism as being considered too good to be true.
In contrast, in optical transmission-related problems, the 99\% accuracy -- when deciding which bit was transmitted -- is typically the minimum required by state-of-the-art transponders operating at pre-forward error correction (FEC) bit error rate (BER) ranging between $10^{-3}$ -- $10^{-2}$. In modern transponders, this threshold is even higher, e.g., $2\cdot10^{-2}$ -- $4\cdot10^{-2}$, and we ought to push the BER tolerance to higher values to improve the operating margin of the system performance. Consequently, we inherently have to impose stricter conditions on our NN architectures: in the case of optical communications, the learning quality must be higher than in the other fields. Unfortunately, efficient NN structures that meet such ultra-high accuracy requirements are often more vulnerable to various problems and pitfalls.

The severe overfitting that is frequently encountered while dealing with optical channel equalization is a critical element addressed in this work. Furthermore, because we are dealing with a high-accuracy problem, local minima can play an important role, since the NN can stop learning if the current local minima yield an almost zero gradient. Arguably, the overfitting and local minima are the main issues capping the equalization capacity of the NN-based DSP elements (when we neglect other technical issues such as the high complexity of NN-based components).
Generally, overfitting occurs when models or procedures that violate a ``parsimony principle'' are used. For example, this happens when more elements than required by the task at hand or more complicated approaches than necessary are employed. In other words, it occurs ``when using a model that includes irrelevant components (excessive degrees of freedom)''~\cite{hawkins2004problem}.

Overfitting in optical channel equalization is the state in which an NN starts to interpret the structure of the training set with such detail that it effectively models the noise or some spurious periodicity present in the transmission data instead of identifying the true inverse transfer function (the deterministic part) of the channel. 
In this context, the ``barriers'' that preclude the NN learning process should be considered with a wider meaning than just the well-known performance gap between the training and validation curves. Here we show that learning the noise characteristics in the training dataset, typically produces the ``jail window'' pattern in equalized 2D constellations, and this effect degrades the learning \textit{while we can still have a small gap between the training and testing loss curves}. This result indicates that the manifestation of overfitting and local minima in our problems can be intricate and unusual. Additionally, a small gap (between training and validation) can be observed when the NN learns the possible periodicity in the dataset, but it is a clear deviation from the ``true purpose'' of the NN-based equalizer.  

Another major issue that is frequently identified in the literature, is the overestimation of the performance and computational complexity (CC) of the NN equalizers. We devote the entire Sec.~\ref{sec:complex} to discuss this. Therein, we show that the pseudo-random bit sequence (PRBS) order and digital-to-analog converter (DAC) memory may lead to an overestimation of the system performance. Furthermore, we also show that CC can be misinterpreted and exaggerated if certain points are overlooked. In this case, we demonstrate that the number of NN parameters is not a precise indicator of the true CC. In addition, when pruning and quantizing an NN-based equalizer, special care must be taken to how the NN is pruned and quantized; otherwise, overestimation can occur in the CC analysis.

In summary, we investigate typical caveats and pitfalls that may occur when using machine learning-based methods for impairment compensation in coherent optical communication systems. We introduce and discuss the most widely used metrics to quantify the performance and complexity of NN-based equalizers. Our main goal is to present some form of a guide, providing the reader with an intuitive understanding of the key pitfalls when using NNs in optical communications. We illustrate the potential issues that can occur when using some current practices proposed in the literature in this field. We also hope that this paper can serve as an introduction to this fast-growing field of high practical importance.

\section{Experimental and Numerical Setup and NN-based Equalizer Characteristics}

Our results will be further predominantly demonstrated and confirmed by the data obtained in the extensive numerical modelling of rather general transmission systems.  However, we also verified key conclusions using the (relatively) smaller set of the experimental data for specific set-ups. We would like to stress that our conclusions, findings, and design propositions are quite general and apply to various similar optical transmission systems.

\begin{figure}[htbp]
\centering
\includegraphics[width=0.5\textwidth]{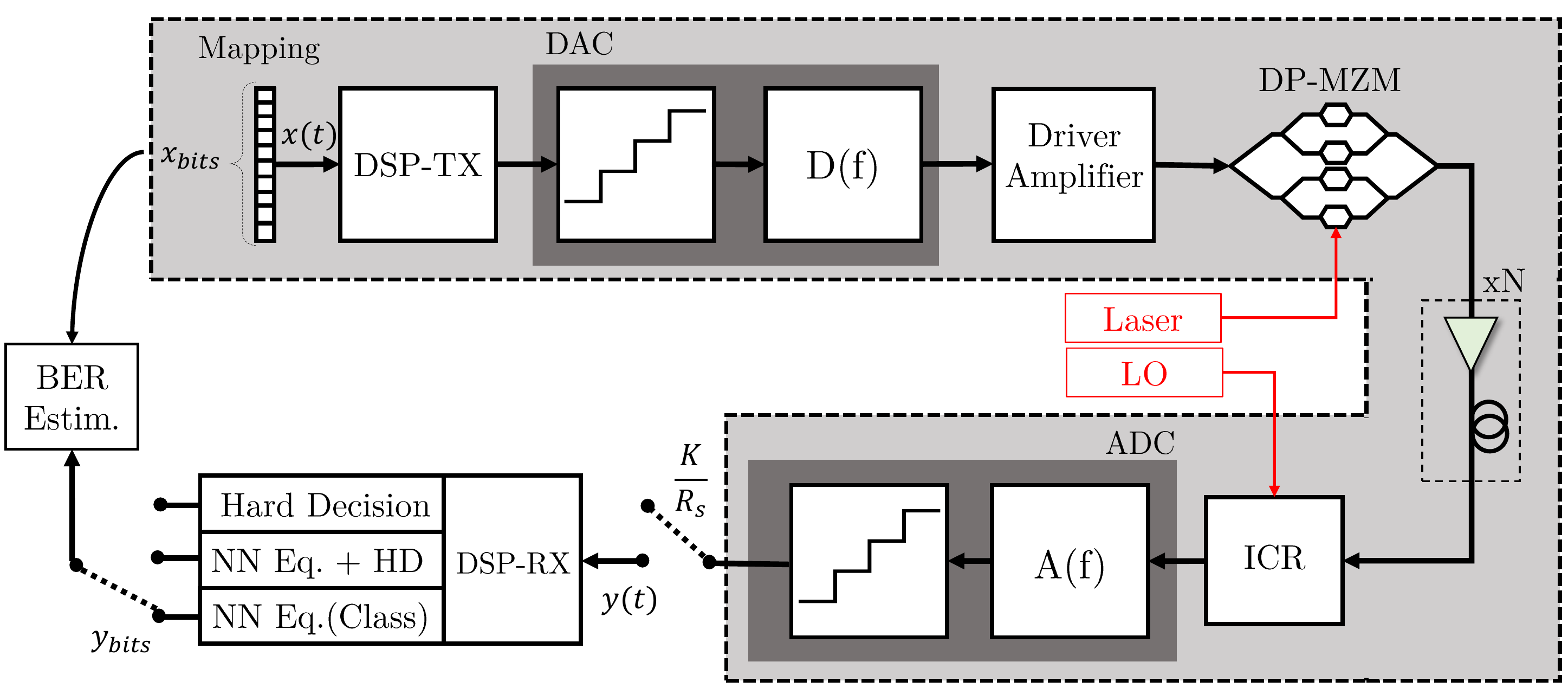}
\caption{Experimental setup used to analyze the performance of NN equalizers; $D(f)$ is the DAC electrical transfer function, $A(f)$ is the ADC electrical transfer function, and $x(t)$ and $y(t)$ are the transmitted/received time sequences of symbols, respectively. The input of the NN is the soft output of the regular DSP just before the decision module.}
\label{setup}
\end{figure}

The setup used in our experiment is depicted in Fig.~\ref{setup}. At the transmitter (TX) side, a dual-polarization (DP) 16-quadrature amplitude modulation (16-QAM) 34.4~GBd symbol sequence was mapped out of data bits generated by a $2^{32} - 1$ PRBS. Then a digital root-raised cosine (RRC) filter with a roll-off factor of 0.1 was applied to limit the channel bandwidth to 37.5~GHz. The resulting filtered digital samples were resampled and uploaded to a DAC operating at 88~GSamples/s.  DAC outputs were amplified by a four-channel electrical amplifier that drove a Mach-Zehnder modulator in the phase/quadrature of DP, modulating the continuous waveform carrier produced by an external cavity laser at $\lambda = 1.55 \, \mu$m. 
The resulting optical signal was transmitted over the 5$\times$50~km spans of standard single-mode fiber (SSMF) with an erbium-doped fiber amplifier (EDFA) only. The EDFA noise figure was in the 4.5 to 5~dB range. The parameters of the SSMF at $\lambda = 1.55 \, \mu $m, are: attenuation coefficient $ \alpha = 0.21$~dB/km, dispersion coefficient $D = 16.8$~ps/(nm·km), and effective nonlinear coefficient $\gamma$ = 1.2 (W$\cdot$km)$^{-1}$.

At the receiver (RX) side, the optical signal was converted into the electrical domain using an integrated coherent receiver. The obtained signal was sampled at 50 Gsamples/s by a digital sampling oscilloscope and processed by an offline DSP based on the algorithms described in~\cite{kuschnerov2010data}. First, the bulk accumulated chromatic dispersion (CD) was compensated using a frequency domain equalizer, which was followed by the removal of the carrier frequency offset. Then a constant-amplitude zero autocorrelation-based training sequence was located in the received frames and the equalizer transfer matrix was estimated from it. This equalizer transfer matrix was applied to the signal to remove the remaining linear distortions (due to polarization mode dispersion (PMD), residual CD, filtering), and to demultiplex the signal polarization. The processed signal was corrected for clock frequency and phase offsets. The carrier phase estimation was then achieved with the help of pilot symbols. Thereafter, the resulting soft symbols were used as input for the NN equalizers. Finally, the pre-FEC BER was evaluated from the signal at the NN equalizer output.

A simulator was implemented aiming at mimicking the ideal transmission setup, i.e., using ideal TX/RX components. As a consequence, only impairments resulting from fiber propagation and amplified spontaneous emission (ASE) noise originating in the EDFA are considered in the numerical simulations\footnote{We considered the propagation of a single-channel signal filtered by an RRC filter with 0.1 roll-off factor and with an upsampling rate of 8 samples per symbol over different transmission systems.}. The propagation of the optical signal along the optical fiber was simulated by solving the Manakov equations using the split-step Fourier method (with a resolution of $1$ km per step) \cite{agrawal2013nonlinear}. Each fiber span was followed by an optical amplifier with the noise figure $\text{NF}=4.5$~dB, which fully compensates for fiber losses and adds ASE noise. At the receiver, after full electronic CD compensation (CDC) by the frequency-domain equalizer and downsampling to the symbol rate, the received symbols were normalized to the transmitted ones. 

The BER was estimated from the received symbols following three different approaches: i) applying a hard decision strategy only; ii) first, equalizing the signal using an NN equalizer in the regression task and then applying a hard decision, or iii) using an NN classifier (the classification is used only in the section where we compare its performance with the regression). These approaches are schematically depicted in Fig.~\ref{setup}. The NN input mini-batch shape can be defined by three dimensions \cite{freire2021performance}: $(B, M, 4)$, where $B$ is the mini-batch size, $M$ is the memory size defined through the number of neighbors $N$ as $M = 2N + 1$, and $4$ is the number of features for each symbol, referring to the real and imaginary parts of two polarization components. The objective of NN is to recover the real and imaginary parts of the $k$-th symbol in one of the polarizations, so that the shape of the NN output batch can be expressed as $(B,2)$.

In general, for the regression task -- within the NNs considered in this paper, we incorporate the mean squared error (MSE) loss function estimator and the classical Adam algorithm for the stochastic optimization step with the default learning rate set to 0.001~\cite{gulli2017deep}. Standard training was carried out for up to 1000 epochs with a batch size of 2048, which has proven to be high enough to reach convergence for the considered transmission scenarios. Furthermore, the total dataset used consisted of $2^{18}$ symbols for the training dataset, $2^{18}$ symbols for the validation dataset, and $2^{18}$ independently generated symbols for the testing phase. All three are generated with a different random seed. The training dataset is used to update the weights of our NN model; the validation one is to monitor the overfitting of the learned model and to trigger the early stopping, and the testing one gives us the final measurement with a never-seen dataset after the training has been done. In this paper, all results showing the Q-factor after equalization are obtained by using the testing dataset, and all results that show the Q-factor over the epochs, are obtained by using the validation datasets.

\begin{figure}[htbp]
\centering
\includegraphics[width=0.45\textwidth]{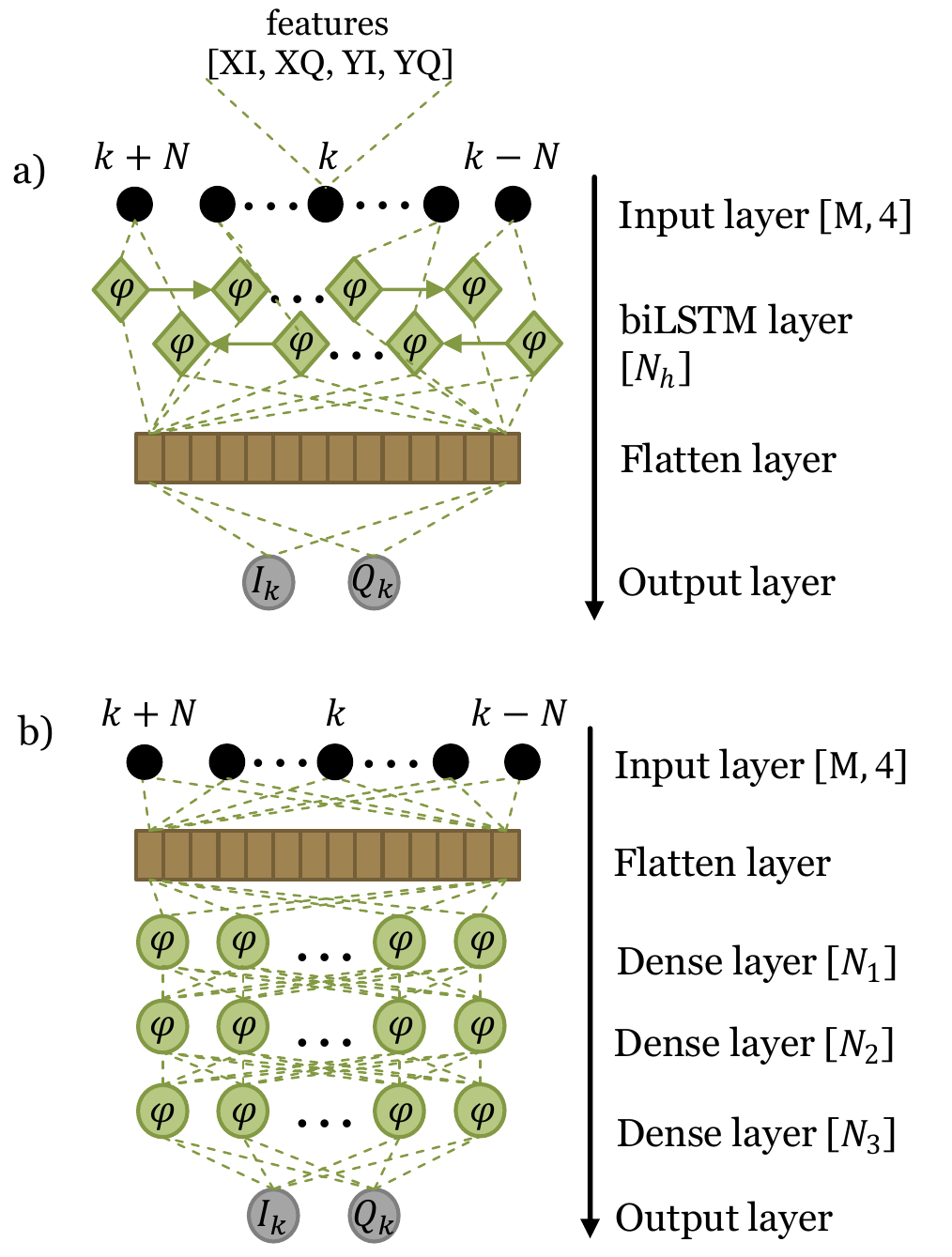}
\caption{The schematics of different NN architectures considered in our paper: a) biLSTM equalizer with $N_h$ hidden units and b) MLP equalizer having three hidden layers, with $N_1$, $N_2$, and $N_3$ neurons in each consecutive layer, respectively. The input has a memory equal to $M=2N+1$ and 4 features representing the real (I) and imaginary (Q) parts of both X and Y polarizations. The function $\varphi$ represents the activation function which in our case is the ``tanh''. }
\label{fig:NN_MODELS}
\end{figure}

The training dataset was shuffled at the beginning of every epoch to avoid overfitting caused by learning the connections between the neighboring training pairs~\cite{eriksson2017applying}. All simulated datasets were generated using the Mersenne twister generator~\cite{matsumoto1998mersenne} with different random seeds, which guarantees a cross-correlation below 0.004 between the training and testing datasets, meaning that the symbols are virtually independent.

\begin{table*}[htbp]
  \centering
  \caption{Summary of the best performance after 1000 training epochs for different QoT metrics and transmission setups. (Simulation results of single-channel transmission)}
    \resizebox{\textwidth}{!}{\begin{tabular}{|c|c|c|c|c|c|c|c|c|c|}
\hline
\multirow{2}{*}{Scenarios}                                                               & \multicolumn{3}{c|}{MI {[}bits/symbol{]}} & \multicolumn{3}{c|}{EVM {[}$\%${]}} & \multicolumn{3}{c|}{Q-factor {[}dB{]}} \\ \cline{2-10} 
                                                                                         & Ref          & biLSTM       & MLP         & Ref        & biLSTM     & MLP       & Ref         & biLSTM      & MLP        \\ \hline
\begin{tabular}[c]{@{}c@{}}Case i) 20×80km; SSMF; 5dBm;  8-QAM; 34.4GBd\end{tabular}  & $2.93$       & $2.99$       & $2.90$      & $22.7$     & $5.9$      & $8.3$     & $7.82$      & $10.72$     & $8.37$     \\ \hline
\begin{tabular}[c]{@{}c@{}}Case ii) 15×100km; SSMF; 4dBm; 16-QAM; 28GBd\end{tabular} & $3.83$       & $3.99$       & $3.89$      & $19.7$     & $7.4$      & $9.4$     & $7.42$      & $10.48$     & $7.89$     \\ \hline
\begin{tabular}[c]{@{}c@{}}Case iii) 30×50km; SSMF; 5dBm; 16-QAM; 64GBd\end{tabular} & $3.84$       & $3.97$       & $3.87$      & $19.4$     & $8.1$      & $10$     & $7.48$      & $9.15$      & $7.65$     \\ \hline
\begin{tabular}[c]{@{}c@{}}Case iv) 10×60km; SSMF; 3dBm; 64-QAM; 30GBd\end{tabular}  & $5.68$       & $5.98$       & $5.92$      & $10.4$     & $5$        & $7.9$     & $7.1$      & $12.75$     & $8.95$     \\ \hline
\end{tabular}}

    \label{Table_Metrics}
\end{table*}

Finally, since the goal of this work is to demonstrate possible pitfalls and overestimation scenarios of the NN-based equalizers, we tried to use the same architecture and hyperparameters throughout the paper. In some sections, those parameters are altered for some specific purpose that we will be clearly highlighted. In general, when using the multilayer perceptron (MLP), we considered three hidden layers with $[N_1=481/N_2=31/N_3=263]$ neurons in each layer, respectively. When using the bidirectional long-short-term memory (biLSTM) equalizer, the number of hidden units ($N_h$) was set to 226. Both NN models are illustrated in Fig.~\ref{fig:NN_MODELS}. The standard number of taps used was $N = 25$: this is the maximal memory size estimation for the scenarios that we will address. The memory effect is important because, even though we compensate for the effect of CD electronically, we still have to mitigate the impact of the coupling between the nonlinearity and the CD, which occurs along with the fiber transmission. Since the aim of this work is not to propose a new NN equalizer, we do not focus on optimizing the NN structure for different transmission setups~\cite{freire2021performance}. In contrast, we used the same NN structure in different scenarios. Nonetheless, the chose the NN structure to be complex enough to be capable of dealing with the different levels of nonlinearity while allowing us to clearly demonstrate various caveats and pitfalls that can occur in the coherent optical channel equalization task\footnote{As discussed in~\cite{freire2020complex,freire2021performance}, the step of hyperparameter optimization is crucial on the process of designing a NN equalizer with good performance, and to tackle this we use and suggest the usage of the Bayesian Optimization tool in~\cite{klein2017fast}, which is more efficient than other traditional power-hunger techniques.}.

\section{Quality of Transmission Metrics} \label{Sec:QOT}

There exists a variety of performance indicators that can be used to assess the quality of $M$-ary QAM transmission systems: bit error rate (BER), mutual information (MI), Q-factor, signal-to-noise ratio (SNR), effective SNR, and error vector magnitude (EVM), to mention the most used examples.
 
The ultimate quality of transmission (QoT) metric in digital communications is the post-FEC BER or the Q-factor\footnote{BER entirely defines the Q-factor in this case, and we use it to employ dB-s instead of a linear scale.}, because all real transmissions occur with close to 0 post-FEC BER. The MI can also be a highly valuable metric because it estimates the achievable spectral efficiency when matched with the post-FEC BER. When the decoder is not available (e.g. when measuring channel equalization using received soft symbols), the post-FEC metric is often omitted in favor of the pre-FEC metric, which is a common performance measure for an uncoded system.

The pre-FEC BER can vary depending on the decision technique (hard or soft decision, HD and SD) utilized after the equalization of the soft symbols. This metric accurately predicts the post-FEC BER for HD-FEC with optimal interleaving. Adopting such a metric, however, can result in an incorrect spectral efficiency estimate, which is especially evident at low code rates (see~\cite{Alex01} for details). However, when dealing with NN equalizers, the pre-FEC BER/Q is commonly used because obtaining the post-FEC BER/Q would require integrating the equalized symbols into the rest of the DSP chain, which is not worth the work for initial performance evaluations.
The SNR and EVM are the least accurate QoT measurements. However, these two measurements can also be utilized to provide a qualitative comparison of different transmission regimes. It is worth noting that these metrics are related to channels with known statistics, such as the additive white Gaussian noise (AWGN) channel. The correlations between the different QoT metrics in nonlinear channels, on the other hand, are not always known, and adopting extrapolations based on linear theories should be done with caution.

\begin{figure*}[t!]
   \subfloat[\label{a1} $(9.4\%/3.12 \frac{\mathrm{bits}}{\mathrm{symbol}} / 7.87\mathrm{dB})$]{%
      \includegraphics[ width=0.3\textwidth]{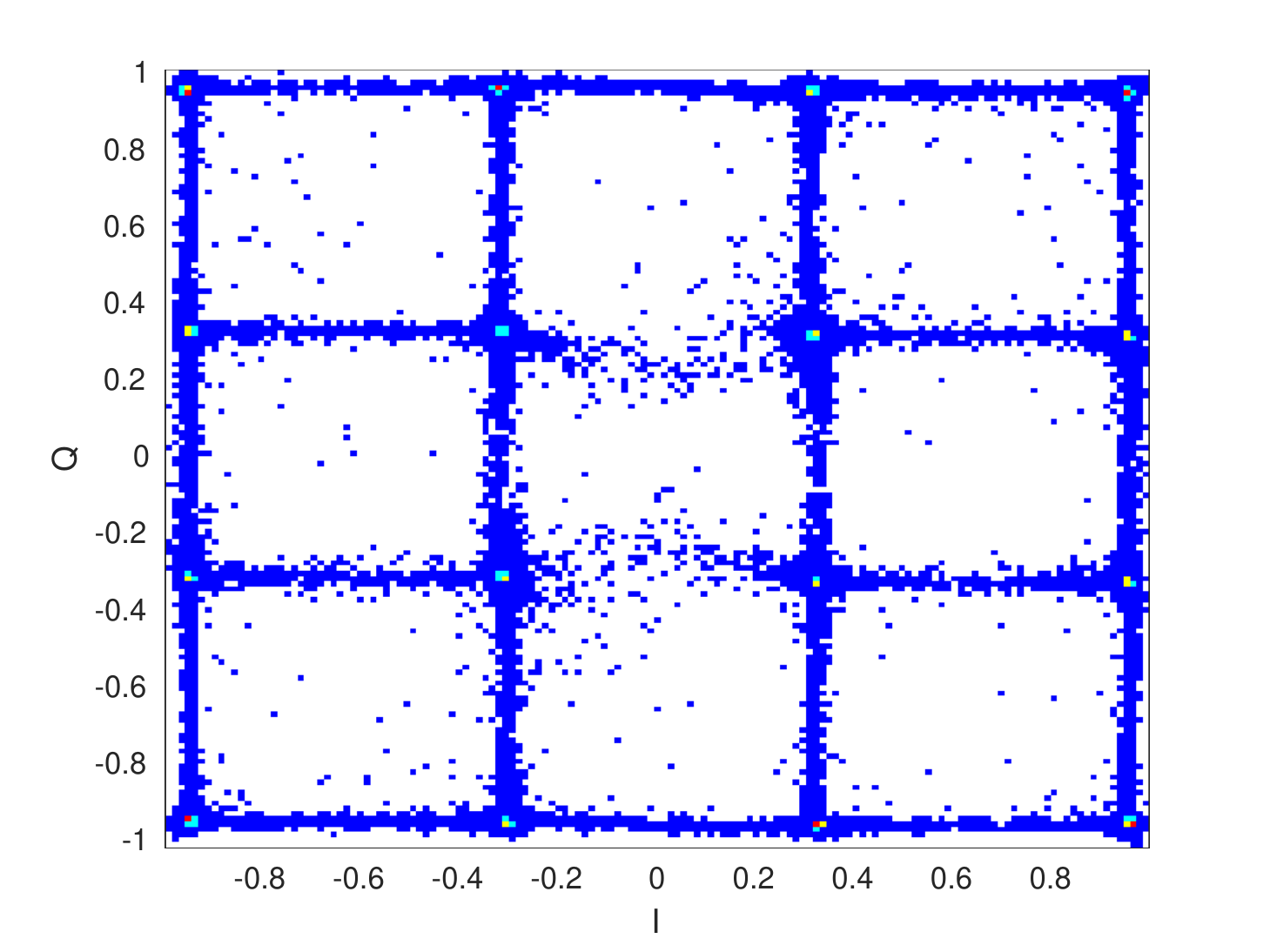}
      }
\hspace{\fill}
   \subfloat[\label{a2} $(9.8\%/3.25 \frac{\mathrm{bits}}{\mathrm{symbol}} / 7.89\mathrm{dB})$]{%
      \includegraphics[width=0.3\textwidth]{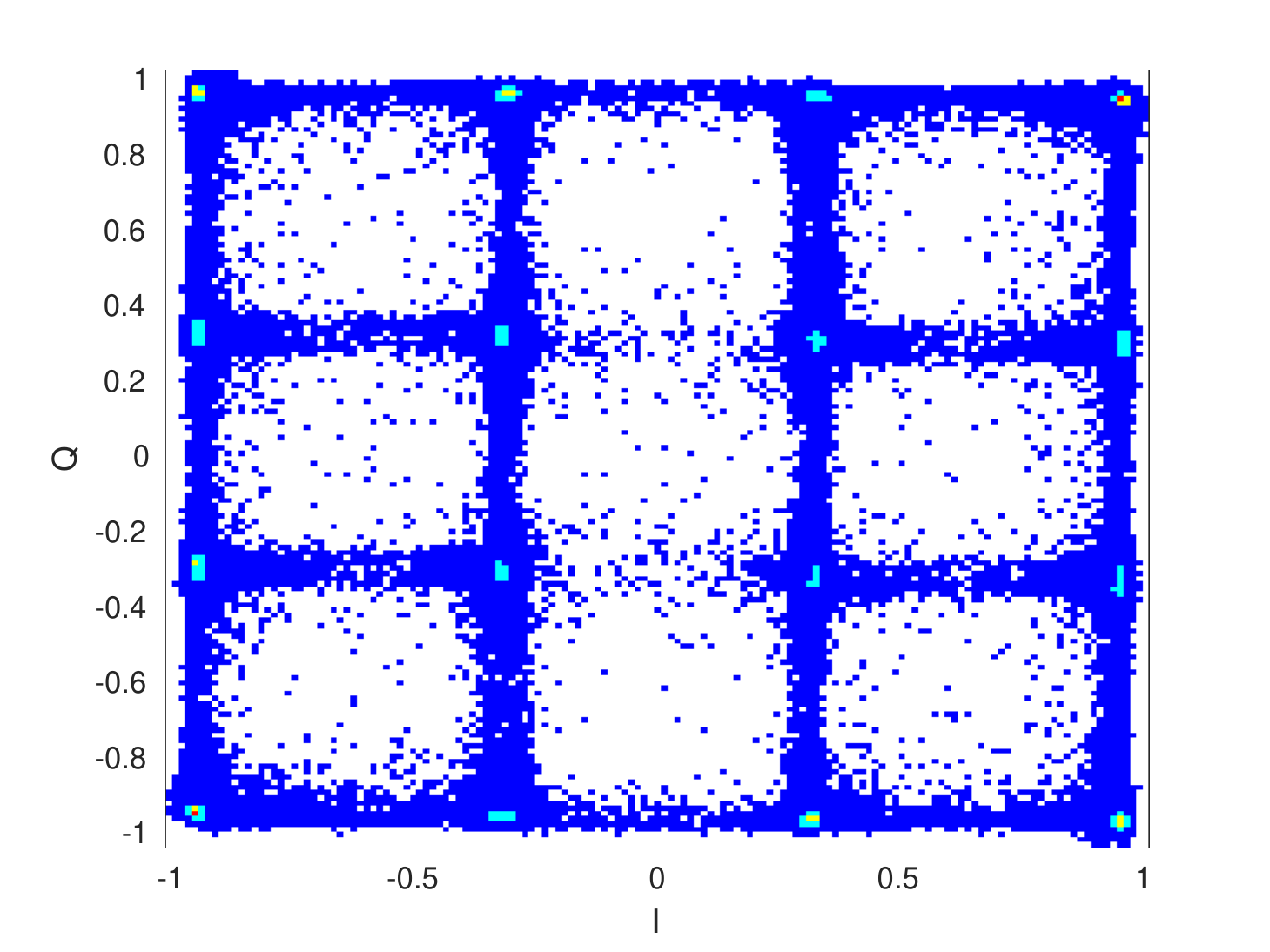}}
\hspace{\fill}
   \subfloat[\label{a3} $(17.27\%/3.89 \frac{\mathrm{bits}}{\mathrm{symbol}} / 7.84\mathrm{dB})$]{%
      \includegraphics[ width=0.3\textwidth]{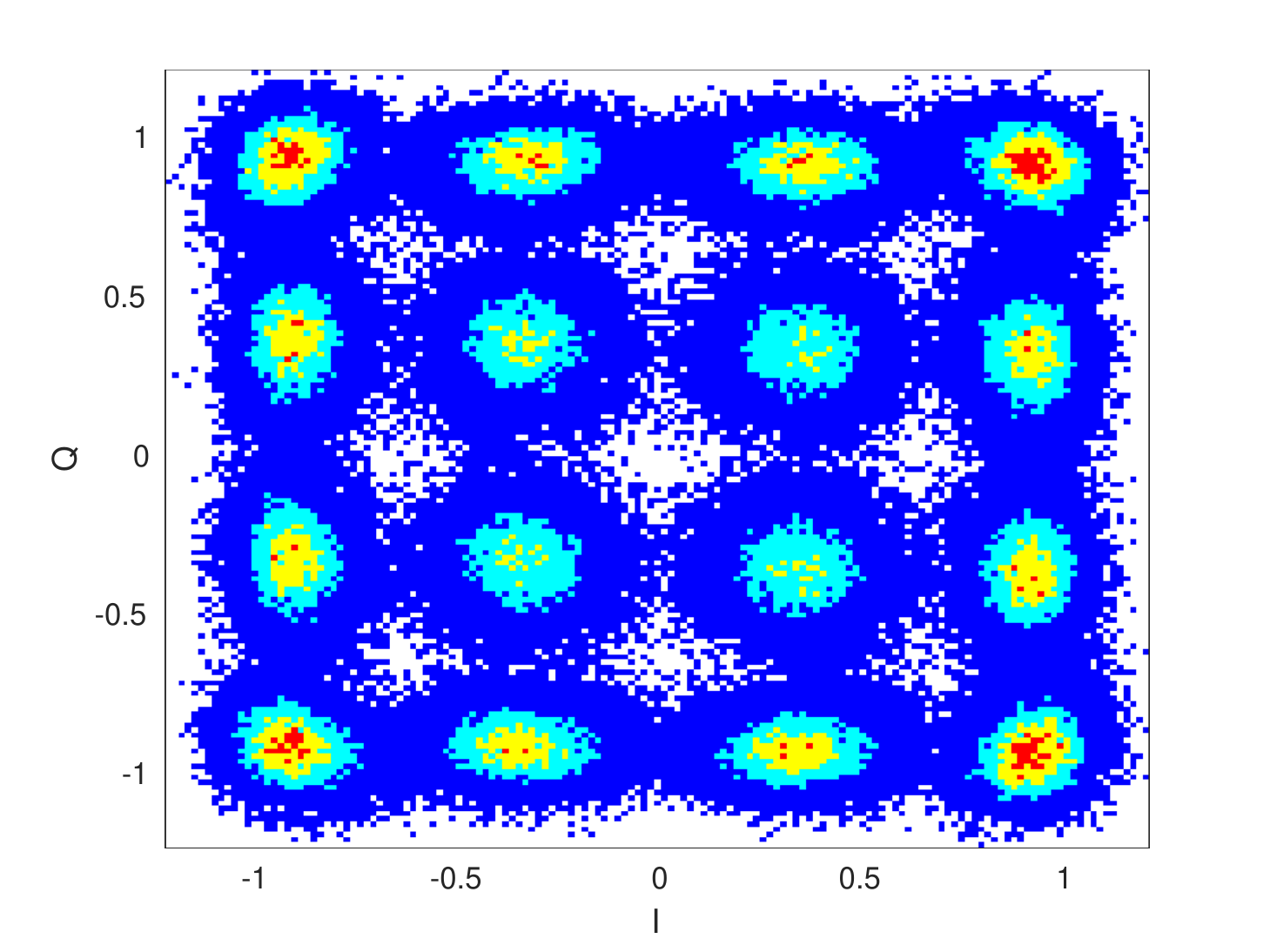}}
   \vskip\baselineskip
      \subfloat[\label{b1} $(7.9\%/5.91 \frac{\mathrm{bits}}{\mathrm{symbol}} / 8.94\mathrm{dB})$]{%
      \includegraphics[ width=0.3\textwidth]{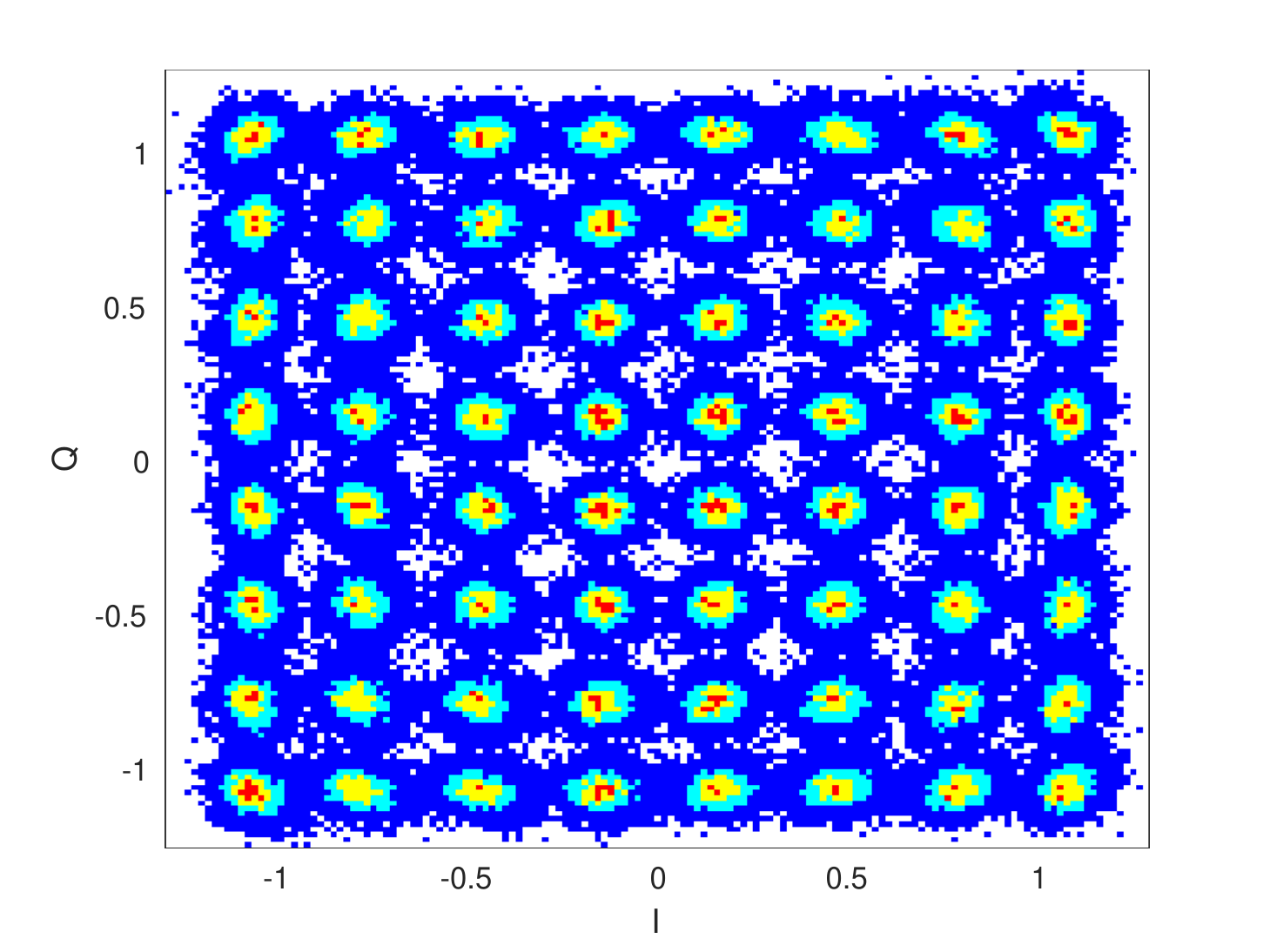}}
\hspace{\fill}
   \subfloat[\label{b2} $(7.93\%/5.91 \frac{\mathrm{bits}}{\mathrm{symbol}} / 8.95\mathrm{dB})$]{%
      \includegraphics[width=0.3\textwidth]{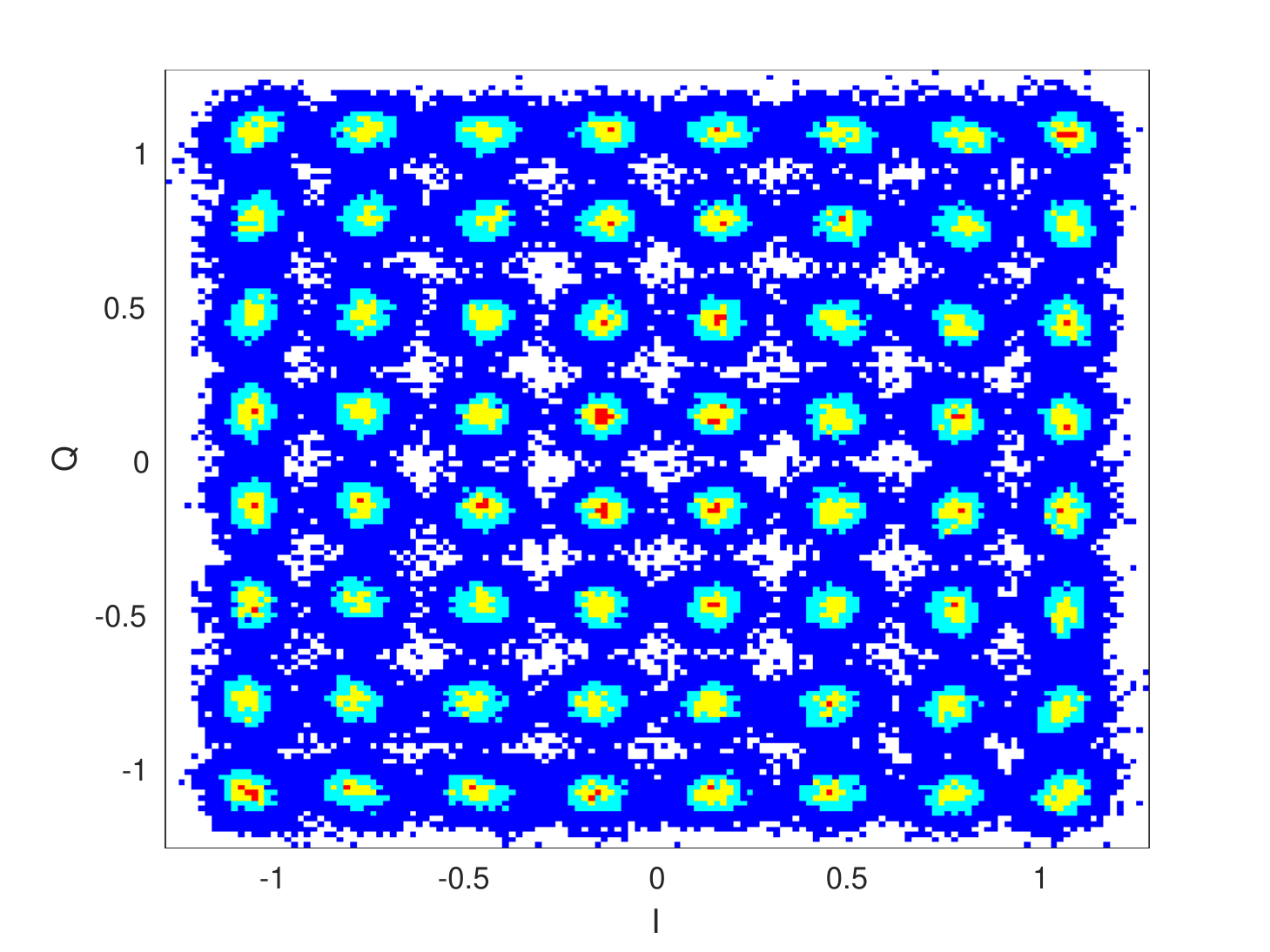}}
\hspace{\fill}
   \subfloat[\label{b3}$(7.92\%/5.92 \frac{\mathrm{bits}}{\mathrm{symbol}} / 8.94\mathrm{dB})$]{%
      \includegraphics[ width=0.3\textwidth]{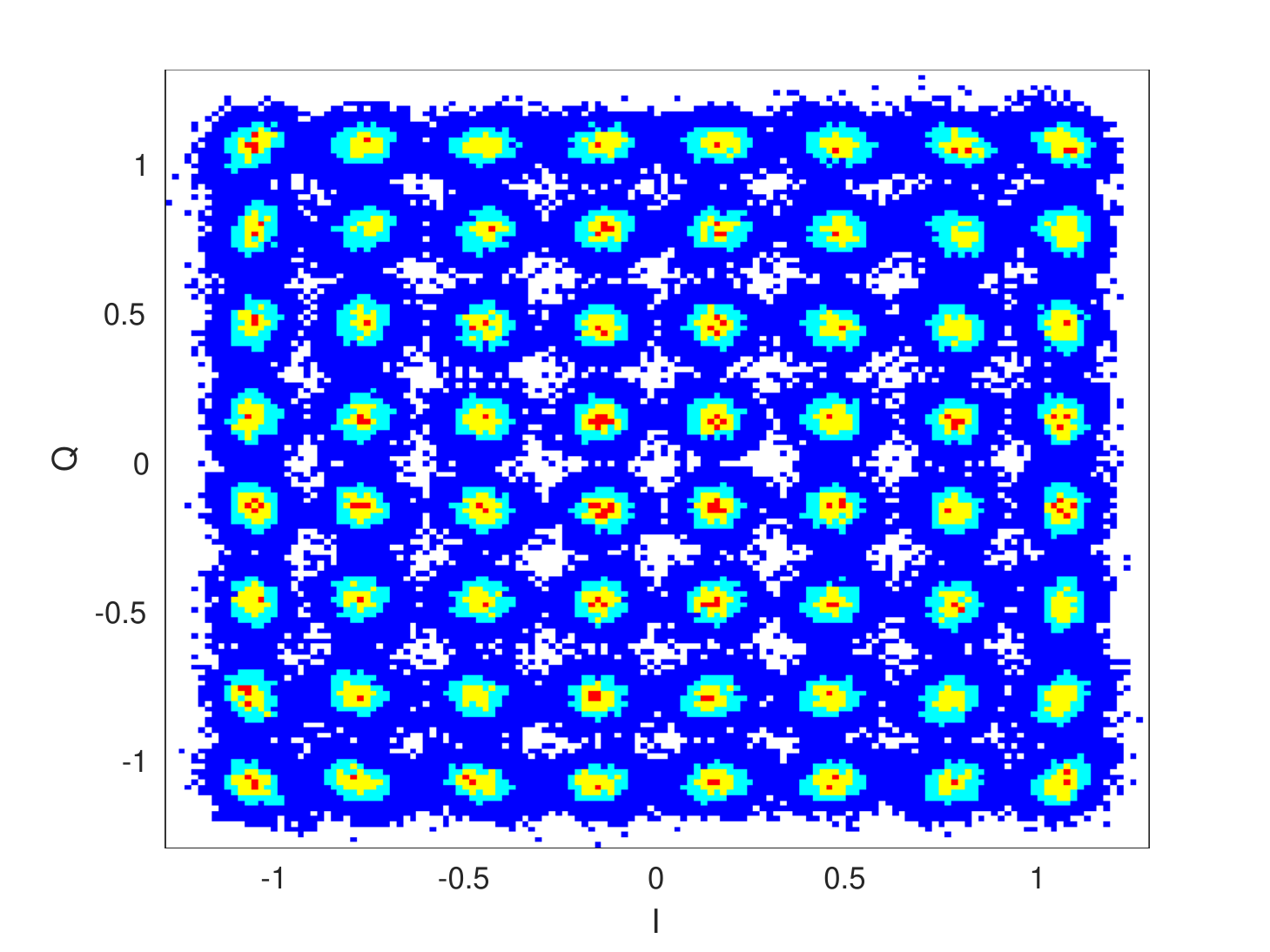}
      }
\caption{Constellation diagrams after the MLP-based equalization, corresponding to the best performance when using the different QoT metrics. (a) and (d) lowest EVM; (b) and (e) best Q-factor; (c) and (f) highest MI. The simulated transmission scenarios are: $15\times100$~km, 4~dBm, 28~GBd 16-QAM (a)--(c); $10\times60$~km, 3~dBm, 30~GBd, 64-QAM (d)--(f). In each constellation, the QoT metrics were presented as (EVM/ MI/ Q-factor).}
\label{Fig_constellation_QOT}
\end{figure*}

Although the pre-FEC BER is an important QoT metric, it does require the transmission of a known pattern, such as a training sequence, through the system for continuous performance monitoring\footnote{Note that in practical terms, the pre-FEC BER is usually derived from the post-FEC BER assuming that the FEC can correct all errors.}. As a result, the effective SNR (ESNR) and EVM gained popularity insofar as they lent themselves well to the study of unknown symbol sequences. Furthermore, as previously stated, these QoT metrics can still provide an accurate estimation of the BER when the system errors are primarily caused by optical AWGN, i.e., when fiber transmission is close to the linear regime and nonlinear effects (arising in TX/RX components and non-additive noise) are almost negligible. For such metrics, it is inherently assumed that the reception is non-data-aided and that a quadratic-QAM signal constellation is used. The ESNR, EVM, and Q-factor can be calculated using the following expressions that are valid for a Gaussian-distributed signal~\cite{mckinley2004evm,schmogrow2011error,freude2012quality, shafik2006extended}:
 \begin{equation}
    \textrm{EVM}_{RMS} = \left[\frac{1/N \sum^{N}_{n=1} |y_n -x_n|^2}{1/N \sum^{N}_{n=1} |x_n|^2}\right]^{\frac{1}{2}} \! \! \!,
 \end{equation}
 \begin{equation}
     \textrm{SNR} \approx \left[\frac{1}{\textrm{EVM}_{RMS}}\right]^{2} \! \! \! ,
 \end{equation}
 \begin{equation}\label{Eq.Q-factor}
     Q = \sqrt{2}~\textrm{erfc}^{-1}(2 \textrm{BER}) ,
 \end{equation}
where $y_n$ is the normalized $n$-th symbol in the stream of measured symbols, $x_n$ is the ideal normalized constellation point of the $n$-th symbol (i.e. a symbol from the $M$-QAM alphabet), $N$ is the total number of symbols in the constellation, and $\text{erfc}^{-1}$ is the inverse complementary error function. We typically consider these metrics in dB, using the relation: $T[\textrm{dB}] = 20 \log_{10}(T)$, where $T$ is the QoT metric under investigation.

From a statistical perspective, the BER depends on the particular decision mechanism utilized at the receiver side, whereas the MI gives the effective transmission capacity regardless of the decision process. However, because we work with the received soft symbols, the MI cannot be conveyed explicitly. One option to derive the MI value is to estimate it by assuming a single-input single-output AWGN channel, which yields a suboptimal estimate giving out the MI's lower bound. This lower bound to the MI, $I(X;Y)$, can be expressed as~\cite{eriksson2017characterization, eriksson2015four, catuogno2018non}:
\begin{equation} \label{MI_equation}
    I(X;Y) =  \EX \!  \left[  \log_2\left( \frac{p(y|x_k)}{\sum^{M \! F-1}_{i=0}p(i)p(y|x_k)} \right)  \! \right]\! ,\!
\end{equation}
where $p(i)$ is the probability distribution of each $k$-th QAM alphabet symbol, and $p(y|x_k)$ defines the conditional probability of the received constellations given the $k$-th QAM input symbol. Then we can use the multivariate Gaussian distribution estimator \cite{rousseeuw1999fast,kramer2016scikit} to calculate $p(y|x_k)$ of the transmitted-received complex symbols, which, ultimately, gives as the lower bound for the MI via Eq.~(\ref{MI_equation}).

Now, as we have established the framework for the most relevant QoT metrics and addressed the assumptions used in their computation, we will look at how those metrics may be used to evaluate the performance of the NN equalizer. The main purpose of this section is to raise awareness of the fact that, depending on the modulation format, the NN structure, and the level of noise, the NN can produce the effect of \textit{squeezing the constellations}: we name this effect the ``jail window'' pattern. We notice that numerous other works \cite{jarajreh2014artificial,schaedler2019deep,zhang2019functional,liu2018ols, liu2018effective, Jerart, Kotlyar:21, ming2021ultralow} have also reported the ``jail window'' type constellations after the NN-based equalization, often without paying attention to the true meaning and consequences of this phenomenon. We clarify this effect and explain possible related misinterpretations. The effect of ``jail window'' results from the regression task carried out by NN equalizers based on the MSE loss; the effect occurs because the ultimate goal of such NNs is to minimize the Euclidean distance between recovered and transmitted symbols. However, we emphasize that the ''jail window'' constellation forms \textit{violates the Gaussian channel assumption used in the computation of some metrics mentioned above}. Indeed, this effect reduces the accuracy of all Gaussian-assumption-based metrics (e.g., the ESNR) besides the Q-factor calculated directly from the BER obtained via direct error counting. Thus, when using these inaccurate Gaussian-assumption metrics, we can obtain false results indicating that the NN performs well while, in reality, the true gain provided by the ``jail window'' constellation is highly overestimated. The ``jail window'' effect can also be explained by the mismatch between the true transmission performance metric (the BER) and the metric that is minimized by the NN training (the MSE loss), such that we have a disagreement between the objective function and the actual NN result; however, the BER itself cannot be used as a NN loss function inasmuch as it is non-differentiable. Further investigation of the effect of the ``jail window" is given further in Section~\ref{sec:batch}.

To illustrate the metric-related problem, we have tested two types of equalizer, biLSTM and MLP, in four different numerical transmission setups: i) 20$\times$80~km SSMF when transmitting 8-QAM at 34.4~GBd; ii) 15$\times$100~km SSMF when transmitting 16-QAM at 28~GBd; iii) 30$\times$50~km SSMF when transmitting 16-QAM at 64~GBd; iv) 10$\times$60~km SSMF when transmitting 64-QAM at 30~GBd; the launch power was set to 3~dB higher than the power level for the best performance without NN equalizer. Clearly, when a nonlinear equalizer is employed, the optimal launch power should increase. The scenarios and NN-equalization results for different metrics are summarized in Table.~\ref{Table_Metrics}. 

Here, we note that Table~\ref{Table_Metrics} shows the best value of each QoT metric after running the process over 1000 epochs, but the best value did not occur at the same epoch number for all metrics. For example, in case ii), 15$\times$100~km SSMF with 16-QAM at 28~GBd, we observed that, for the epoch corresponding to the lowest EVM (9.4\%), the MI and Q-factor values were, respectively, 3.12 bits/symbol and 7.87~dB; for the epoch corresponding to the highest MI (3.89 bits/symbol), the EVM and Q-factor values were, respectively, 17.27\% and 7.84~dB; and for the epoch leading to the highest Q-factor (7.89~dB), the MI and EVM values were, respectively, 3.25 bits/symbol and 9.8\%. This result clearly highlights that the particular QoT metric selected for performance optimization \textit{does impact the eventual result for the system with equalizer} and that the differences can be even more accentuated if another transmission setup/ NN architecture is used. Indeed, depending on the quality metric, the constellation after the equalization changes. Fig.~\ref{Fig_constellation_QOT} shows the constellation corresponding to the maximum of each metric for cases ii) and iv), where, in the first case, the ``jail window'' pattern does appear, but in the second it does not. For case ii), we can see the ``jail window'' with thin lines connecting the constellation points for the best EVM. The same ``jail window'', but with somewhat thicker lines, can be observed in the case leading to the best Q-factor. Contrarily, the traditional Gaussian-type constellations are observed in the case leading to the best MI. For case iv), the NN did not generate the ``jail window'' pattern. In this case, the minimum in the EVM and the maximum in the MI and Q-factor were achieved at approximately the same epoch. It is important to note that the Q-factor in all three cases did not change significantly, but the improvement in the EVM with the ``jail window'' caused a high decrease in the MI estimation from 3.89 bits/symbol (the epoch leading to the best MI) down to 3.12 bits/symbol (the epoch leading to the best EVM). This result is the consequence of the non-Gaussian shape of the constellation obtained, which violates the applicability conditions of Eq.~(\ref{MI_equation}) used to measure the MI. Furthermore, Fig.~\ref{Fig_constellation_QOT} points to one of the repercussions of the ``jail window'' pattern: the NN continued to minimize the Euclidean distance between the prediction and the labels, but this process no longer decreases the BER, and so we departed from the main goal of equalization. Our observation was that after the epoch of the highest MI (Fig.~\ref{Fig_constellation_QOT} (c)), the NN began to converge to the ``jail window'' and stopped further improving the Q-factor, which is why Fig.~\ref{Fig_constellation_QOT} (a) and (c) have roughly the same BER / Q-factor but Fig.~\ref{Fig_constellation_QOT} (a) has a much smaller EVM than Fig.~\ref{Fig_constellation_QOT} (c).

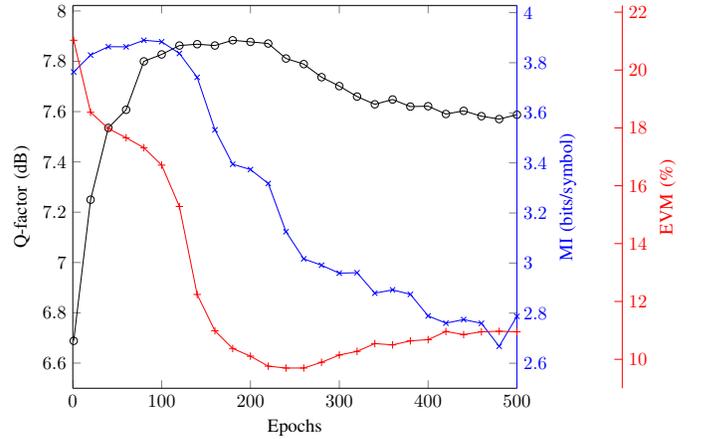
\begin{figure}
    \centering
    \begin{tikzpicture}[scale =0.7]
 \pgfplotsset{every axis/.style={ymin=0}}
\begin{axis}[ scale only axis, xmin=0,xmax=500,ymin=6.5, axis y line*=left, xlabel=Epochs, ylabel= Q-factor (dB)] 
    \addplot[black, mark=o, draw] coordinates {
(1,6.68845501922480)(20,7.24974907298051)
(40,7.53521842841758)(60,7.60726963006983)
(80,7.79972349478278)(100,7.827308812812049)
(120,7.8626511493709)(140,7.86801241681208)
(160,7.8627412609684)(180,7.88364327717876)
(200,7.87714982244067)(220,	7.87068119312296)
(240,7.81111686916215)(260,7.78874663847014)
(280,7.73686778729198)(300,7.70137839017115)
(320,7.65989665224997)(340,	7.62901197490569)
(360,7.64817388897688)(380,	7.61997285202092)
(400,7.62152811329577)(420,	7.59036649441872)
(440,7.60295215256797)(460,	7.58182085236657)
(480,7.57028642727748)(500,	7.58761533540844)
    };
\end{axis}
\begin{axis}[blue, scale only axis, xmin=0,xmax=500,ymin=2.5, axis y line*=right, axis x line=none, ylabel= MI (bits/symbol) ]
    \addplot[blue, mark=x] coordinates {
(1,3.76299080269063)(20,3.83060948410661)
(40,3.86403259354261)(60,3.86344614387967)
(80,3.88971958638048)(100,3.88373315030330)
(120,3.83714336884747)(140,3.74175532504123)
(160,3.53196551841602)(180,3.39471531855698)
(200,3.37355142656119)(220,	3.31781623958562)
(240,3.12507352882730)(260,3.01674966273786)
(280,2.99126525893307)(300,2.95956220527901)
(320,2.96154677914929)(340,	2.87953307530238)
(360,2.89285181754312)(380,	2.87503655468939)
(400,2.78895262357922)(420,	2.76002685057378)
(440,2.77424868208471)(460,	2.75937756597852)
(480,2.66757765384557)(500,	2.78717364467958)

    };
\end{axis}
\begin{axis}[red, scale only axis, xmin=0,xmax=500 ,ymin=9 , axis y line*=right, axis x line=none, ylabel= EVM (\%)]
\pgfplotsset{every outer y axis line/.style={xshift=2cm}, every tick/.style={xshift=2cm}, every y tick label/.style={xshift=2cm} }
    \addplot[red ,mark=+] coordinates {
    (1,21.0290143470631)(20,18.5398223076965)
    (40,17.9774603785965)(60,17.6598750143717)
    (80,17.3151557327763)(100,16.7135856531226)
    (120,15.2825980425308)(140,12.2407129924269)
    (160,10.9843730024594)(180,10.3741036515722)
    (200,10.1078619520657)(220,9.76243345739911)
    (240,9.69817885988436)(260,9.69903059638985)
    (280,9.89667755274356)(300,10.1478411243438)
    (320,10.2737232301297)(340,10.5433507249692)
    (360,10.4988148617355)(380,10.6375708842544)
    (400,10.6802173722950)(420,10.9611258541815)
    (440,10.8547663633067)(460,10.9503709402940)
    (480,10.9696041949233)(500,10.9501337720337)
    };
\end{axis} 

\end{tikzpicture}
    \caption{Evolution of Q-factor (simulations), MI, and EVM over the training epochs when using the MLP equalizer in the $15\times100$~km, 4~dBm, 28~GBd 16-QAM system. }
    \label{fig:Evolution_QoT}
\end{figure}

This effect shows that the usage of QoT metrics mentioned above for the regression task can be misleading, resulting in an underestimation of the true achievable MI and converging to a non-optimal equalizer's structure. Ultimately, in Fig~\ref{fig:Evolution_QoT} we present the evolution of those three metrics over the epochs for case ii), to visualize the previous statements, where we can see that the overall behavior and the best values are different for different metrics. 

Table~\ref{Table_Metrics} also demonstrates that the NN equalizer may lead to significant improvement of the EVM when compared to the linear equalization only, but without rendering the same corresponding improvement in the Q-factor. To better illustrate this effect, we can use the following expression to estimate the BER from EVM after the equalization \cite{fatadin2016estimation}:
\begin{equation} \label{estiamteBER}
  \textrm{BER} = \kappa \frac{1-M^{-1/2}}{1/2 \log_2(M)} \, \textrm{erfc} \! \left[\sqrt{\frac{3/2}{(M-1)\textrm{EVM}_{RMS}^2}}\, \right],
\end{equation} 
where $M$ is the cardinality of the modulation format and $\kappa$ is the correction factor. Now, we estimate how much the result obtained through Eq.~(\ref{estiamteBER}) deviates from the one calculated through the direct bit error counting. We start by using the EVM and BER before the NN to calculate the correction factor $\kappa$: for case ii), the correction factor is $\kappa = 1.076$, which means that the BER calculated via (\ref{estiamteBER}) using the reference value of the EVM (before the NN equalization) is a suitable QoT estimator, almost matching the true BER value after the respective conversion. However, this equation shows how overestimated the EVM can be if the ``jail window'' is present after the NN equalization. For the MLP equalizer, by using Eq.~(\ref{estiamteBER}) for case ii), we obtain a Q-factor estimate of 13.62~dB while, in reality, it is only 7.89~dB. Additionally, in case iii), the Q-factor estimated using Eq.~(\ref{estiamteBER}) is 13~dB, whereas the true Q-factor is just 7.65~dB. However, in case iv), where the ``jail window'' is absent (see Fig.~\ref{Fig_constellation_QOT}) the Q-factor estimated through the EVM is equal to 9.4~dB, whereas the true one is 9.2~dB, showing a good match. Interestingly, even in cases where the ``jail window'' is absent, the Q-factor calculated through the EVM can be considerably overestimated. For instance, in case ii) and using the biLSTM equalizer, the Q-factor estimated through the EVM is 15.6~dB, whereas the true value is 10.7~dB.
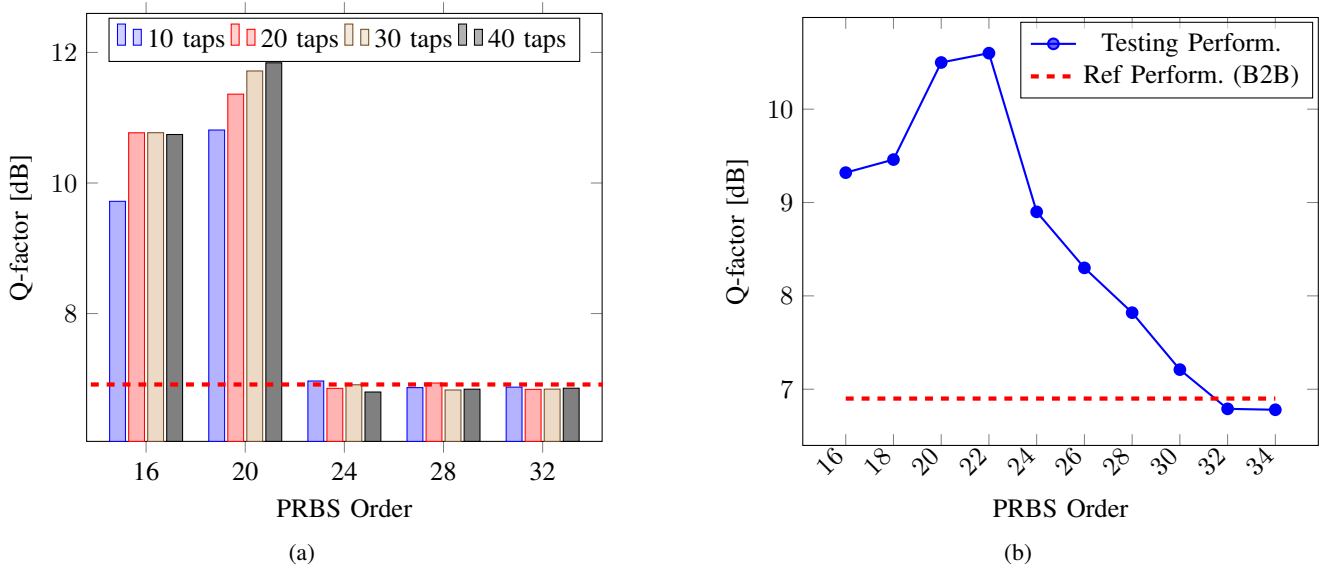
\begin{figure*}[t!]
   \begin{subfigure}[b]{0.475\textwidth}
            \centering
    \begin{tikzpicture}
\begin{axis}[
    ybar,
    enlargelimits=0.15,
    ylabel={Q-factor [dB]},
    xlabel={PRBS Order},
    symbolic x coords={ 16, 20, 24, 28, 32},
     ybar=1.2pt,
     bar width=6pt,
    legend style={at={(0.5,0.99)} ,fill=white, fill opacity=0.6, draw opacity=1,text opacity=1,
    anchor=north,legend columns=-1},
    nodes near coords align={vertical},
    ]

\addplot coordinates {(16, 9.7195) (20, 10.8115) (24, 6.966925234) (28, 	6.865596405	) (32,  6.873085086)};
\addplot coordinates {(16, 10.7693) (20, 11.36106) (24, 6.854409877) (28, 6.935843741) (32, 6.838656526)};		
\addplot coordinates {(16, 10.77042) (20, 11.71604) (24, 6.909002322) (28, 6.829440729) (32,6.842353348)};		
\addplot coordinates {(16, 10.74287) (20, 11.84105) (24, 6.798382844) (28, 6.841428577) (32, 6.856270449)};
\draw [red] ({rel axis cs:0,0}|-{axis cs: 16, 20, 24, 28, 32}) -- ({rel axis cs:1,0}|-{axis cs:  16, 20, 24, 28, 32}) node [pos=0.33, above] {KPI};
\legend{10 taps, 20 taps, 30 taps, 40 taps}
\draw[dashed,ultra thick, red](-2cm,0.1cm) -- (7cm,0.1cm);%
\end{axis}
\end{tikzpicture}
    \caption{}
       
    \end{subfigure}
      \hfill
       \begin{subfigure}[b]{0.475\textwidth}
            \centering
    \begin{tikzpicture}
\begin{axis}[legend style={at={(0.99,0.99)} ,fill=white, fill opacity=0.6, draw opacity=1,text opacity=1,,anchor=north east},
    ylabel={Q-factor [dB]},
    xlabel={PRBS Order},
xtick={16,18,20,22,24,26,28,30,32,34},
x tick label style={rotate=45,anchor=east}]
\addlegendentry{Testing Perform.}
\addplot[mark=*,thick,blue] coordinates {
(16,9.32) (18,9.46) (20,10.5) (22,10.6) (24,8.9)((26,8.3)
(28,7.82)(30,7.21)(32,6.79)(34,6.78)

};

\addlegendentry{Ref Perform. (B2B)}
\addplot[dashed,ultra thick, red] coordinates {
(16,6.9) (20,6.9) (21,6.9) (22,6.9) (23,6.9) (24,6.9)  (28,6.9)(29,6.9)(30,6.9)(31,6.9)(32,6.9)(33,6.9)(34,6.9)
};

\end{axis}
\end{tikzpicture}
    \caption{ }
        
    \end{subfigure}
        \caption{Q-factor's dependencies on the PRBS order for (a) the different values of input memory (the different number of taps is highlighted with different bar's color) and different random seeds for training and testing with dataset sizes of $2^{18}$ symbols; (b) for the same random seed for training and testing, but taken from different chunks, with using 40 taps and different training dataset sizes per PRBS order as indicated in Table \ref{Table_PRBS}. The red dashed lines for both panels show the threshold, over which we arrive at the overestimation issue. (Simulation results)}\label{Fig:PRBS_Results}
\end{figure*}

Finally, the results in Table~\ref{Table_Metrics} also show that when using the biLSTM equalizer, the ``jail window'' phenomenon occurs rarer compared to when we use the MLP equalizer. However, it can still persist for the biLSTM in some scenarios, as will be pointed out later in section~\ref{sec:batch}. Therefore, we can conclude that to avoid the overestimation of the system's performance, the results should be presented in terms of BER or Q-factor derived through Eq.~(\ref{Eq.Q-factor}). Otherwise, the constellations after equalization should be assessed to ensure that it is still approximately Gaussian so that the other QoT metrics provide a good approximation of the true performance.

\section{PRBS Order Impact in Systems' Performance}

There has been a plethora of empirical evidence that deep enough NNs can memorize random labels, even when using considerably large datasets \cite{zhang2021understanding}. Thus, in principle, by sufficiently increasing the size of a NN, we can always reduce the training error to small enough values, even though the task of learning a completely random sequence is meaningless.

Unfortunately, when transmitting the commonly-used PRBS (say, of orders 7, 9, 11, 15, 20, and 23), the benefits rendered by the NN-based equalization can be overestimated. Indeed, the NN may learn the PRBS generation rules themselves, instead of estimating the inverse of the transmission channel model, and this naturally results in a sharp equalizer's performance degradation when truly random data (say, obtained from live traffic) is transmitted. This overfitting effect is particularly relevant because the NN-based equalizers are often trained using PRBS-based datasets, even in experiments.

When dealing with NNs, two key PRBS-related issues must be addressed. First, the PRBS has a periodicity that is determined in terms of symbols by the PRBS order and the modulation format cardinality. The order 20 PRBS, for example, has a period of $2^{20}-1$ bits, and if we use a 64-QAM signal with 6 bits per symbol representation, the symbol periodicity will be around 174k symbols\footnote{The true period of symbols will be 6 times longer in this case, because this is when the bit sequence is aligned to the symbol boundaries. However, because the NN maps the input symbols onto a multidimensional space, the NN may also trace the information on bit periodicity. As a result, we define symbol periodicity as the number of symbols that contain the entire bit periodicity.}. To avoid NN learning such a pattern in this circumstance, the size of the training dataset used to train the NN must be less than the aforementioned quantity. This topic was investigated in Ref.~\cite{eriksson2017applying}, where an MLP classifier was used to check the overestimation of PRBSs of orders 7 and 15 in the 4-pulse amplitude modulation (PAM) transmission system using a large training dataset of size $2^{19}$. According to the findings of that work, the NN was able to produce a reasonable result when tested on another PRBS sequence, but when tested on a fully random signal, the system's performance degraded dramatically because the NN did not learn the channel equalization but instead learned the PRBS periodicity.

A very simple and popular approach for the generation of PRBS is to use a linear feedback shift register (LFSR) with particular initialization and feedback. This approach can lead to a serious limitation since, if the NN's input is broad enough to catch the whole inputs used by the LFSR to construct the current symbol (typically its size is equal to the memory), the NN will be able to learn the PRBS model properties instead of performing the genuine nonlinearity mitigation task. More recently, in Ref.~\cite{yang2020overfitting}, this issue was mathematically studied and tested for an MLP classifier for both on-off keying (OOK) and 4-PAM transmission. In that Ref., the authors used the PRBS of order 20 and generated $2^{19}$ bits, a sequence that is shorter than the generator periodicity. The bits were converted into symbols and fed into the NN classifier. The authors then created an input with the neighbor symbols but removed the current symbol to be classified. If the current symbol cannot be learned from adjacent symbols, the NN cannot correctly decide which bit was transmitted, and this situation corresponds to $\text{BER}= 0.5$. Otherwise, the BER should be significantly lower than 0.5, indicating that the NN can understand the symbol generation and mapping rules. In Ref.~\cite{yang2020overfitting}, the authors found that, in the case of OOK signals, the input memory length of 17 was enough to recover the symbols. In the case of transmission of 4-PAM signals, the same behavior was observed once the input had 9 taps. In the case of 4-PAM, some fewer taps were needed because each symbol carries 2 bits, whereas the OOK signals carry only 1 bit per symbol. In order to avoid the issues resulting from the limited PRBS length, instead of LSFR, the Mersenne Twister random sequence (MTRS) generator should be used, which can provide the sequences with a much longer period than that of the LFSR.

In this section, differently from the two previous works~\cite{eriksson2017applying,yang2020overfitting}, we evaluate both aforementioned issues using the recurrent equalizer (biLSTM). The biLSTM's likelihood of learning the deterministic time correlation due to the PRBS order is by far superior to that of the MLP equalizer. Additionally, we consider the transmission of a 64-QAM modulation format, which increases the number of bits per symbol, therefore enhancing the PRBS-related problems in the NN-based equalization. 

We generated data from 10 transmission runs with PRBS orders 16, 18, 20, 22, 24, 26, 28, 30, 32, and 34. AWGN was added to the RX input in a back-to-back scenario, so that all data sets after the hard decision had the same Q-factor (6.9~dB). Since the only source of signal degradation is a random noise coming from the AWGN added, the NN should not provide any performance improvement, as the NN is a nonlinear deterministic function. Hence, any Q-factor improvement when employing the NN results from the NN's learning of the PRBS generation rule. 

We perform two tests to evaluate the impact of the PRBS order on the performance of NN. First, we study the impact of the training data set and PRBS symbol periodicity. Fig.~\ref{Fig:PRBS_Results} shows the Q-factor as a function of the PRBS order when $2^{18}$ symbols are used to train the NN. A different seed is used to generate another sequence of $2^{18}$ symbols (with the same PRBS order), which are used to test the NN. The analysis of Fig.~\ref{Fig:PRBS_Results} shows a clear improvement in the Q-factor for PRBS orders 16 and 20. This result confirms that the NN can learn the PRBS periodicity when the training data sets ($2^{18}$ symbols) are larger than the symbol periodicity. Moreover, Fig.~\ref{Fig:PRBS_Results} also shows that increasing the number of taps enables achieving an even better Q-factor (the NN could train faster); this behavior was also observed in Ref.~\cite{yang2020overfitting}. For the PRBS orders higher than 24, the training data set becomes smaller than the symbol periodicity, and, consequently, the NN is no longer capable of learning the PRBS generation rules: the values drop below the threshold (the dashed line).

\begin{table}[htbp] 
  \centering
  \caption{Summary of parameters such as periodicity and training/testing dataset size per PRBS order used in our study with 64-QAM.}
  \resizebox{0.5\textwidth}{!}{
\begin{tabular}{|c|c|c|c|c|}
\hline
PRBS Order & \begin{tabular}[c]{@{}c@{}}Periodicity\\ Bits\end{tabular} & \begin{tabular}[c]{@{}c@{}}Periodicity\\ Symbols\end{tabular} & \begin{tabular}[c]{@{}c@{}}Training\\ Dataset Size\end{tabular} & \begin{tabular}[c]{@{}c@{}}Testing\\ Dataset Size\end{tabular} \\ \hline\hline
$16$       & $2^{16}-1$                                                 & $ \approx 10k$                                  & $5k$                                                            & $5k$                                                            \\ \hline
$18$       & $2^{18}-1$                                                 & $ \approx43k$                                 & $20k$                                                           & $20k$   
              \\ \hline
$20$       & $2^{20}-1$                                                 & $ \approx174k$                                 & $87k$                                                           & $87k$   
\\ \hline
$22$       & $2^{22}-1$                                                 & $ \approx699k$                                 & $262k$                                                           & $262k$   
              \\ \hline
$24$       & $2^{24}-1$                                                 & $ \approx2.79M$                                & $1M$                                                          & $1M$                               
\\ \hline
$26$       & $2^{26}-1$                                                 & $ \approx11.18M$                                & $1M$                                                           & $1M$                               
\\ \hline
$28$       & $2^{28}-1$                                                 & $ \approx44.7M$                                & $1M$                                                          & $1M$                                                          \\ \hline
$30$       & $2^{30}-1$                                                 & $ \approx178.9M$                                & $1M$                                                          & $1M$                                                          \\ \hline
$32$       & $2^{32}-1$                                                 &  $ \approx 715M$                                 & $1M$                                                           & $1M$                                                          \\ \hline
$34$       & $2^{34}-1$                                                 &  $ \approx 2.8B$                                 & $1M$                                                           & $1M$                                                          \\ \hline
\end{tabular}
}
\label{Table_PRBS}
\end{table}
\begin{figure*}[t!]
    \centering
    \includegraphics[width=0.85\textwidth]{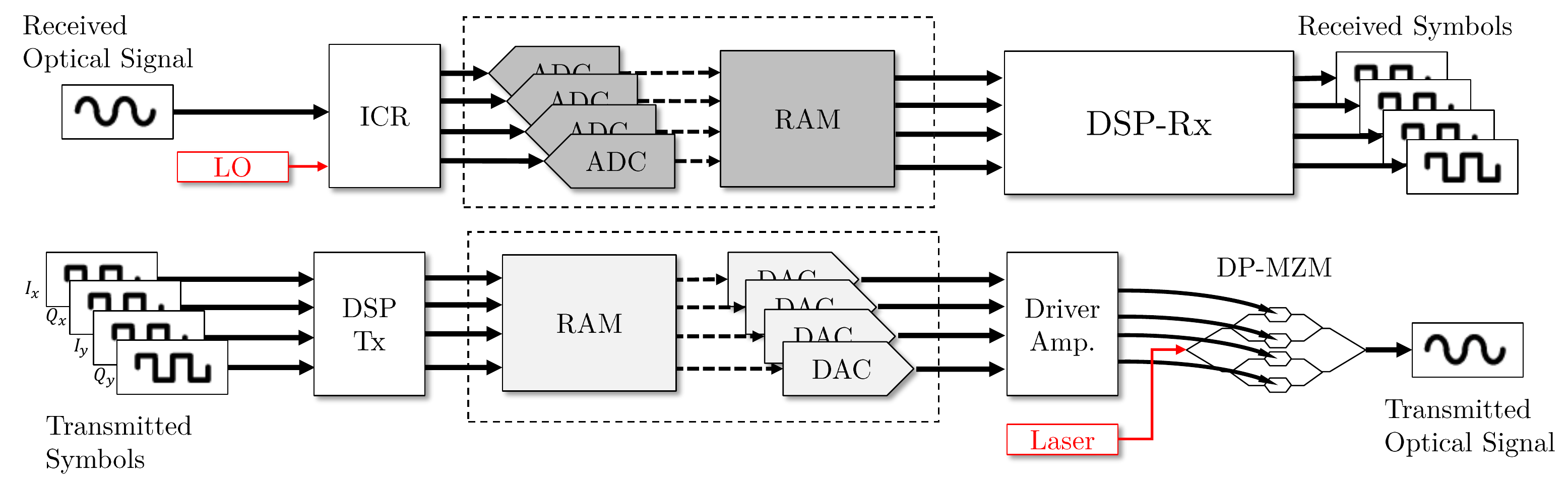}
    \caption{A simplified block diagram illustrating the components of a typical transmitter and receiver. The DAC/ADC RAM memory is highlighted to emphasise its role in the transmission chain.}
    \label{Fig:DAC_Model}
\end{figure*}

At this point, it is worthwhile to question for which dataset size the NN would learn the data generation rule for the higher-order PRBS. To address this issue, we have conducted a second round of tests, where we trained the NN using up to $1M$ symbols for the PRBS orders ranging from 16 to 34. The main results of this study are reported in Fig.~\ref{Fig:PRBS_Results}(b). Note that to guarantee that the NN learns the PRBS generation rule instead of the symbol periodicity, we generated the training and testing data sets with the same seed, but for training and testing, we selected different data blocks with lengths shorter than the PRBS periodicity. Table~\ref{Table_PRBS} provides the periodicity for each PRBS order, and the training/testing data set sizes for our study. The number of taps was set equal to 40 and the NN equalizer was trained for more than 5000 epochs. As before, if the Q-factor increases above the reference (red line in Fig.~\ref{Fig:PRBS_Results} (b) is the B2B level), this indicates that the NN was able to learn the data generation rule. Fig.~\ref{Fig:PRBS_Results} (b) shows that the Q-factor increases with PRBS order up to 22. This behavior can be explained by the fact that PRBS orders 16 and 18 were trained with 5k and 20k symbols, respectively. This training data set is too small to fully train the NN, but we cannot increase it further because of the periodicity constraint given in the third column of Table~\ref{Table_PRBS}. 
 
When the PRBS order is increased, the amount of training data required to learn the respective PRBS generation rule also increases. However, increasing the size of the training data beyond $1M$ becomes impractical (the training takes too long). Thus, the Q-factor curve in Fig.~\ref{Fig:PRBS_Results} (b) starts decreasing for PRBS orders above 22, indicating that the NN's capacity to recover the PRBS generation rule becomes progressively worse. In this case, the $1M$ training data size is insufficient (and/or the NN complexity is insufficient to learn the PRBS generation rule of the highest order). Thus, when increasing the PRBS order from 24 to 30, we observe a decrease in the Q-factor until the point where the NN completely ceases to learn the PRBS generation rule (for the PRBS orders equal to or higher than 32).
 
We emphasize again that the MTRS should always be used in simulations, instead of the LSFR, because it renders virtually infinite PRBS lengths. Therefore, an undesirable system performance overestimation can be avoided. When dealing with experimental data, even larger PRBS orders (e.g. $\geq$ 32) should be used with caution, since, depending on the training dataset size, modulation format, and input memory, overestimation can still occur therein.

\section{DAC/ADC Memory Impact on Training NN Equalizers}

As described in the previous section, the two most influential factors related to the data quality are the dataset size and its variability (the absence of spurious periodicity, bias, etc). Fig.~\ref{Fig:DAC_Model} shows the typical layout of a lab-style optical transmitter and a coherent receiver. The Tx-DSP writes the signal samples into the DAC memory, and the signal is transmitted cyclically. On the Rx side, the output of the ADC is written into memory and processed subsequently by the Rx-DSP. When dealing with experimental data, there are several factors to take into account in order to ensure proper data quality for the NN training. First, the DAC and ADC memory is clearly limited in size, which puts an upper limit on the length of the signal to be transmitted. This applies primarily to capturing the buffers built into real-time transponders, where the memory resource is very precious and, hence, the available memory is usually rather limited. However, the same argument applies, in principle, to lab equipment, where, however, the limits are not so stringent. The other factors to consider are the sampling frequency of DAC and ADC devices and the symbol rate of the transmitted signal. For the NN training, we are interested in the sequences of symbols or bits; for a given DAC memory, the number of symbols the DAC can hold depends on the sampling frequency and symbol rate. Finally, an additional constraint can emerge due to the Rx DSP architecture. Often, to achieve synchronization, the DSP implementation assumes that the data are composed of frames of equal length. Furthermore, all frames may be assumed to contain the same payload data to facilitate the alignment of received and reference sequences. In this case, only the data obtained from a single received frame can be taken up for the NN's training and testing. In this section, we will look into how the aforementioned properties impact the training of NN-based equalizers and data variability quality.

\begin{figure}[htbp]
    \centering
    \includegraphics[width=0.45\textwidth]{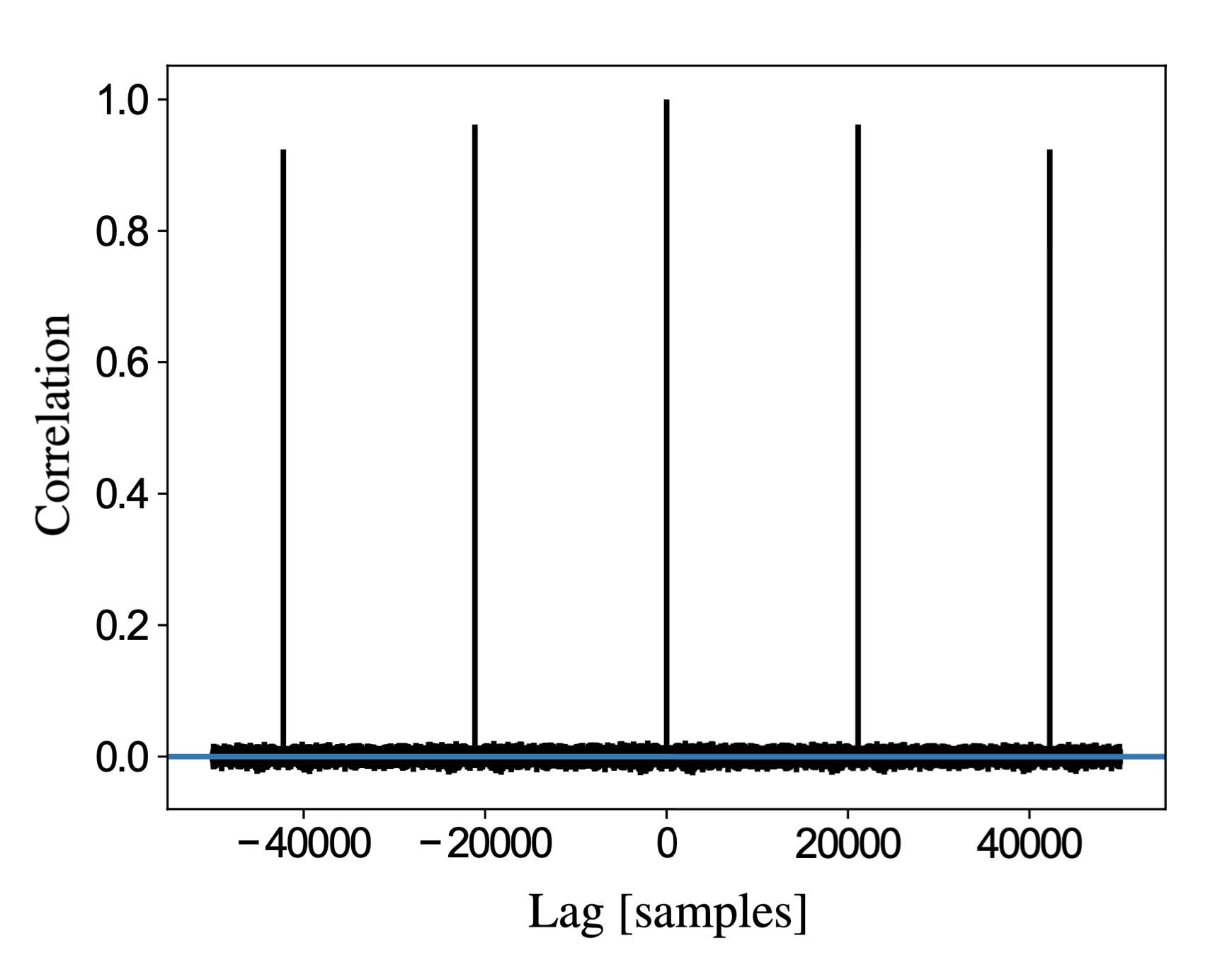}
    \caption{The auto-correlation diagram of the received experimental signal shows the impact of the DAC memory on the signal's periodicity.}
    \label{Fig:ACorr}
\end{figure}

Consider an exemplary scenario where we have a DAC / ADC with a memory equal to 512k samples per channel, operating at the sampling frequency of 80~GSample/s. Assume that the DSP requires about 10 frames holding identical data to achieve a proper synchronization and evaluate BER. Then, the number of samples per frame is around 52k samples. Let us further assume that our transmission symbol rate is 34.4~GBd. This leads to the number of effective symbols 52k/(80/34.4) $\approx$ 22k, which are available for NN training. This can be easily verified by applying the autocorrelation function for the received symbols: Fig.~\ref{Fig:ACorr} shows the autocorrelation of the experimental data that was produced with a DAC specification close to the ones mentioned above. As it can be seen, the difference in the peaks is around 22k, which is the same value that we calculated.

\begin{figure}[htbp]
    \centering
 \begin{subfigure}{.49\linewidth}
  \centering
    \includegraphics[width=\linewidth]{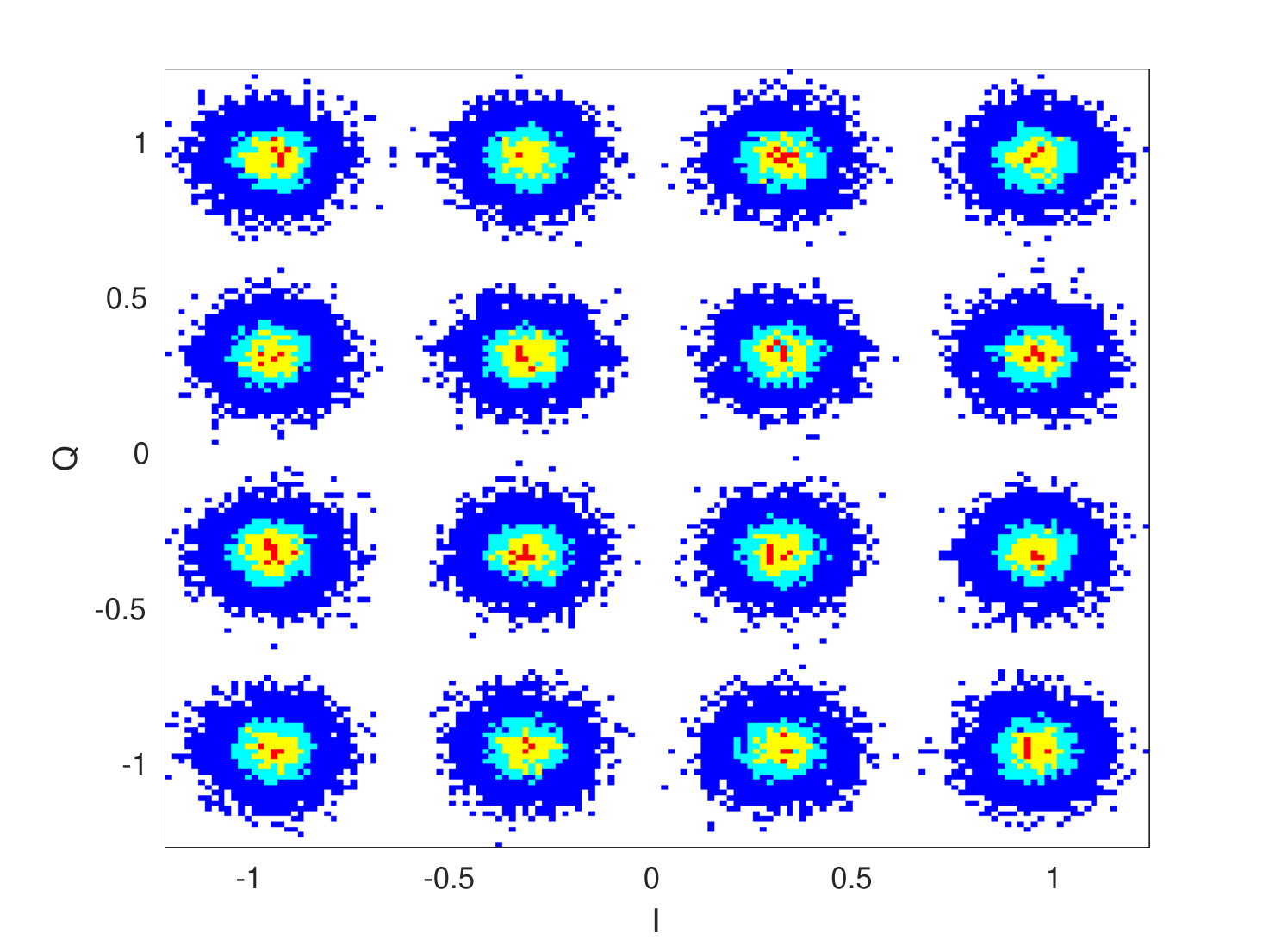}
    \caption{Same random seed}
    \label{fig:Same random seed}
\end{subfigure}
 \begin{subfigure}{.49\linewidth}
  \centering
    \includegraphics[width=\linewidth]{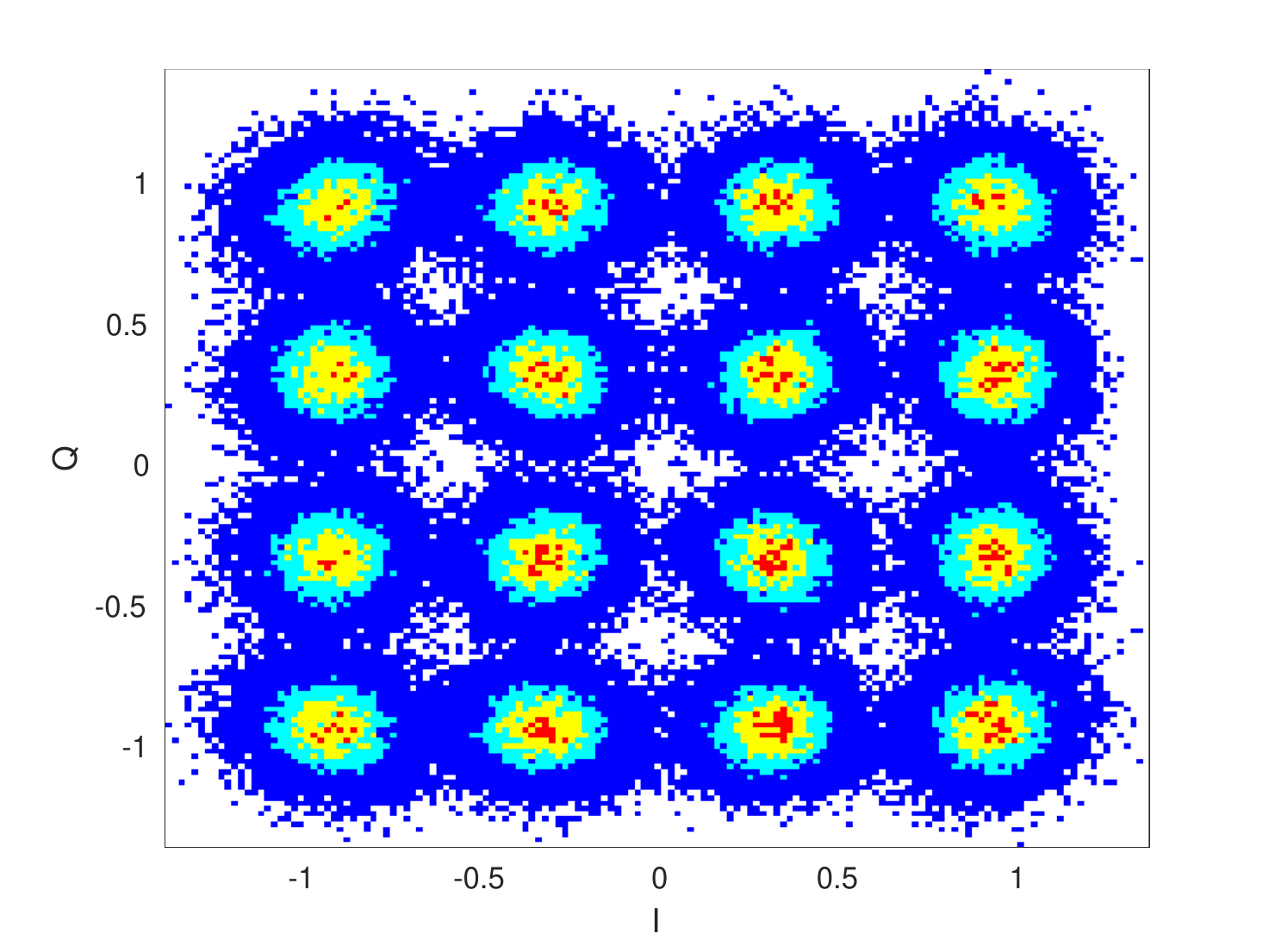}
    \caption{Different random seed}
    \label{fig:Different random seed}
\end{subfigure}
  \caption{16-QAM signal constellation after the NN-based equalization when (a) training and testing are carried out using different datasets but generated with the same random seed, and when (b) training and testing are run using datasets generated with different random seeds.} 
    \label{Fig:Overstimation}
\end{figure}

With this information, two main concerns can be raised in terms of the use of machine learning in systems employing DAC/ADC. First, our having a system that is unintentionally biased towards one subset of symbols can result in poor model performance when the latter is validated on a different subset. This fact can ultimately lead to the overfitting of the NN-based equalizer. Second, as shown in Fig.~\ref{Fig:ACorr}, even when we use a PRBS of order 32 or even completely random data, we cannot remove the periodicity in the data since the DAC will repeat just a portion of the PRBS. Because of this, \textit{the same transmission trace cannot be used for training and testing} even if we select non-overlapping chunks. In Fig.~\ref{Fig:Overstimation}, we show the constellations after NN equalization using different chunks of the same transmission trace, Fig.~\ref{Fig:Overstimation}(a), and when using the transmission traces with different random seeds, Fig.~\ref{Fig:Overstimation}(b). From these constellations, it is evident that using the same random seed to train and test the NN provides a constellation of outstanding quality (the respective Q-factor is equal to 13~dB). Nevertheless, the NN trained with the same random seed (the one that produced the picture in Fig.~\ref{Fig:Overstimation}(a)) cannot generalize to perform the channel equalization: when we use it with another random seed, we see that the resulting equalized constellation is noticeably degraded, and the NN's true performance in terms of Q-factor is just 8.66~dB, well below the previous overestimated value. This fact clearly demonstrates that, in this scenario, the system has overfitted over the training sequence pattern. Therefore, when reporting results using the same random seed for training and testing, you may overestimate the performance of your equalizer. 

To avoid overfitting, we have generated our NN inputs with the following procedure. First, we record 60 measurement traces, each with $2^{18}$ symbols, using a different random seed to generate each trace. After that, we split the entire data set into two parts: 50 traces were taken for training purposes and the remaining 10 traces were saved for testing. From these traces, we then generated the window vector inputs for each symbol to be recovered. Afterward, we concatenated all these window vectors to generate our training and testing datasets. Finally, after having $\approx 13$M window vectors in the training and $\approx 2$M window vectors in the testing datasets, we select $2^{20}$ random input vectors from the overall $13M$ in each epoch while training the NN. To show the impact of not having enough variability in the training dataset, we compared the NN equalizer performance trained using our aforementioned dataset generation solution with the case where we trained the NN with $2^{20}$ vectors generated by just one random seed. To evaluate the performance of the equalizers, we tested both trained models with $2^{18}$ input vector taken from $2M$ (the testing set) that were never used in the training.

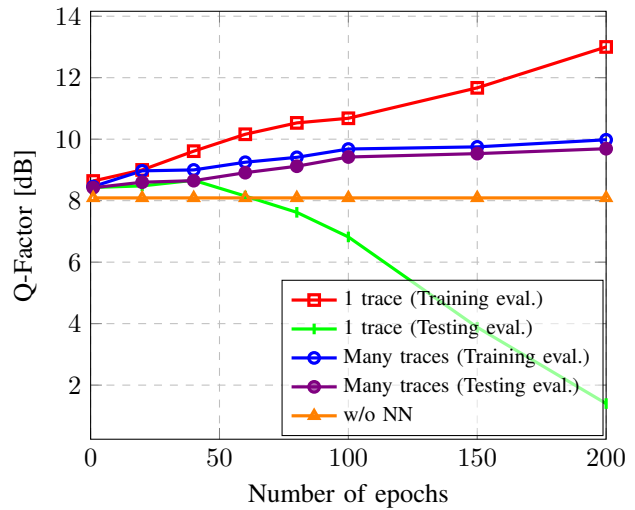
\begin{figure}[htbp]
    \centering
      \begin{tikzpicture}
    \begin{axis} [
        xlabel={Number of epochs},
        ylabel={Q-Factor [dB]},
        grid=both,        
    	xmin=0, xmax=200,
        legend style={legend style={ at={(0.995,0.19)},anchor= east}, legend cell align=left,fill=white, fill opacity=0.6, draw opacity=1,text opacity=1},
    	grid style={dashed}]
        ]
      \addplot[color=red, mark=square, very thick]
    coordinates {(1,8.64)(20,9)(40,9.61)(60,10.16)(80,10.53)(100,10.68)(150,11.67)(200,13)(250,13)(300,13)(350,13)(400,13)(450,13)(500,13)};
    \addlegendentry{\footnotesize{1 trace (Training eval.)}};
    
    \addplot[color=green, mark=|, very thick]
  coordinates {(1,8.43)(20,8.48)(40,8.663)(60,8.15)(80,7.62)(100,6.82)(150,3.88)(200,1.4)(250,8.663)(300,8.663)(350,8.663)(400,8.663)(450,8.663)(500,8.663)};
    \addlegendentry{\footnotesize{1 trace (Testing eval.)}};
    
    \addplot[color=blue, mark=o, very thick]   
    coordinates {(1,8.46)(20,8.97)(40,9)(60,9.25)(80,9.41)(100,9.68)(150,9.75)(200,9.98)(250,10.22)(300,10.27)};
    \addlegendentry{\footnotesize{Many traces (Training eval.)}};
    
    \addplot[color=violet, mark=*, very thick]     
    coordinates {(1,8.42)(20,8.6)(40,8.65)(60,8.91)(80,9.12)(100,9.42)(150,9.53)(200,9.69)(250,9.84)(300,9.78)};
    \addlegendentry{\footnotesize{Many traces (Testing eval.)}};

    \addplot[color=orange, mark=triangle, very thick]     
    coordinates {(1,8.09)(20,8.09)(40,8.09)(60,8.09)(80,8.09)(100,8.09)(150,8.09)(200,8.09)(250,8.09)(300,8.09)(350,8.09)(400,8.09)(450,8.09)(500,8.09)};
    \addlegendentry{\footnotesize{w/o NN}};
  \end{axis}
    \end{tikzpicture}
    \caption{Q-factor versus training epochs for the experiment with 16-QAM 5$\times$50~km SSMF, 34.4~GBd, 6~dBm power, using the biLSTM equalizer. These curves refer to the training and testing performance of the NN when using one trace or when using multiple traces (our solution is described in the main text) for training.}
    \label{Fig:Overstimation_q_factor}
\end{figure}

The results of our comparison are summarized in Fig.~\ref{Fig:Overstimation_q_factor}, and several conclusions can be readily drawn from that figure. First, for the case where we trained the NN with only one trace, the traditional overfitting appears: the training curve keeps growing while the testing curve bends down after some point as the model does not generalize. Because the model's variability was only for 22k points, it overfitted quickly, and the maximum Q-factor of the testing dataset was just 8.66~dB. On the other hand, when we used our multiple trace training solution, we see that both the training and testing datasets' curves grow simultaneously, indicating the generalization capability of the equalizer. Using our method, we were able to reach a maximum Q-factor of 9.69~dB using the testing dataset, which is almost 1~dB higher than the value obtained when training the NN with one single trace. Note that our solution uses the different parts of the training dataset, which benefits not only from the diversity of different symbols picked from different random seeds but also from the fact that noise (which is different for each trace) adds diversity to the dataset as well. Heuristically, we expect this noise to ``smear out'' each data point, making it difficult for the network to properly match individual data points, and therefore reducing the overfitting \cite{bishop1995neural}. However, as it has been observed in several previous works \cite{an1996effects,neelakantan2015adding,809097, 6796498, yin2015noisy, rifai2011adding,brown2003use }, the noise injection to various parts of the NN during the back-propagation training can remarkably improve NN's generalization capability, and the latter observation fully complies with the result achieved with our solution, Fig.~\ref{Fig:Overstimation_q_factor}. 

\section{Regression or Classification (Soft Demapping) NN Equalizers: The Design Dilemma}

First, we mention that the regression versus classification question in designing the most efficient NN equalizer structures\footnote{The receiver NN-based classifiers can be understood as soft-demappers implemented with the NNs.} was covered exhaustively in Ref.~\cite{freire2021neural}. Thus, in this section, we do not expose any specific pitfalls attributed to the types of predictive modeling used in the equalizers, referring an interested reader to the aforementioned Ref., but, instead, we review the drawbacks of using specifically the regression or the multi-class classification in the context of coherent optical channel equalization task.

For regression, the MSE loss function can be treated as a simplification of the true likelihood measurement for the case of optical channel equalization: using the MSE optimization, we assume that the channel noise distribution is additive Gaussian; if the noise distribution is non-Gaussian or signal-dependent, then the MSE-based optimization fails to capture all information in the distorted sequences~\cite{rady2011shannon}, and may not learn the model parameters that optimally maximize the negative log-likelihood that is the cross-entropy between the empirical distribution defined by the training set and the probability distribution defined by the model Ref.~\cite[Chapter 5.5]{Goodfellowbook2016}. In practice, when using MSE-based regression, we try to recover the deterministic part by assuming that the stochastic part has a Gaussian distribution with a signal-independent variance. This approximation is often reasonable, as in many systems the noise-signal interaction is smaller than the transmitter-induced and additive optical amplifier signal-independent noises, and the latter two can often be well approximated as a Gaussian process\cite{binh2019noises}.

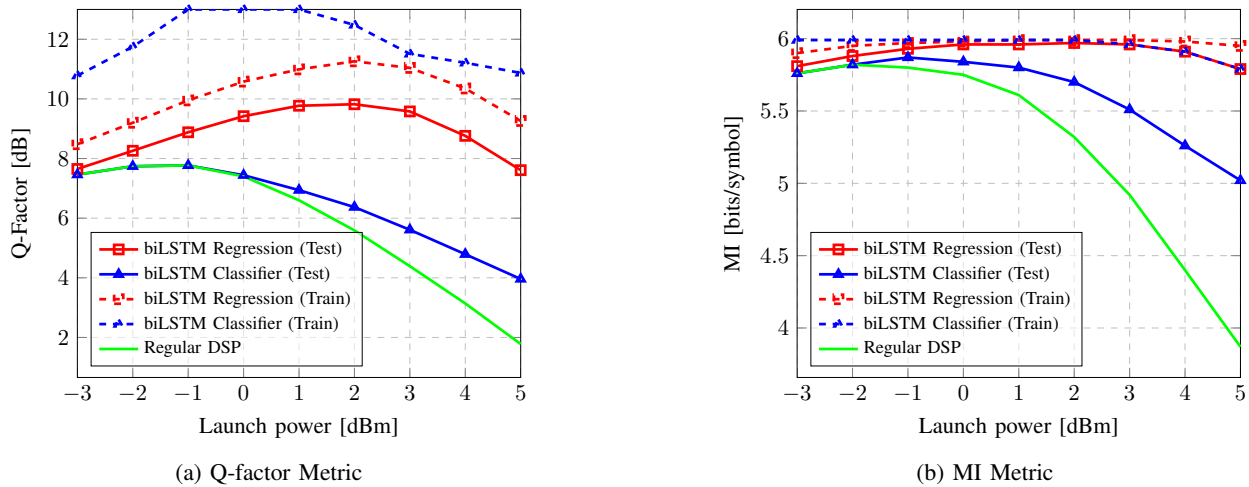
\begin{figure*}[t!] 
    \centering
        \centering
        \begin{subfigure}[b]{0.475\textwidth}
            \centering
                \begin{tikzpicture}[scale=0.86]
       \begin{axis} [
        xlabel={Launch power [dBm]},
        ylabel={Q-Factor [dB]},
        grid=both,   
    	xmin=-3, xmax=5,
    	xtick={-3, ..., 5},
    	ymax =13,
        legend style={legend pos=south west, legend cell align=left,fill=white, fill opacity=0.6, draw opacity=1,text opacity=1},
    	grid style={dashed}]
        ]
          \addplot[color=red, mark=square, very thick]
    coordinates {
         (-3,7.65)(-2,8.26)(-1,8.88)(0,9.42)(1,9.77)(2,9.82)(3,9.58)(4,8.76)(5,
7.61)
    };
    \addlegendentry{\footnotesize{biLSTM Regression (Test)}};

    \addplot[color=blue, mark=triangle, very thick]        coordinates {
         (-3,7.46)(-2,7.74)(-1,7.77)(0,7.44)(1,6.94)(2,6.37)(3,5.61)(4,4.79)(5,
3.96)
    };
    \addlegendentry{\footnotesize{biLSTM Classifier (Test)}};
    
              \addplot[color=red, mark=square,dashed, very thick]
    coordinates {
              (-3,8.48)(-2,9.20)(-1,9.95)(0,10.58)(1,10.99)(2,11.26)(3,11.04)(4,10.35)(5,9.26)
    };
    \addlegendentry{\footnotesize{biLSTM Regression (Train)}};

    \addplot[color=blue, mark=triangle, dashed, very thick]        coordinates {
         (-3,10.78)(-2,11.75)(-1,13)(0,13)(1,13)(2,12.48)(3,11.51)(4,11.21)(5,
10.87)
    };
    \addlegendentry{\footnotesize{biLSTM Classifier (Train)}};
    
    \addplot[color=green, very thick]         coordinates {
         (-3,7.46)(-2,7.74)(-1,7.77)(0,7.40)(1,6.60)(2,5.59)(3,4.39)(4,3.14)(5,
1.78)

    };
    \addlegendentry{\footnotesize{Regular DSP}};

    \end{axis}
    \end{tikzpicture}
    \caption{Q-factor Metric}
    \label{LSTM_Q} 
        \end{subfigure}
        \hfill
        \begin{subfigure}[b]{0.475\textwidth}
            \centering
                \begin{tikzpicture}[scale=0.86]
      \begin{axis} [
        xlabel={Launch power [dBm]},
        ylabel={MI [bits/symbol]},
        grid=both,   
    	xmin=-3, xmax=5,
    	xtick={-3, ..., 5},
        legend style={legend pos=south west, legend cell align=left,fill=white, fill opacity=0.6, draw opacity=1,text opacity=1},
    	grid style={dashed}]
        ]
          \addplot[color=red, mark=square, very thick]
    coordinates {
         (-3,5.81)(-2,5.88)(-1,  5.93)(0,5.96)(1,5.96)(2,5.97)(3,5.96)(4,5.91)(5,5.79)
    };
    \addlegendentry{\footnotesize{biLSTM Regression (Test)}};

    \addplot[color=blue, mark=triangle, very thick]        coordinates {
          (-3,5.76)(-2,5.82)(-1,5.87)(0,5.84)(1,5.8)(2,5.7)(3,5.51)(4,5.26)(5,
5.02)
    };
    \addlegendentry{\footnotesize{biLSTM Classifier (Test)}};
    
              \addplot[color=red, mark=square,dashed, very thick]
    coordinates {
              (-3,5.90)(-2,5.95)(-1,5.97)(0,5.98)(1,5.99)(2,5.99)(3,5.99)(4,5.9805)(5,5.95)
    };
    \addlegendentry{\footnotesize{biLSTM Regression (Train)}};

    \addplot[color=blue, mark=triangle, dashed, very thick]        coordinates {
         (-3,5.99)(-2,5.99)(-1,5.99)(0,5.99)(1,5.99)(2,5.99)(3,5.96)(4,5.91)(5,5.79)
    };
    \addlegendentry{\footnotesize{biLSTM Classifier (Train)}};

    \addplot[color=green, very thick]         coordinates {
         (-3,5.76)(-2,5.82)(-1,5.80)(0,5.75)(1,5.61)(2,5.32)(3,4.92)(4,4.40)(5,
3.87)

    };
    \addlegendentry{\footnotesize{Regular DSP}};

    \end{axis}
    \end{tikzpicture}
    \caption{MI Metric}
    \label{LSTM_MI} 
        \end{subfigure}
  \caption{Performance metrics' comparison for the regression- and classification-based equalizers, showing the impact of overfitting that can be seen when comparing the training and testing learning curves. The analyzed setup considered a 20$\times$50 km SSMF link and a single-carrier-DP 64-QAM signal with 30 GBd and 0.1 RRC pulse shape (simulation results).}
  \label{fig:classvsregression} 
\end{figure*}

For the classification, the cross-entropy loss (CEL) -- the most common loss function used in the classification tasks \cite{zhang2018generalized} -- is the most suitable loss function concerning its meaning in information theory\cite{Georg2021book}, effectively representing any type of noise statistics. However, there are two major drawbacks associated with this loss function, emerging specifically from the machine learning-related perspective, which can make the training of such a classifier a troublesome task. First, regardless of the corresponding inaccuracy in the target space, the CEL penalizes the misclassification between the two classes (i.e. between any two constellation points in our problem) with the same ``cost'' value: the penalization ignores the spatial proximity of the labels, reflecting only the fact that the constellation point has been misclassified. However, typically, the account of the misclassification ``type'', i.e. when the cost of all errors is not equal, can be (and typically is) quite beneficial for the efficient NN's training. The cost of making a mistake can be determined by the projected and actual classes of an example \cite{pazzani1994reducing, gutierrez2015ordinal}. Each class represents a distinct notion that can be identified using a NN in the conventional classification task, for example, when we classify different kinds of coordinate objects. However, when it comes to the problem of optical equalization, each class contains more information than just a label. In other words, the different classes correspond to different point positions in the constellation, and each class has its nonlinear distortion level. The additional information about each label can be obtained by using the physical nature of each nonlinear level, i.e., the aforementioned distortion level that depends on the constellation point's power.

To better understand this question, consider the task of classifying symbols in, say, a 16-QAM constellation. In this problem, the misclassification between classes that share the same decision boundary should cost less than the misclassification of the ones that do not share any decision boundary: in the first case, one naturally shares some symbols due to the noise-induced clouds' spreading and overlap, while the second case is a ``serious error''. However, using the CEL, we will only capture errors on the target class: it discards any notion of errors that you might consider ``false positive'' and does not care how predicted probabilities are distributed other than the predicted probability of the true class (since we deal with one-hot encoded vectors), implying that only the predicted probability associated with the label influences the value of the CEL. Thus, from the standpoint of machine learning, we can say that there is a natural ordering among the labels of the target variable. The training difficulties that the standard CEL can bring in classification with natural ordering problems are discussed in more detail in Ref.~\cite{lathuiliere2019comprehensive}.

The second classification disadvantage is the magnitude of the gradients that arise during the training process with the CEL. The CEL surfaces, according to Ref.~\cite{bosman2020visualising}, present fewer local minima than the square error-based losses (SEL), the type to which the MSE loss belongs. The CEL, on the other hand, has stronger gradients than the SEL, which leads to a stronger tendency of overfitting in CEL-trained systems, leading to the SEL's having a better generalization property in almost all scenarios examined in Ref.~\cite{bosman2020visualising}. This result was explained thereby assuming that the CEL loss surface is more prone to sharp minima (narrow valleys) than the SEL's surface, making the overfitting considerably ``easier'' to happen for the former systems. Additionally, it was also shown that such classification-based systems can suffer from the gradient vanishing problem. For instance, Ref.~\cite{9662308} shows the gradient vanishing when using the softmax with categorical-CEL, and the same for the sigmoid with the binary-CEL was reported in Ref.~\cite{end_to_end_MSE}.
We observed the same tendency in our coherent channel equalization problem, when the MSE-based system generalized better than the categorical CEL systems, due to overfitting, sharp local minimal, and ultimately the gradient vanishing problem that is observed in the CEL-based learning.

Finally, we can point out one more disadvantage of classification-based equalizers, which is more relevant to their use in real transmission systems. When utilizing the classification, the equalization must have a predefined number of outputs that correspond to the constellation's cardinality. This means that the classifier model's operation is dependent on the modulation format on which it was trained. In other words, a classifier's practical implementation (e.g., in hardware) would limit the device's applicability to a particular modulation format only, which greatly reduces the device's flexibility and adaptability. However, as shown in Refs.~\cite{freire2021transfer}, when we use regression, we can easily adapt the model to work with different modulation formats. As a result, regression-based models are much more flexible (reconfigurable) than classification models, allowing us to use the former under conditions different from the ones in which the regression models were trained.

\begin{figure}[ht!]
\centering
    \includegraphics[width=\linewidth]{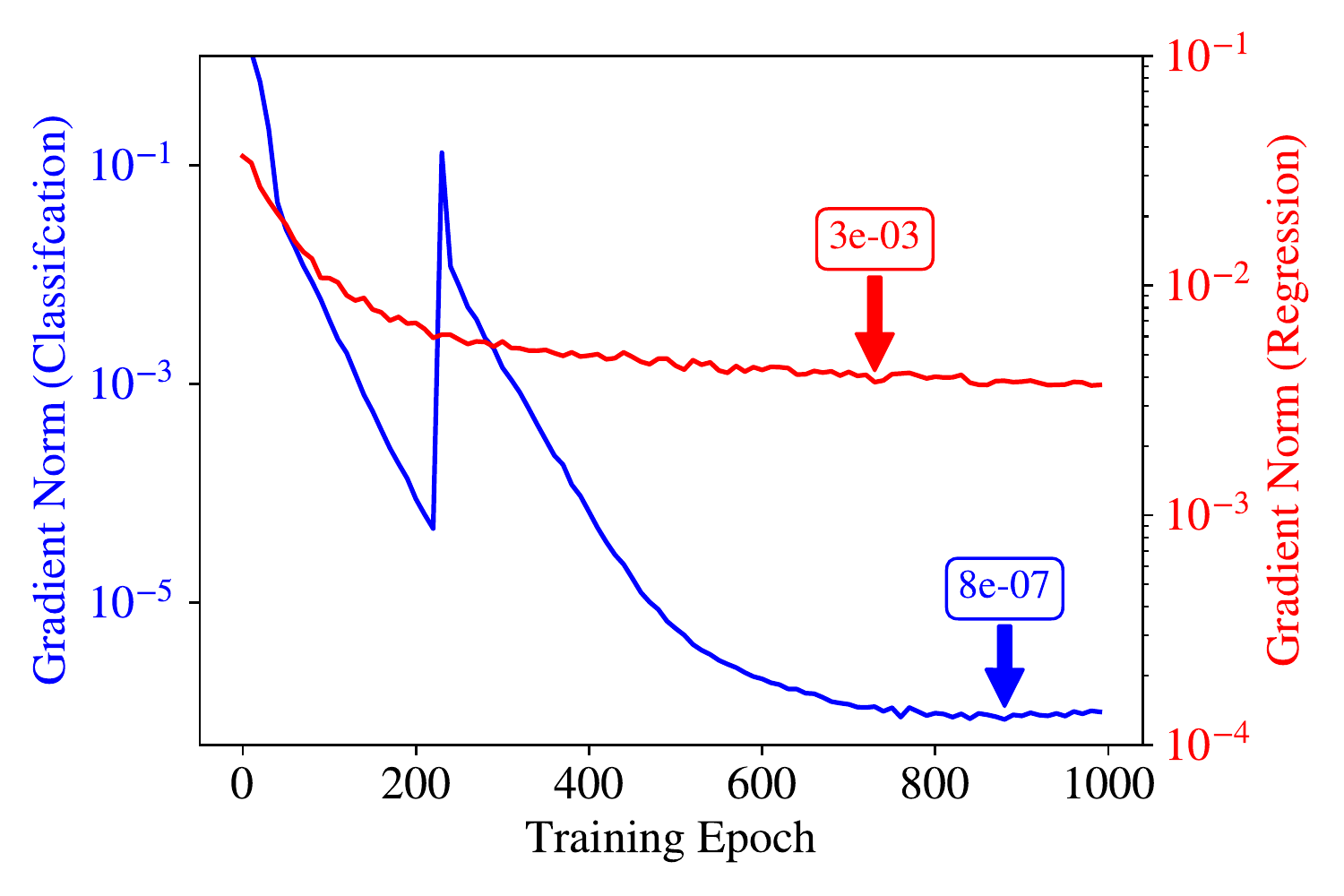}
\caption{Gradient norm over the training epochs for the Regression model (red) and Classifier model (blue), to highlight the gradient vanishing problem and sharp loss-landscape local minima. }
\label{fig:normgard}
\end{figure}

Now, since we have covered the key points related to the regression and classification tasks, we will look at how an equalizer (or a soft-demapper)  with the same architecture (applied to the data from the same transmission setup) performs, and how the result depends on whether the regression or classification task is employed. In this test, a single channel DP signal of 30~GBd is transmitted over 20$\times$50~km employing 64-QAM (used in Figs.~\ref{fig:classvsregression} and~\ref{fig:normgard}). The optical launch power is varied from -3 dBm to 5 dBm. We used a biLSTM equalizer with 208 hidden units and 66 neighboring symbols for the input (the memory). Note that both regression and classification architectures have the same number of layers, inputs, hidden units, and they were trained on the same data sequences. However, for fairness purposes, we have optimized the mini-batch size and learning rate for each equalizer individually, because we observed that using the same ones in regression and classification was causing even stronger overfitting after a few epochs for the latter, which is one of the classification drawbacks.

Fig.~\ref{fig:classvsregression} shows the Q-factor and MI dependencies for training and test datasets when we use the same biLSTM equalizer for regression and classification (of course, the output layers of NN structures are different for the two tasks). The Q-factor curves for the training and testing of the classifiers exhibit a considerable discrepancy, as shown in Fig.~\ref{LSTM_Q}, indicating that the classification model is overfitted. This noticeable difference in the classification task persisted even after we had optimized the learning rate and mini-batch size. However, in the case of regression, the training and testing output curves behave almost identically, in contrast to the classifier's results. It follows that for our test scenarios, the regression model based on the MSE generalizes considerably better, which is consistent with the findings from Ref.~\cite{bosman2020visualising}, where the strong gradients in the classification task loss function impact learning further: the learning state gets ``trapped'' in a sharp local minimum of loss landscape. Furthermore, we point out that after the equalization, the regression equalizers resulted in a higher Q-factor than the achieved value for the classifiers, owing to the better generalization of the regression-based NN structures on the optical datasets. The maximum regression-based NN's Q-factor on the testing dataset was 9.82~dB at 2~dBm, while the maximum Q-factor for the classification was 7.74~dB at -1~dBm. Besides the Q-factor, it is also instructive to show the MI for both tasks as well. The reason for this is that the classification loss function, the CEL, is directly related to the MI metric, therefore, displaying Q-factor after a hard decision can limit the entire potential of the model\cite{freire2021transfer}. Looking at the MI values in Fig.~\ref{LSTM_MI}, we can see that the classifier's training performance was overfitted, yielding nearly the maximum MI attainable for each power when the training dataset was used. However, in the case of regression, the training and testing curves followed the same trend, indicating a much better generalization capability of the equalizers. The maximum MI on the testing dataset of the regression NN was 5.97 bits/symbol (note that this value is only a MI's lower bound), and the one for the classification was 5.87 bits/symbol (but this MI value is exact). We can then conclude that in typical optical transmission tasks, the machine learning drawbacks associated with conventional classification ``overpower'' the regression-related shortcomings related to statistical limitations.

Aside from the overfitting, we also want to highlight the consequence of the second classification drawback, regarding the gradient vanishing in the training process.  According to Ref.~\cite[Chapter 8.2.2]{Goodfellowbook2016}, a practical way to demonstrate that the local minima is potentially the cause of the learning problem is to verify that the gradient norm shrinks to some ``insignificant'' values along with the training. As in Ref.~\cite{freire2021neural}, we depicted in Fig.~\ref{fig:normgard} the gradient norm of the last layer for the biLSTM equalizer over the epochs to show the effects of the loss function on the behavior of the gradient norm. Unlike Ref.~\cite{freire2021neural} that presented this plot for the MLP architecture, we now show it for the biLSTM equalizer, basically observing the same trend as reported for the MLP equalizer in a completely different transmission setup.  From Fig.~\ref{fig:normgard}, it is clear that after just 200 epochs, the gradient norm for the categorical-CEL case drops from $1.25$ to $7 \times 10^{-5}$ and, with continued training, goes even lower to $8 \times 10^{-7}$. For the regression case using MSE, the gradient continues to stably decrease over the epochs, reaching a minimum of around $0.003$. Therefore, we can affirm that we potentially have sharp local minima in categorical CEL for the optical channel equalization task, since, as described in Ref.~\cite[Chapter 8.2.2]{Goodfellowbook2016}, the gradient norm shrinks to a negligible value (on the order of $10^{-7}$) during training.

\begin{table*}[htbp]

  \centering
  \caption{Summary of biLSTM and MLP performance dependence on the mini-batch size for the different constellation cardinalities. Simulation results.}
    \resizebox{\textwidth}{!}{
\begin{tabular}{|c|c|c|c|c|c|c|c|c|c|c|c|c|c|}
\hline
                      &                                                                           & \multicolumn{6}{c|}{\begin{tabular}[c]{@{}c@{}}MLP Equalizer\\ Q-factor (Different Batch Sizes)\end{tabular}}                                                                             & \multicolumn{6}{c|}{\begin{tabular}[c]{@{}c@{}}biLSTM Equalizer\\ Q-factor (Different Batch Sizes)\end{tabular}}                                                                          \\ \cline{3-14} 
\multirow{-2}{*}{QAM} & \multirow{-2}{*}{\begin{tabular}[c]{@{}c@{}}Q-factor \\ Ref\end{tabular}} & $8$ & $16$ & $32$ & $64$ & $128$ & $2048$ & $8$ & $16$ & $32$ & $64$ & $128$ & $2048$ \\ \hline\hline

 $8$ &    $8.65$ dB &
          $8.62$ dB &
          $8.70$ dB &
          $8.65$ dB &
          $8.74$ dB &
          $8.82$ dB &
          $9.44$ dB &
          $10.08$ dB &
          $10.25$ dB &
          $10.51$ dB &
          $10.71$ dB &
          $11.28$ dB &
          $13$ dB \\ \hline
         
        $16$ &
          $7.22$ dB &
          $6.91$ dB &
          $7.02$ dB &
          $7.09$ dB &
          $7.17$ dB &
          $7.28$ dB &
          $7.72$ dB &
          $8.58$ dB &
          $9.14$ dB &
          $9.72$ dB &
          $10.10$ dB &
          $10.43$ dB &
          $11.57$ dB \\ \hline
         
        $32$ &
          $5.05$ dB &
          $4.47$ dB &
          $4.57$ dB &
          $4.69$ dB &
          $4.80$ dB &
          $5.01$ dB &
          $5.58$ dB &
          $7.02$ dB &
          $7.66$ dB &
          $8.36$ dB &
          $8.69$ dB &
          $8.75$ dB &
          $8.79$ dB \\ \hline
         
        $64$ &
          $3.15$ dB &
          $2.67$ dB &
          $2.73$ dB &
          $2.95$ dB &
          $2.96$ dB &
          $3.06$ dB &
          $3.79$ dB &
          $5.78$ dB &
          $5.87$ dB &
          $6.29$ dB &
          $6.47$ dB &
          $6.51$ dB &
          $6.78$ dB \\ \hline
         
        $128$ &
          $1.53$ dB &
          $0.77$ dB &
          $0.90$ dB &
          $1.10$ dB &
          $1.30$ dB &
          $1.36$ dB &
          $2.09$ dB &
          $3.54$ dB &
          $3.92$ dB &
          $4.09$ dB &
          $4.13$ dB &
          $4.14$ dB &
          $4.27$ dB \\ \hline
\end{tabular}
}
    \label{table:BatchStudy}
\end{table*}

Finally, we conclude this section with a brief discussion of the loss function study in the task of optical channel equalization/soft demapping. Even though the MSE is a good fit when we deal with a Gaussian noise model making the NN learn effectively and deliver attractive Q-actor gains, in some scenarios, the stochastic gradient descent (SGD) algorithm continues to decrease the MSE, but this does not translate into the BER improvement, which means that we fall into a local minimum. This is visually observed when the ''jail window'' constellation pattern appears, which indicates a mismatch between the MSE loss and the BER metric. We believe that the ''jail window'' appears for local minima in learning, and although it can produce seemingly good BER results, it is not the ultimate equalization. In the next section, we present a scenario where we clearly show that when the ``jail window'' disappears due to optimization of the mini-batch size, the final Q-factor performance increases for the model without the ``jail window''\footnote{Note that this is not always the case, increasing the mini-batch size will not necessarily eliminate the ''jail window'' pattern.}. To match the QoT metric of our transmission and the loss function of our learning algorithm, we can claim that the CEL is the most suitable loss function for communication applications~\cite{Georg2021book} since by minimizing cross-entropy, we maximize the mutual information of the system, and therefore no mismatch between the loss function and the QoT metric appears. However, as presented in this section, we observed other machine learning-related problems attributed to the learning process, namely SGD learning. We note that the landscape of such a loss function has very sharp local minima, and the gradients tend to vanish in the early learning stage due to our high accuracy requirements in optical channel equalization tasks. Therefore, even though the CEL does not have a statistical limitation for the noise likelihood, it poses many learning difficulties that can ultimately make its performance worse than that of the regression NNs based on MSE.  Here, we also stress that several works have also acknowledged that both regression and classification have disadvantages \cite{li2016online,mukherjee2021joint, liu2018joint, liu2017deep, wu2021joint,chen2019joint,lin2017focal}.

So we can confidently say that, for optical communication, we still do not have the ultimate answer to the question of what would be the best loss function, as each candidate has drawbacks. We believe that by looking at different fields we can potentially find some other possible candidates, but this required further investigation. In computer vision, it was discovered that when an image is processed, the majority of the pixies describe the image background, and only a few pixels express the objects in the image. This resulted in inefficient training because most parts of the image correspond to ``an easy prediction'' (which means that they can be easily labeled as background by the detector) and therefore offer little relevant learning. Although individually they provide tiny contributions to the loss value, when we combine those contributions, they can overwhelm the loss and computed gradients, resulting in a degraded model's prediction performance. It happens because easy predictions (detections with high probabilities or, in our context, the correct classifications following a simple hard decision) account for a large share of inputs. To address this issue, in Ref.~\cite{lin2017focal} Facebook A.I. developed a new modified approach named focal loss (FL), by adding a weighting factor to the CEL function. The FL gives a higher weight to cases that are hardly misclassified: in communications, it would correspond to the cases that has been misclassified after the HD process. We believe that the difficulty of the ``dataset imbalance'' (meaning that just a small fraction of the dataset corresponds to the wrong HD predictions) exists in virtually all high-accuracy communication-related equalization/demapping problems. As a toy example, consider a system where an initial SER after HD is equal to $10^{-3}$. Training the NN-classifier with 100K symbols, in this case, means that only 100 symbols (0.1\%) are the errors that persist after HD and that the model needs to learn from them to mitigate the impairments, while the rest 99900 symbols (99.9\%) of the training dataset corresponds to ``an easy prediction''. Therefore, to improve the performance of classifiers, we expect that a similar focal loss function, as in computer vision, must be created for the communications application. This can be an interesting topic for future research in the field of optical channel equalization and efficient NN-based soft symbol demapping.

\section{Interrelation Between Batch Size, Constellation Cardinality, and Equalization Quality} \label{sec:batch}

The NN training presupposes the use of optimization algorithms, such as SGD or Adam \cite{Kingma2015}, to update the NN parameters (weights) based on the value of the loss function. Traditionally, two training methods are utilized: batch training, in which the algorithm updates the weights after the entire training dataset, and online learning, which is executed after each training sample.

In practice, however, stochastic optimization methods use mini-batches, i.e., the portions of all training examples with the size greater than one, but less than the entire training dataset, to compute the gradients using less memory and update the parameters after each mini-batch\cite{Goodfellowbook2016}. One reason for this is that modern NNs demand such a large amount of data that the computer dynamic memory cannot keep the entire dataset. Notably, the size of a mini-batch is regarded as one of the most important hyperparameters to consider when training an NN. Large mini-batches are well known in deep learning for their ability to significantly speed up the training process, while also providing an accurate approximation of the gradient~\cite{Goodfellowbook2016}.
Small mini-batches, on the other hand, have been shown to have a regularizing impact, preventing overfitting \cite{Wilson2003, Goodfellowbook2016}. The latter phenomenon can be explained by the fact that, while the large mini-batches can indicate the gradient's direction, the algorithm cannot forecast how long this trend would continue, usually causing the training process to take a large update step forward. This leads to unstable learning or falling into local minima. However, small mini-batches, while carrying added noise due to their smaller sampling across the entire dataset, result in small steps and the system can easily converge to a true direction \cite{Wilson2003}. Several studies have been carried out to investigate the effects of mini-batches size on the NNs performing traditional deep learning tasks. Ref.~\cite{Smith2017} proposed using small mini-batches to introduce noise in the gradient estimation and push the gradient away from sharp local minima; the authors of that Ref. also demonstrated that the optimal batch size for the CIFAR-10 dataset is 80. In Ref.~\cite{ShirishKeskar2017}, it was empirically demonstrated that the large mini-batch sizes lead to the convergence at sharp local minima, resulting in a poor NN's generalization, whereas the smaller mini-batches lead to flatter local minima, allowing the NN to achieve a better generalization. However, to the best of our knowledge, no research has been conducted to study the effect of mini-batch size on the performance of NN-based regression equalizers in optical transmission.

\begin{figure}[ht!]
    \centering
    \includegraphics[width=0.75\linewidth]{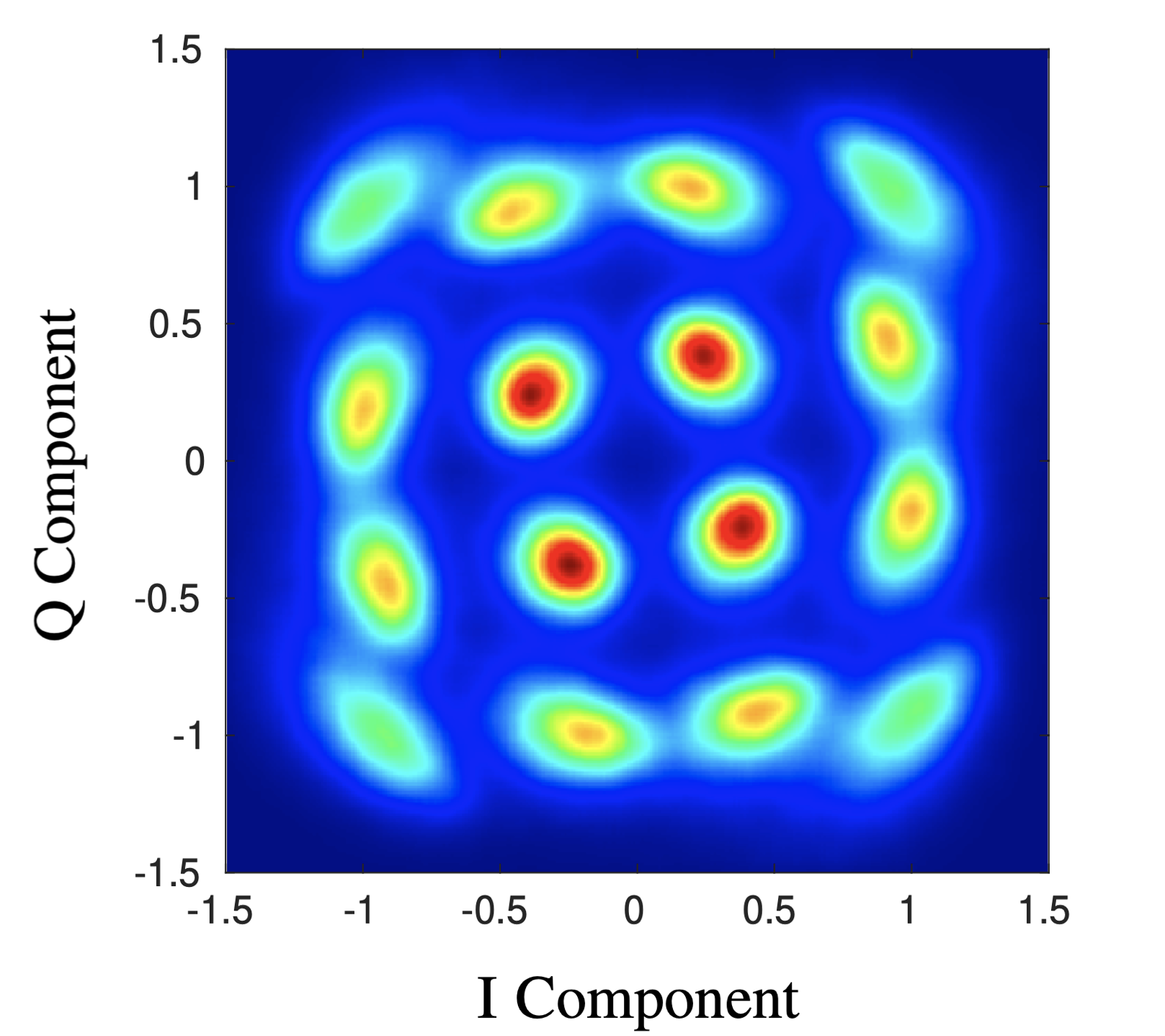}
    \caption{Signal constellation of 16-QAM after metro-link (450 km) simulated optical transmission using a Nonzero Dispersion Fiber \cite{1036395}. We used this highly-nonlinear exemplary system to visually demonstrate the distinct distortion levels at different constellation points: the constellation points at the outer layer are clearly more distorted than the ones closer to the origin. }
    \label{fig:signal_const_batch_size}
\end{figure}

The motivation for this study is that the mini-batch in the NN performing optical channel equalization has more physical meaning than it does in other ``traditional'' machine learning tasks, such as computer vision. Recall that different transmitted symbols in the constellation correspond to different optical field intensities, depending on the modulation format. This can be seen by looking at the 16-QAM received constellation diagram, Fig.~\ref{fig:signal_const_batch_size}, where the outermost points are the most distorted ones. This happens because the fiber nonlinearity is proportional to the cube of the optical field amplitude, so the most distant points (having the largest amplitude) experience the most nonlinear effects. Because the NN's parameters are updated after each mini-batch, the training process can be more efficient if each mini-batch contains training samples that cover the whole range of possible constellation amplitudes, to encompass different distortion levels. If this hypothesis is correct, there should be a relationship between the mini-batch size and the modulation format's cardinality, also affecting the NN's performance. This section reports tests and simulations to explain the aforementioned connections.

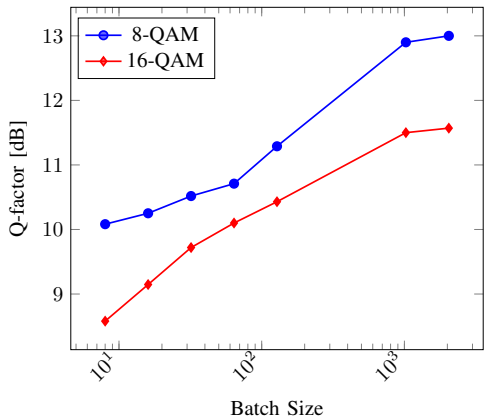
\begin{figure}[ht!] 
    \centering
  \begin{tikzpicture}[scale =0.8]
\begin{axis}[legend style={at={(0.35,0.97)} ,fill=white, fill opacity=0.6, draw opacity=1,text opacity=1,,anchor=north east},
    ylabel={Q-factor [dB]},
    xlabel={Batch Size},
    xmode = log,
x tick label style={rotate=45,anchor=east}]
\addlegendentry{8-QAM}
\addplot[mark=*,thick,blue] coordinates {
(8,10.082) (16,10.251) (32,10.518)(64,10.71)(128,11.289)(1024,12.9)(2048,13)
};

\addlegendentry{16-QAM}
\addplot[mark=diamond*, thick,red] coordinates {
(8,8.58) (16,9.147) (32,9.72)(64,10.1)(128,10.43)(1024,11.5)(2048,11.57)

};

\end{axis}
\end{tikzpicture}
  \caption{Q-factor of 8- and 16-QAM simulated signals equalized by the biLSTM trained with different mini-batch sizes: 8, 16, 32, 64, 128, 1024, and 2048. The simulated transmission setup is described in the main text. }
  \label{fig:singleplot_increaseQfactor} 
\end{figure}

\begin{figure*}[ht!]
\centering

\begin{subfigure}{.23\linewidth}
  \centering
  \stackinset{c}{}{t}{1.2in}{\textcolor{red}{$Q=10.1\mathrm{dB}$}}{\includegraphics[width=\linewidth]{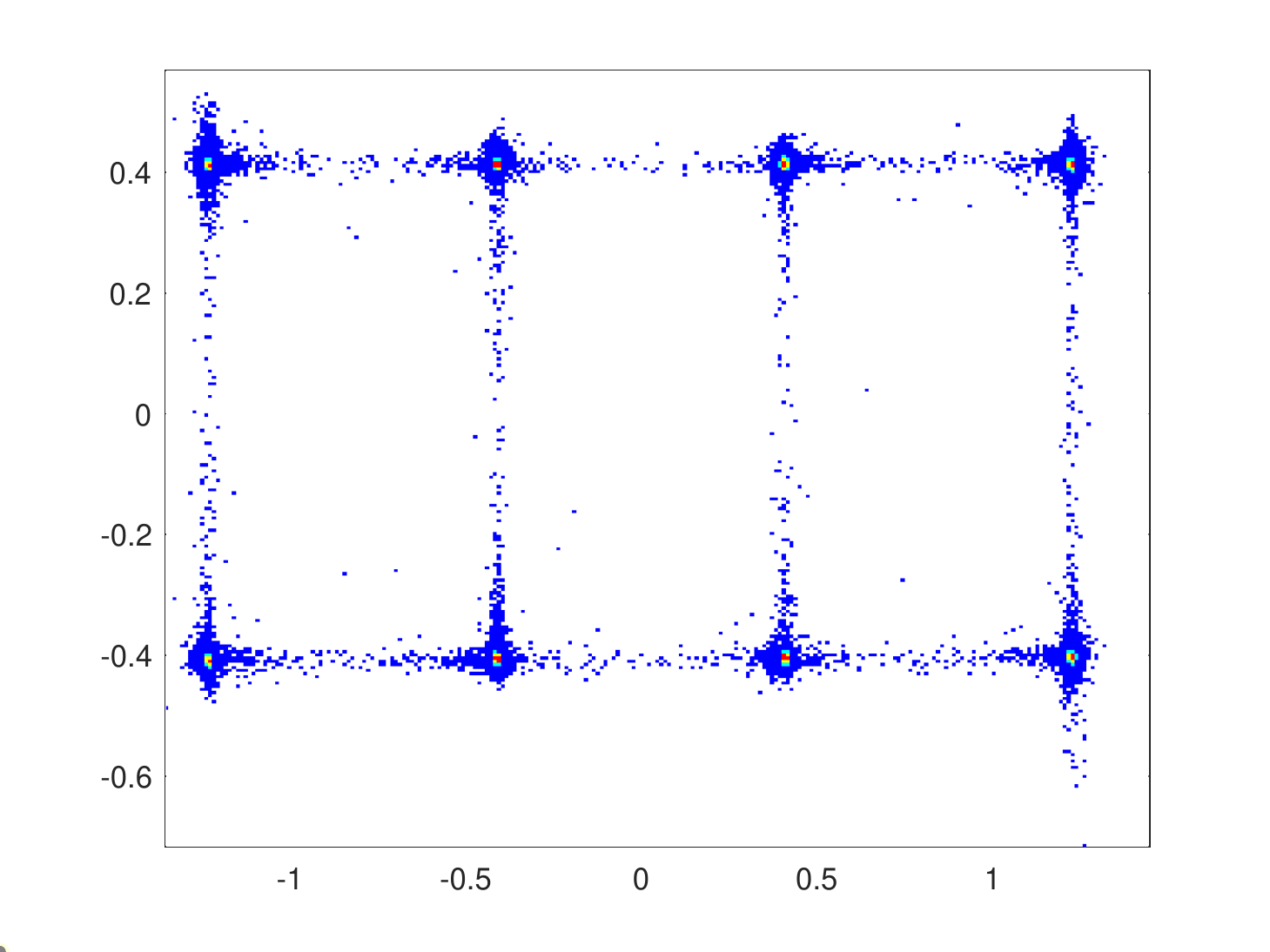}}
  \caption{8QAM - batch size 8}
  \label{fig:8bilstm_batch16}
\end{subfigure}
\begin{subfigure}{.23\linewidth}
  \centering
  \stackinset{c}{}{t}{1.2in}{\textcolor{red}{$Q=10.5\mathrm{dB}$}}{\includegraphics[width=\linewidth]{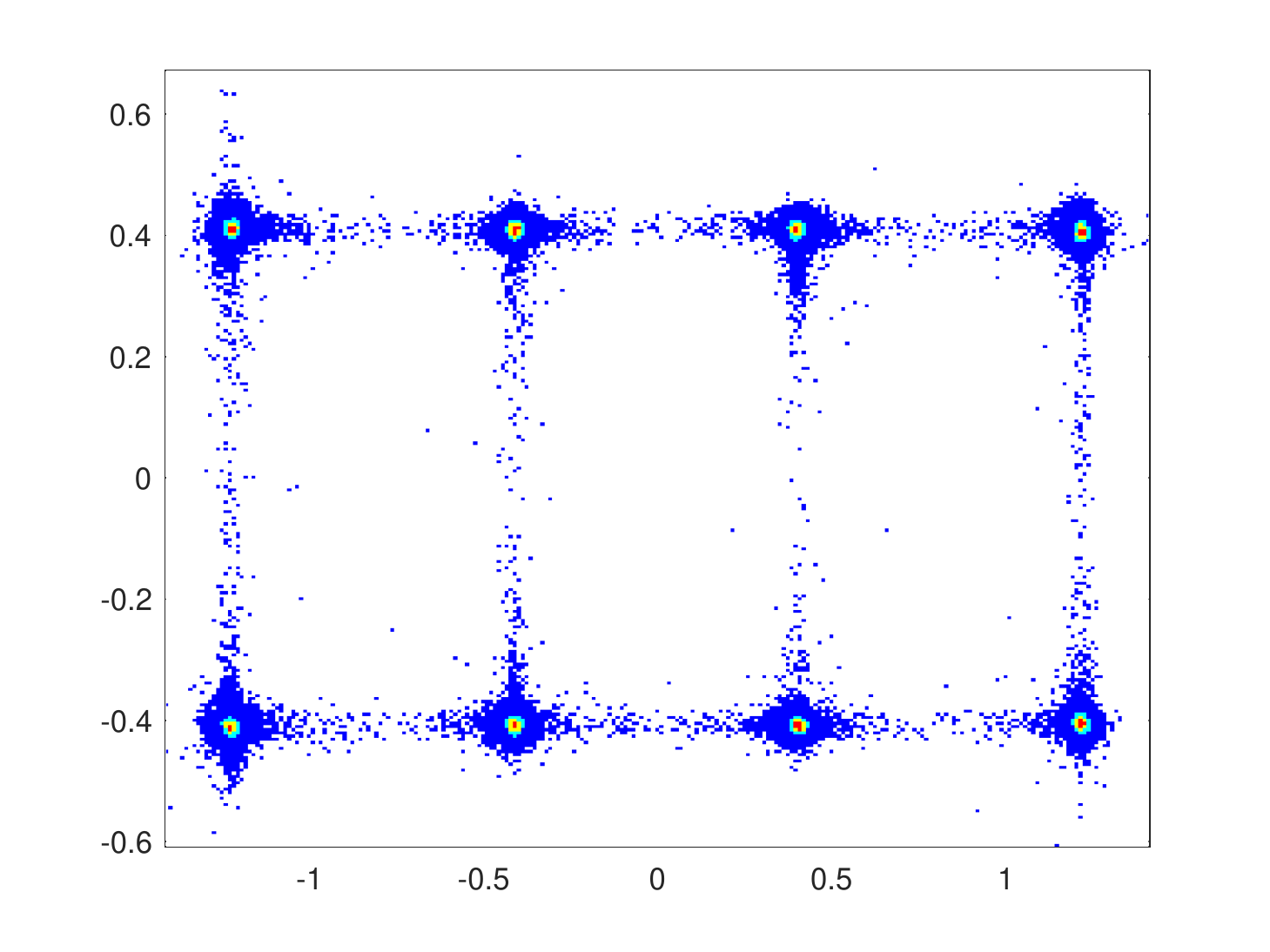}}
  \caption{8QAM - batch size 16}
  \label{fig:8bilstm_batch64}
\end{subfigure}
\begin{subfigure}{.23\linewidth}
  \centering
  \stackinset{c}{}{t}{1.2in}{\textcolor{red}{$Q=11.3\mathrm{dB}$}}{\includegraphics[width=\linewidth]{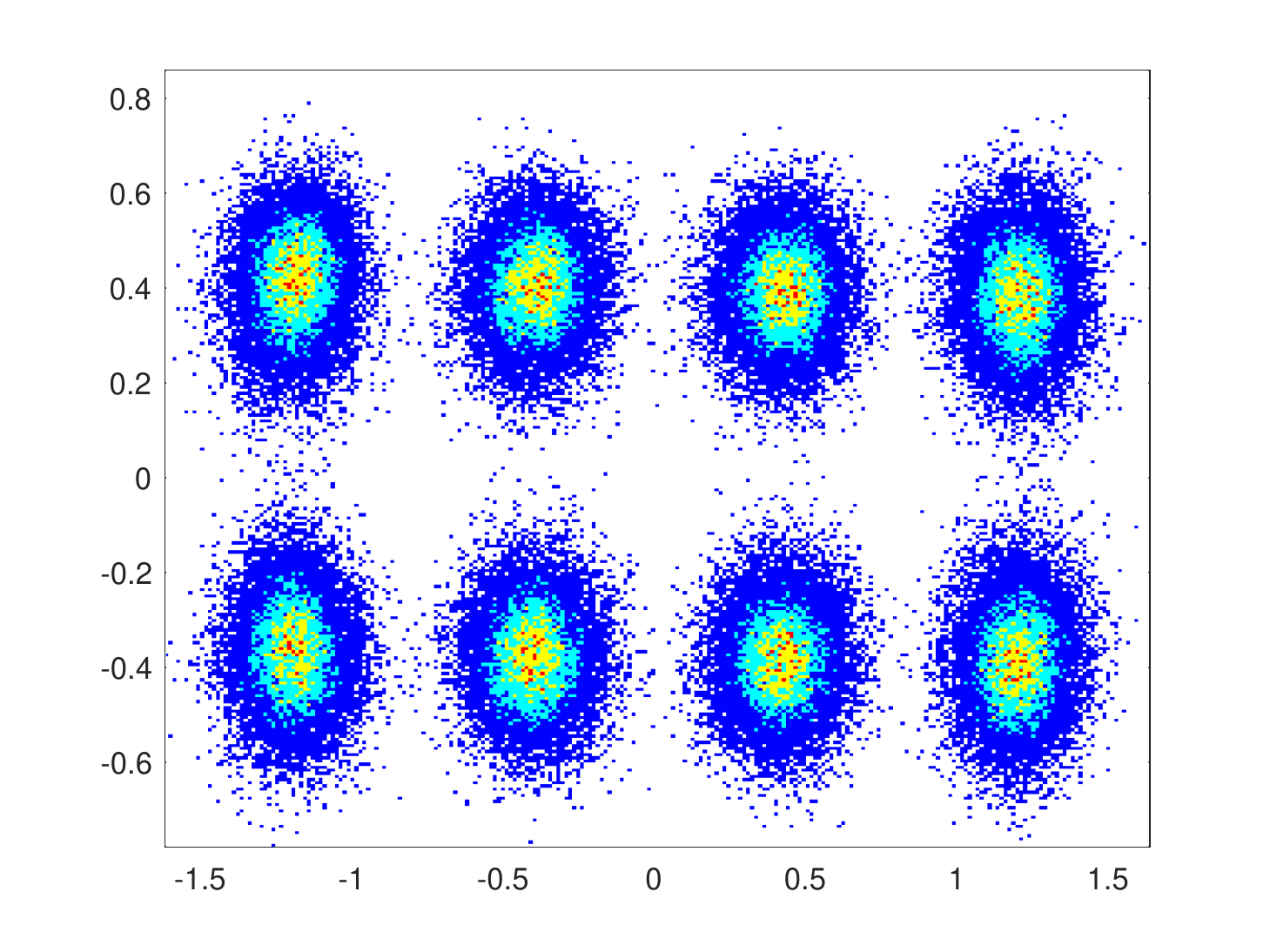}}
  \caption{8QAM - batch size 128}
  \label{fig:8bilstm_batch128}
\end{subfigure}
\begin{subfigure}{.23\linewidth}
  \centering
  \stackinset{c}{}{t}{1.2in}{\textcolor{red}{$Q=13.0\mathrm{dB}$}}{\includegraphics[width=\linewidth]{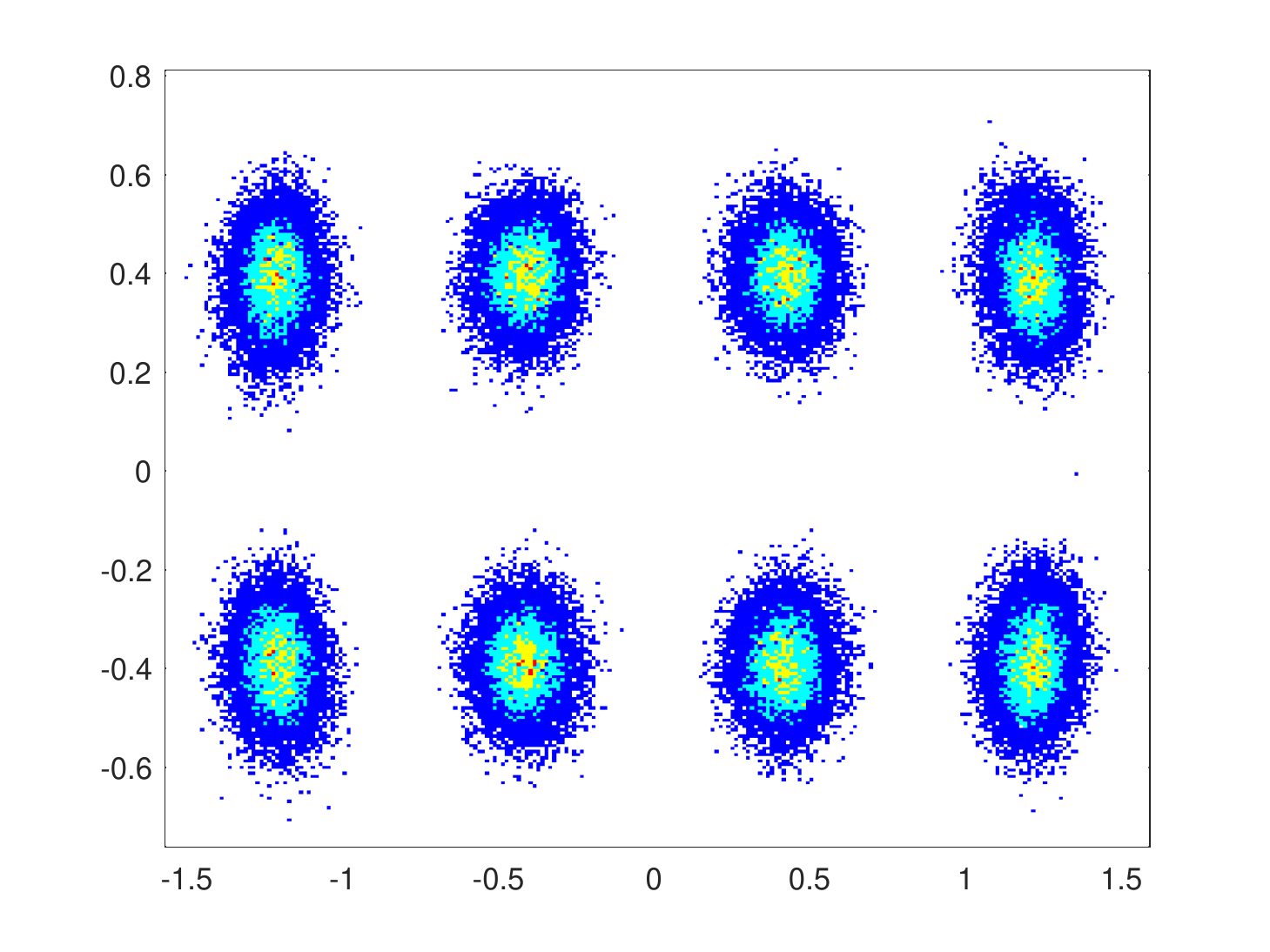}}
  \caption{8QAM - batch size 2048}
  \label{fig:8bilstm_batch2048}
\end{subfigure}

\begin{subfigure}{.23\linewidth}
  \centering
  \stackinset{c}{}{t}{1.2in}{\textcolor{red}{$Q=8.6\mathrm{dB}$}}{\includegraphics[width=\linewidth]{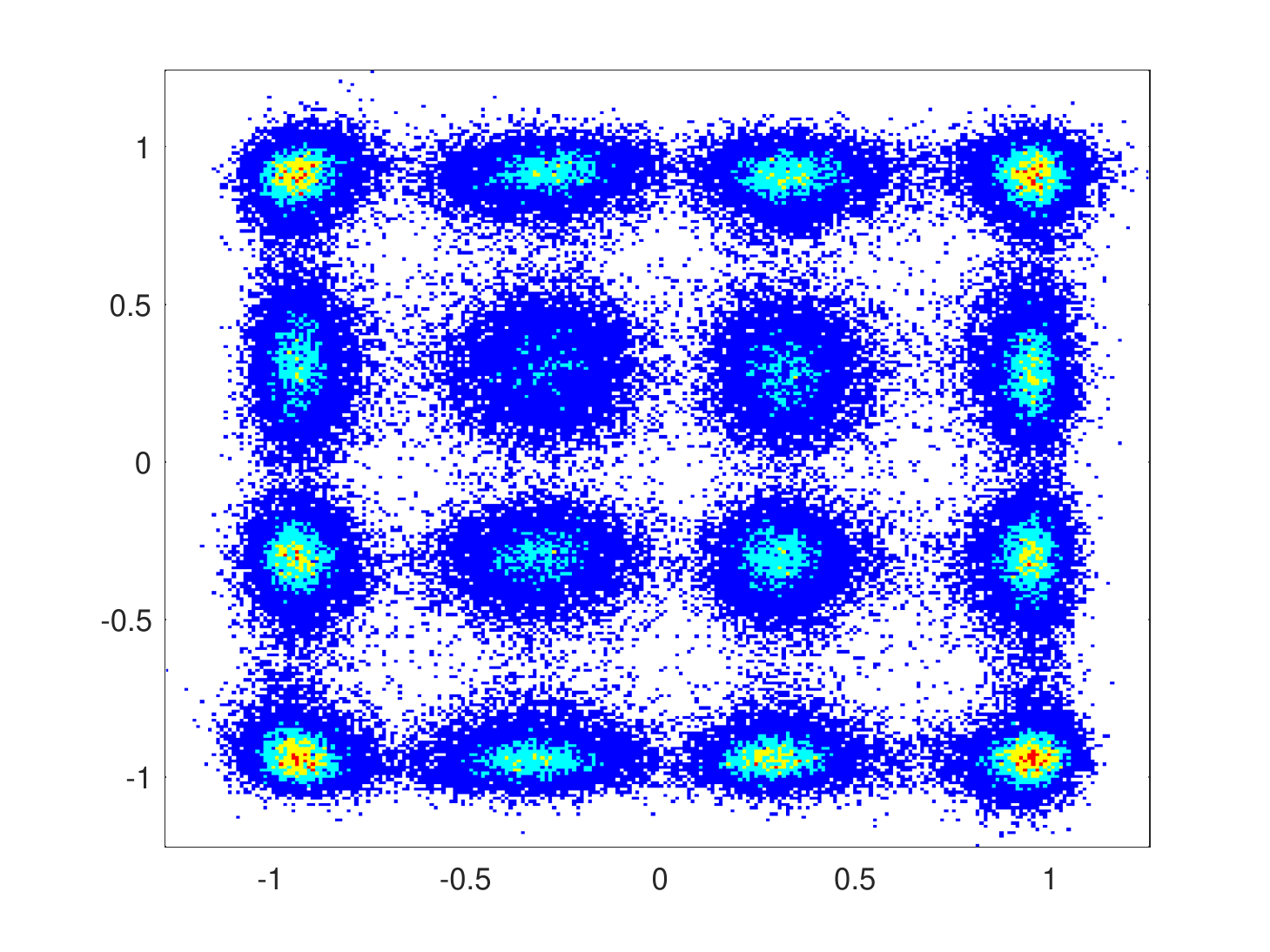}}
  \caption{16QAM - batch size 8}
  \label{fig:16bilstm_batch16}
\end{subfigure}
\begin{subfigure}{.23\linewidth}
  \centering
  \stackinset{c}{}{t}{1.2in}{\textcolor{red}{$Q=9.1\mathrm{dB}$}}{\includegraphics[width=\linewidth]{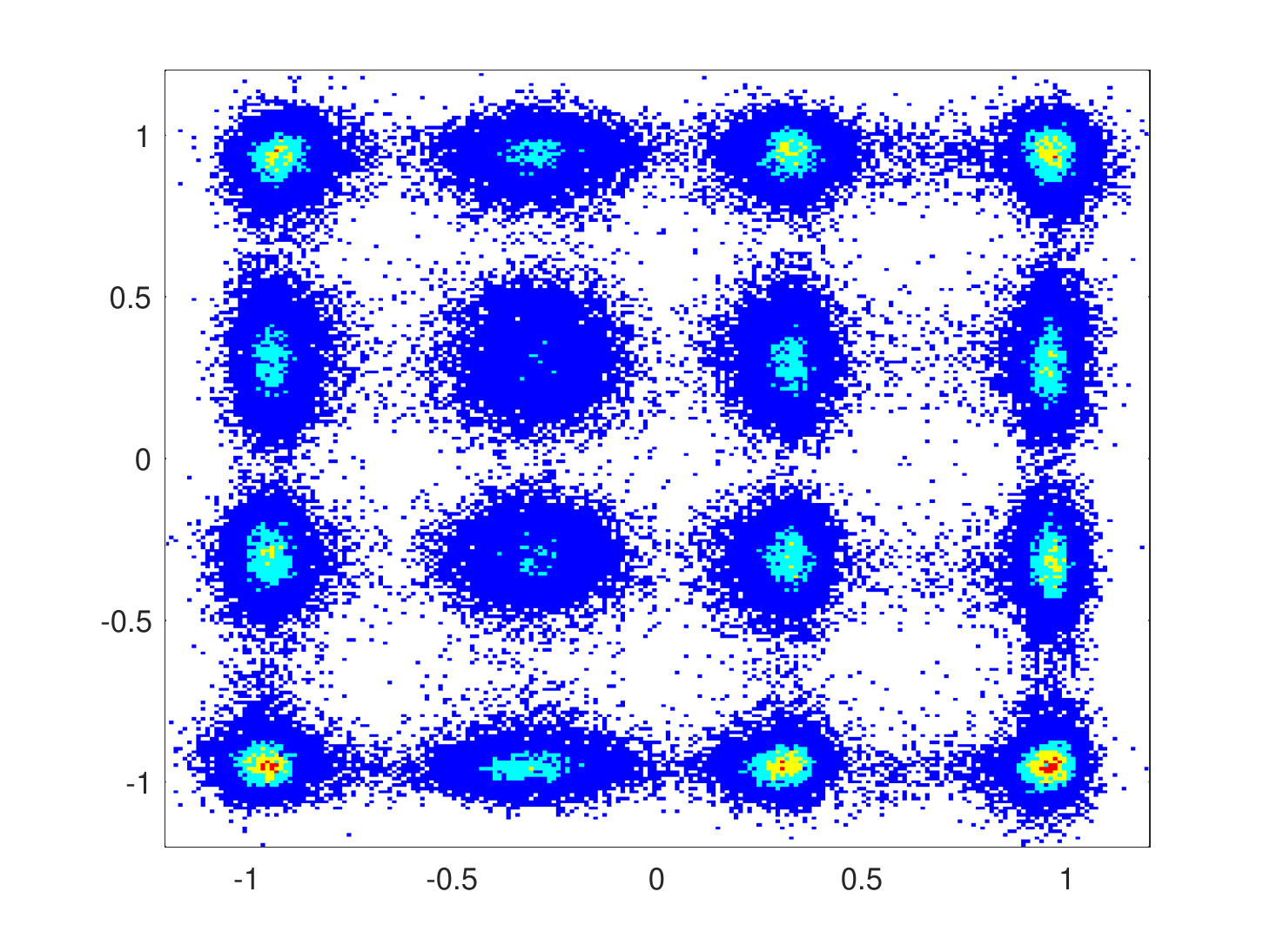}}
  \caption{16QAM - batch size 16}
  \label{fig:16bilstm_batch64}
\end{subfigure}
\begin{subfigure}{.23\linewidth}
  \centering
  \stackinset{c}{}{t}{1.2in}{\textcolor{red}{$Q=10.4\mathrm{dB}$}}{\includegraphics[width=\linewidth]{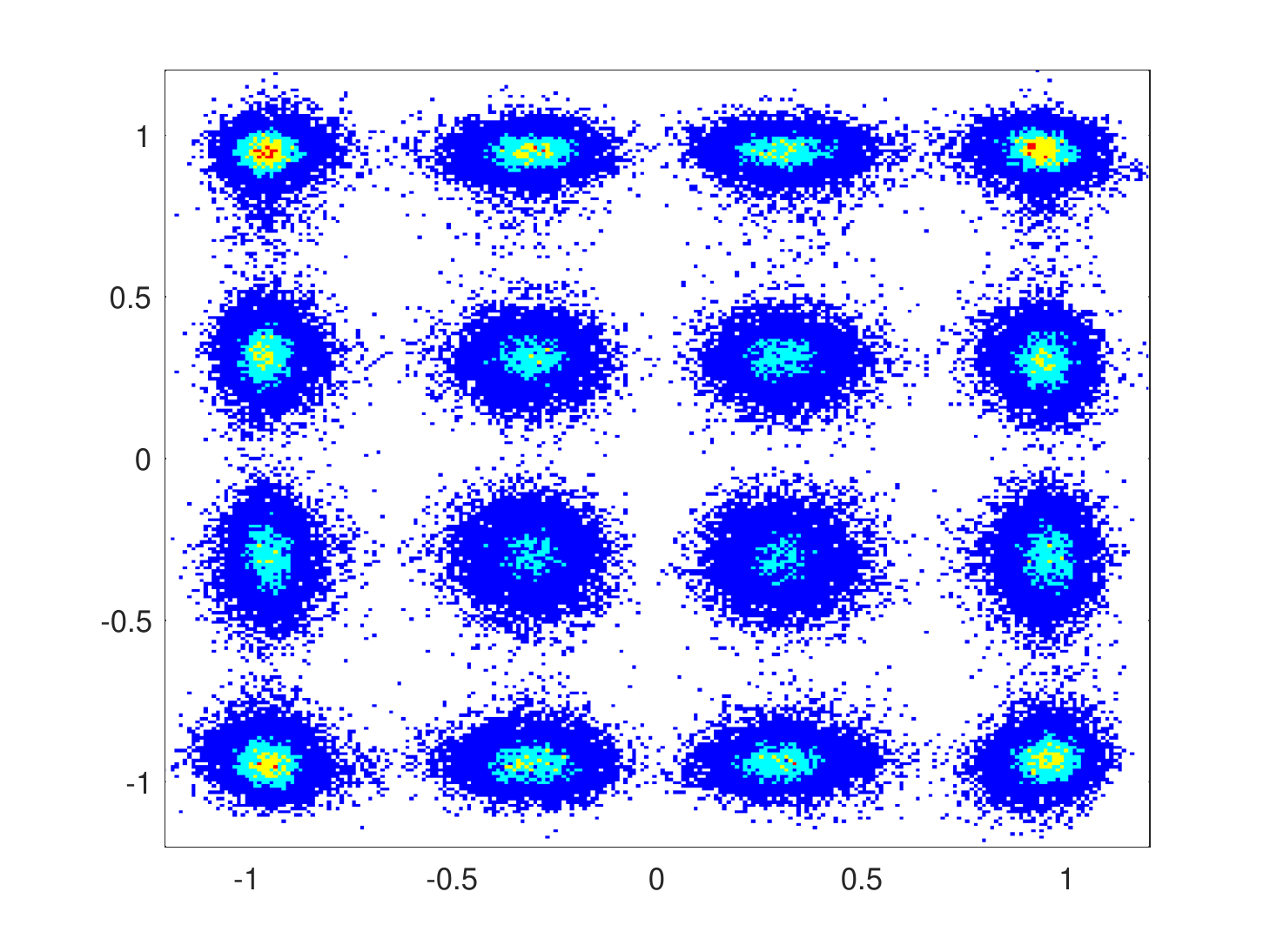}}
  \caption{16QAM - batch size 128}
  \label{fig:16bilstm_batch128}
\end{subfigure}
\begin{subfigure}{.23\linewidth}
  \centering
  \stackinset{c}{}{t}{1.2in}{\textcolor{red}{$Q=11.6\mathrm{dB}$}}{\includegraphics[width=\linewidth]{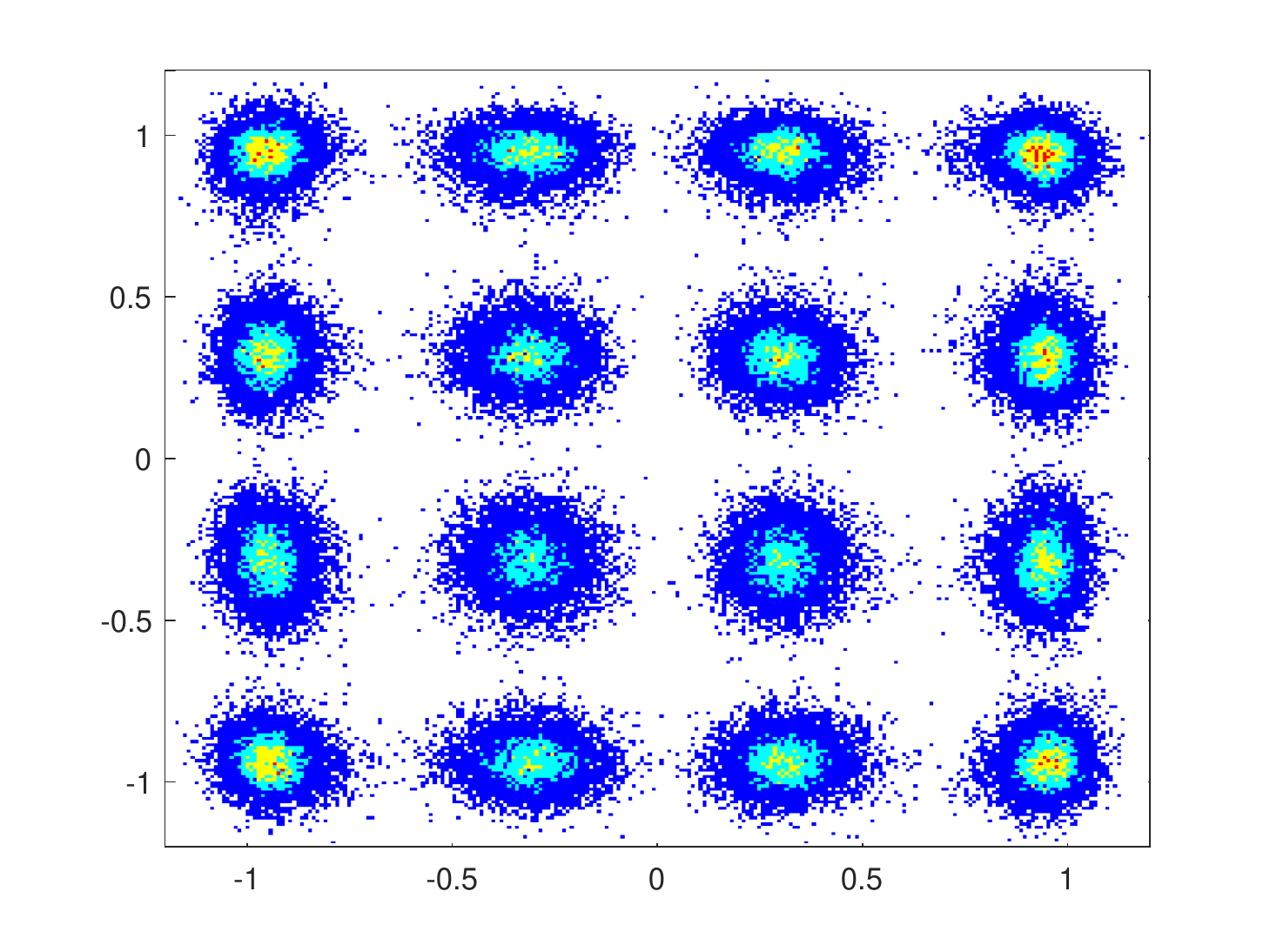}}
  \caption{16QAM - batch size 2048}
  \label{fig:16bilstm_batch2048}
\end{subfigure}

\caption{Signal constellations (simulation results) for different batch sizes after equalization using the biLSTM equalizer for 8- and 16-QAM. The transmission system is described in the text. The Q-factor of each constellation is also shown to highlight the improvement in QoT.}
\label{fig:signal_cosnt_batch_size_pics}
\end{figure*}

Table~\ref{table:BatchStudy} shows the Q-factor of optical signals equalized by the MLP and biLSTM MSE-regression equalizers for a range of modulation formats: 8-, 16-, 32-, 64, 128-QAM\footnote{Note that to generate the 8-QAM constellation we used the standard Matlab function \texttt{qammod}; the same constellation shape was used in Refs. \cite{peng2013bit,kim2020modulation,nadal2015experimental,wang2014capacity}.} and mini-batch sizes: size 8, 16, 32, 64, 128, and 2048. In this section, simulations were performed for the SSMF 16$\times$60~km link at 5~dBm launch power and 34.4~GBd symbol rate. When the size of the mini-batch increases, we observe a progressive growth in the Q-factor after equalization. Let us consider the case of biLSTM, which gives the best equalization quality. When the mini-batch size is increased for the 8- and 16-QAM, the Q-factor after the biLSTM-based equalization dramatically improves. For example, comparing the post-equalization Q-factors in the 16-QAM scenario, for mini-batch sizes 8 and 2048, the 3~dB improvement can be seen for the largest mini-batch over the smallest one. This behavior of mini-batch vs. Q-factor gain for biLSTM is illustrated in Fig.~\ref{fig:singleplot_increaseQfactor} for 8-QAM and 16-QAM. Note that increasing the mini-batch size above 2048 does not provide any further improvement: the Q-factor reaches some kind of plateau (saturates) after the batch size of approximately 1000. When using a higher constellation cardinality, 128-QAM, the improvement rendered by the biLSTM equalizer increases from 3.54  to 4.27~dB when changing from the mini-batch size from 8 to 2048. We note that, from Table~\ref{table:BatchStudy}, for the 128-QAM with the MLP equalization, the difference between the Q-factors corresponding to the lowest and highest mini-batch sizes is approximately 1.3~dB, which is bigger than the difference for the biLSTM. The increase in Q-factor backs up the claim that the NNs can learn more efficiently when given the training samples covering all nonlinearity levels in the constellation.

Together with the improvement in Q-factor (up to some limiting value) following the growth of the mini-batch size, we can see the interrelation between the constellation distribution after equalization and the mini-batch size used in training. The constellations for the 8- and 16-QAM systems for different mini-batch sizes utilizing the biLSTM equalization, are given in Fig.~\ref{fig:signal_cosnt_batch_size_pics}. In this figure, we can observe that utilizing a small batch size for 8-QAM drives the NN-based equalizer to fall into the previously discussed ``jail window'' pattern rather than into the Gaussian-type distribution, but the latter can be obtained when we use larger mini-batch sizes. In the 16-QAM case, the signal constellations after NNs with small mini-batches became noticeably distorted, the ``jail window'' elements are clearly seen. At the same time, for the larger mini-batches, the 16-QAM constellations show little distortion with clear concentrations at the center of each constellation point.

The degradation of constellations from circular clusters into the ``jail window'' pattern when training with small mini-batch sizes, is actually a new and intriguing phenomenon:  to our knowledge, there have been no studies relating the equalizer's performance deterioration to the size of the training mini-batches. Generally, as we have already seen, several factors can contribute to the patterning effect of the ``jail window''. In this section, we addressed the specific situation, where the ``jail window'' occurred solely due to the batch size effect, and so we were able to understand how and why the batch size relates to the ``jail window'' patterning. In a more general case, different contributions amalgamate, resulting in the ``jail window'' pattern; the latter is always an indication that something is not good with the training or with the NN system itself. This is actually evident from the results in Fig.~\ref{fig:signal_cosnt_batch_size_pics} (a) to (d). In case (a), where the learned weights resulted in the ``jail window'', the Q-factor is around 10~dB, while in case (d), where the learned weights did not generate the ``jail window'', we measured the Q-factor to be around 13~dB. Therefore, we expect that when the ``jail window'' pattern is present, even though the performance metric value can be rated as ``satisfactory'',  in the training we have most likely reached just a local minimum, and some better equalization results can be achieved with more appropriate training or by using a modification of the NN architecture.

\begin{figure*}[t!]
\centering

\begin{subfigure}{.48\linewidth}
  \centering
    \includegraphics[width=\linewidth]{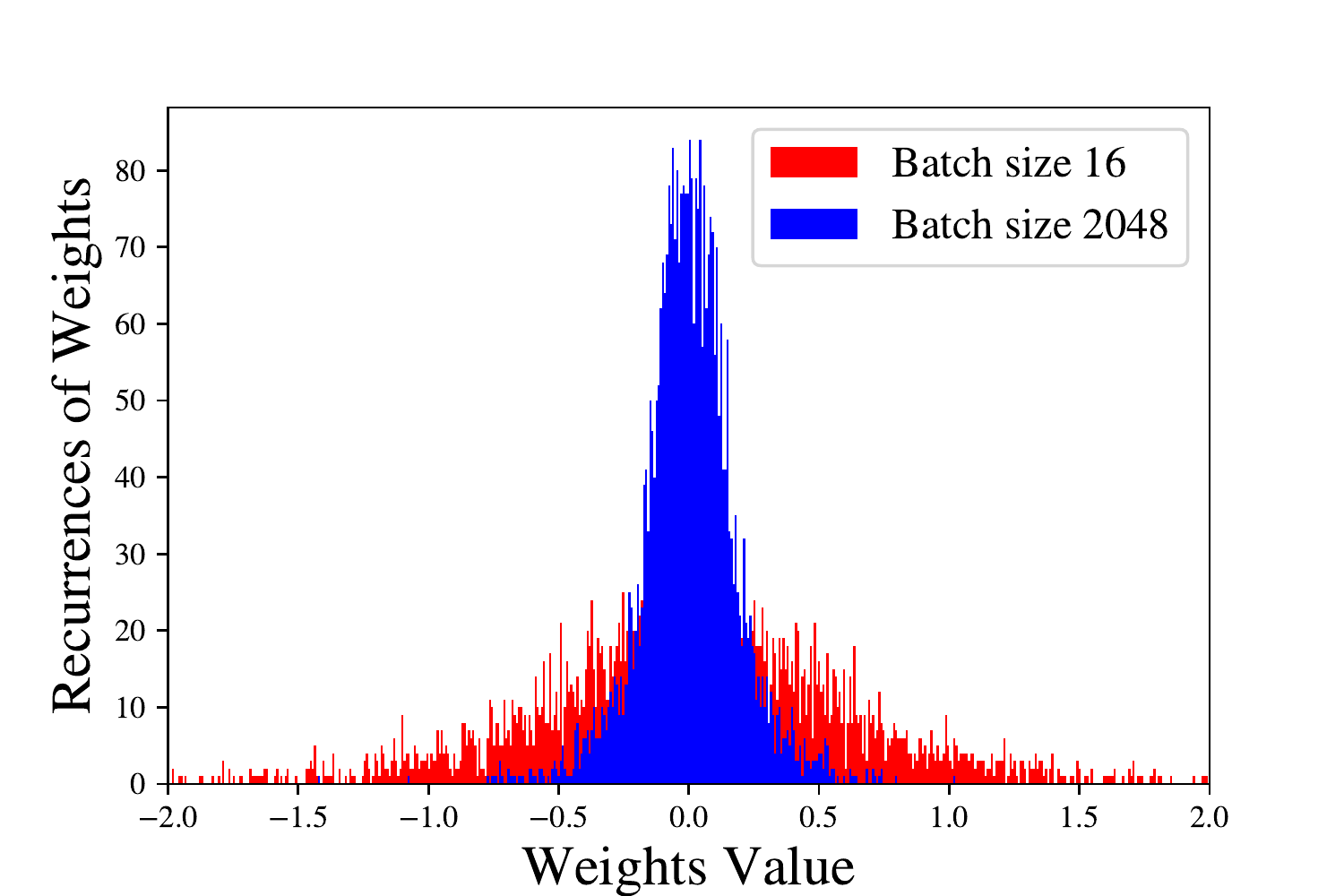}
    \caption{Weight distribution at the forward LSTM layer}
    \label{fig:weight_dist_forward_LSTM}
\end{subfigure}
\begin{subfigure}{.48\linewidth}
  \centering
    \includegraphics[width=\linewidth]{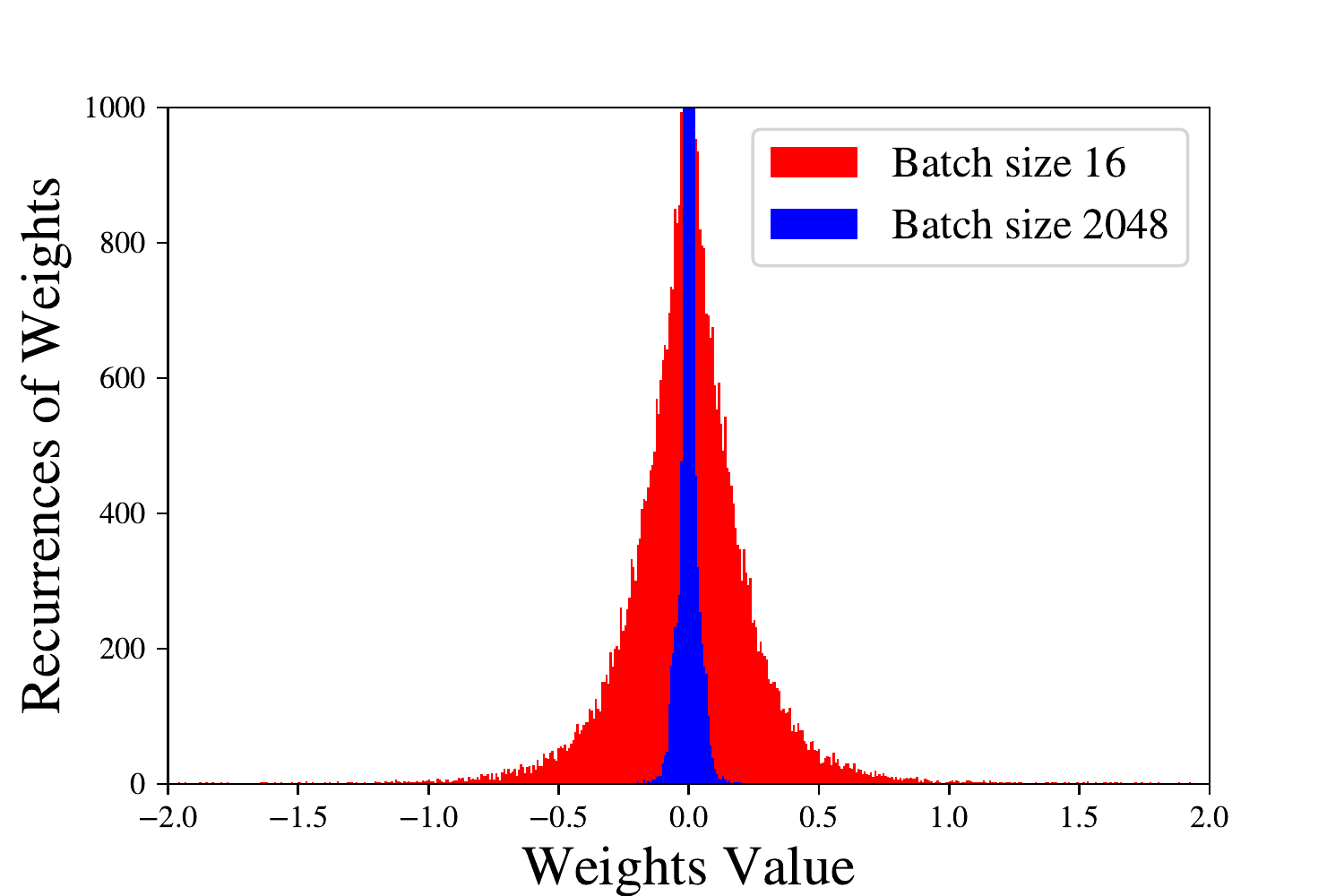}
    \caption{Weight distribution at the final linear layer}
    \label{fig:weight_dist_linear}
\end{subfigure}
\caption{Weight distributions of the biLSTM equalizer's forward and output layers, corresponding to the mini-batch size of 16 (red) and 2048 (blue). }
\label{fig:weight_distribution}
\end{figure*}

To explain further the degradation of signal constellations further, when training the biLSTM equalizer with small mini-batch sizes, we investigated the weight distribution inside the NNs. More precisely, we compared the weight distributions in the last linear layer (output layer) of the biLSTM, and the distribution in the forward LSTM layer, for two values of the training mini-batch size: 16 and 2048, using an 8-QAM system as an example. Fig.~\ref{fig:weight_distribution} shows that the weights, when training with small mini-batches, range over a considerably larger interval than we have when training the NN with larger mini-batches. From Fig.~\ref{fig:weight_dist_linear}, we can see that the final layer weights were significantly saturated when the NNs were trained with small mini-batches\footnote{Well trained NNs usually have the weights' distribution being close to the Gaussian distribution with a small variance. When this variance is too large, say more than one, as we show in Fig.~\ref{fig:weight_distribution}, it usually indicates a saturation of weights, which, in the NN terms, means our disregarding some important input features.}. The saturation is indicated by the presence of large value weights and typically degrades the performance of the NN. At this point, we hypothesize that the presence of large weights makes the output of the NN strongly dominated by a few weights with a large value. This is a widely known phenomenon in deep learning when training NNs to perform regression tasks \cite{brownlee2018better,Goodfellowbook2016}. The NNs featuring large weights tend to have the regression output converging to a few values that are hard-coded from the input. In the case of optical signal equalization, this effect deteriorates the signal constellations into a ``jail window''.

To carry out the analysis further, let us examine the number of outlier weights in the linear layer. As it can be seen in Fig.~\ref{fig:weight_distribution}, the weights greater than one can be considered outliers. We found that in the case of 8-QAM, the real parts of the recovered symbols have twice as many outlier (larger than 1) weights as the imaginary parts (298 for the latter case in comparison with 153 for the former)\footnote{This effect persists in the case of symmetric 16-QAM and other constellations when the ``jail window'' occurs, such that it is not relevant to the asymmetry of the 8-QAM constellations used in our study.} When we increased the outlier weight threshold value criteria from 1 to 4, we found just four weights on the neuron representing the real value of the constellation, while the neuron for the imaginary part had just two. This is the exact ratio between the number of continuous straight ``lines'' (each ``line'' is made up of equalized constellation points) in the amplitudes of the in-phase and quadrature, as seen in Fig.~\ref{fig:8bilstm_batch16}.

We propose the following explanation for the aforementioned patterning effects. When employing small mini-batch sizes with a small cardinality constellation, there exists a noticeable probability that many points in the batch belong to just a single amplitude level. As a result, the NN tends to learn the inverse channel function at this particular nonlinearity level and is more likely to hard-code some inputs to this obvious amplitude output. This issue is less likely to occur with high-cardinality modulation formats since it is less probable that the batch would contain a lot of samples belonging to the same amplitude level. However, even in high modulation formats, we could see the ``jail window'' pattern when using the MLP equalizer, but now with less intensity\footnote{We observed that the ``jail window'' pattern is modulation format dependent. An interesting discussion on the fact that non-linear MSE equalizers realize a stair function can be found in Ref.~\cite{Georg2021book}}. One possible explanation for why the optical signals recovered by the MLP equalizer are more prone to constellation degeneration than in the case of the biLSTM equalizer, is that the feed-forward structure (e.g. the MLP) makes it easier to hard-code input data to the output. At the same time, the recurrent-type structures (e.g. the LSTM type) allow parameters to be shared across the NN model\cite{Goodfellowbook2016}, making the hard-coding effect more difficult to emerge. Here, however, we want to remark on the performance of the MLP and LSTM equalizers. Both Sections \ref{Sec:QOT} and \ref{sec:batch} demonstrated that the LSTM equalization produces a better result than the MLP equalizer. The LSTM, which can be deemed as a nonlinear infinite impulse response (IIR) filter, can represent the inverse of a nonlinear channel more accurately than equalizers that use an MLP structure; the latter is actually a non-linear finite impulse response (FIR) filter\cite{650112}.  IIR filters, unlike FIR filters, do not require that the system has finite memory, giving the LSTM-type equalization an advantage. Theoretically, the MLP may also achieve such a high degree of performance when the input accounts for large enough memory; but, due to the overfitting of the MLP structure\footnote{To recover the nonlinear channel, the MLP would need a complicated structure with large memory. However, such a redundant architecture is equally susceptible to learning the training dataset, which leads to overfitting\cite{freire2021performance}.}, this turns out to be extremely difficult to achieve. On the other hand, because the large-memory handling LSTM structures are simpler than MLPs, the overfitting there does not occur, or at least, is much less pronounced.

After considering the deteriorating effects attributed to the presence of large-value weights, we advocate the use of a well-known regularization technique, $L^2$ regularization, as a feasible way of minimizing the degradation in the equalized signal constellations. The idea behind $L^2$ regularization is to penalize large weights and favor smaller weights throughout the model \cite{brownlee2018better}. From a Bayesian statistics standpoint, the addition of $L^2$ regularization is equivalent to performing a maximum a posteriori estimation with a Gaussian prior; the traditional MSE loss function corresponds to a maximum likelihood estimation when the likelihood has a Gaussian distribution. This means that after the learning, our equalization transfer function (likelihood) has been re-weighted to reflect a prior, making it much more difficult to have hard-coded correlations between the input and output (recall that the hard-coding is one of the reasons causing the ``jail window'' pattern). The following equation represents the $L^2$ contribution to the regularized loss function $L(y,\hat{y})$:
\begin{equation}
    L(y,\hat{y}) = \textrm{MSE}(y,\hat{y}) + \lambda \sum_{i=1}^{n}w_i^2,
\end{equation}
where the first addend is the initial MSE loss function, $\lambda$ is the regularization parameter that must be optimized\footnote{We have tested the regularization parameter from the range between 0.01 and 0.0001, but no drastic difference was observed in terms of Q-factor improvement. In the plots presented in this section we used the parameter value $\lambda=0.001$.}, $n$ is the total number of weights in the NN topology, and $w_i$ is the $i$-th weight in the NN topology.

\begin{figure}[ht!] 
    \centering
 \begin{subfigure}{.48\linewidth}
  \centering
    \includegraphics[width=\linewidth]{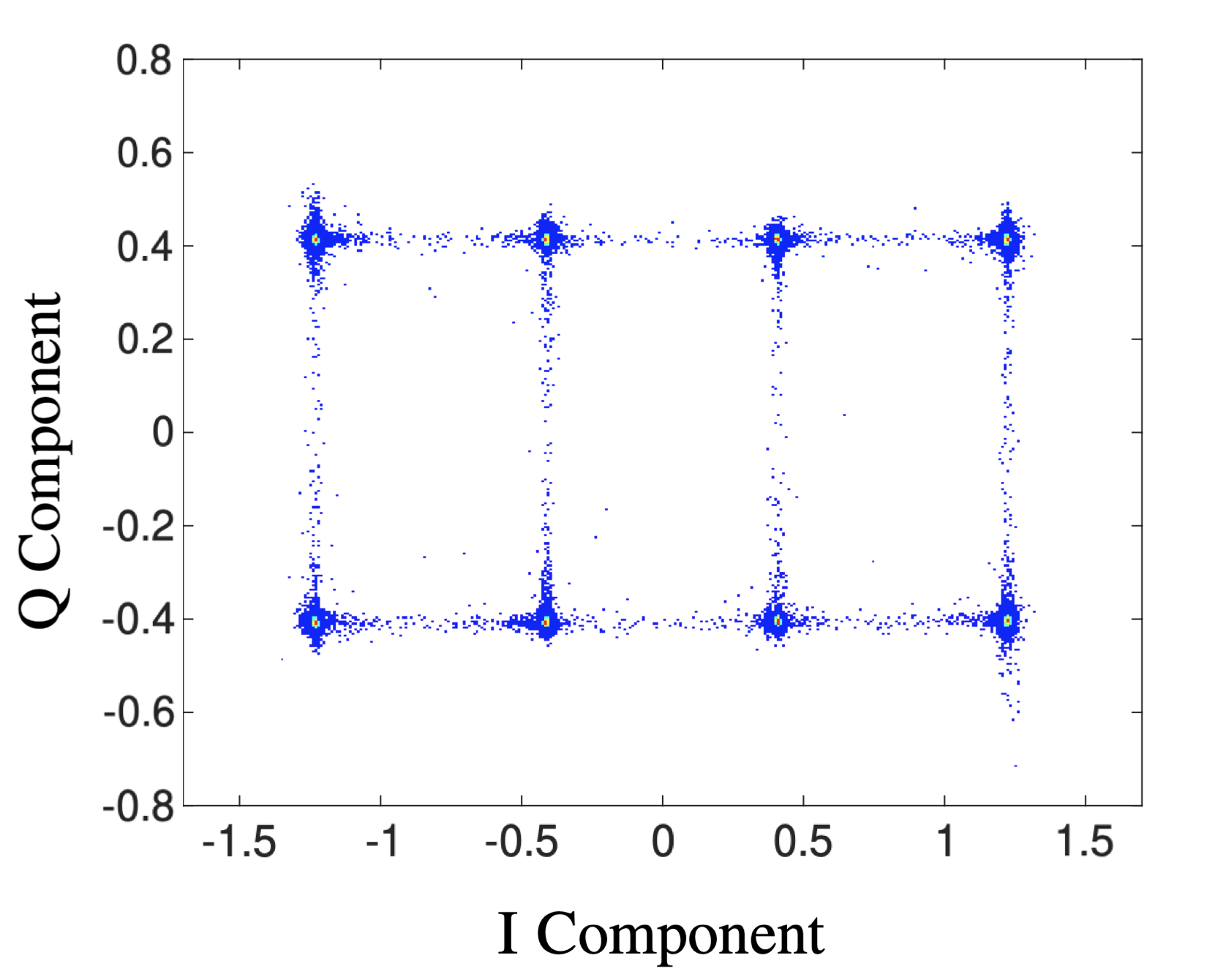}
    \caption{Without regularisation}
    \label{fig:jail_window}
\end{subfigure}
 \begin{subfigure}{.48\linewidth}
  \centering
    \includegraphics[width=\linewidth]{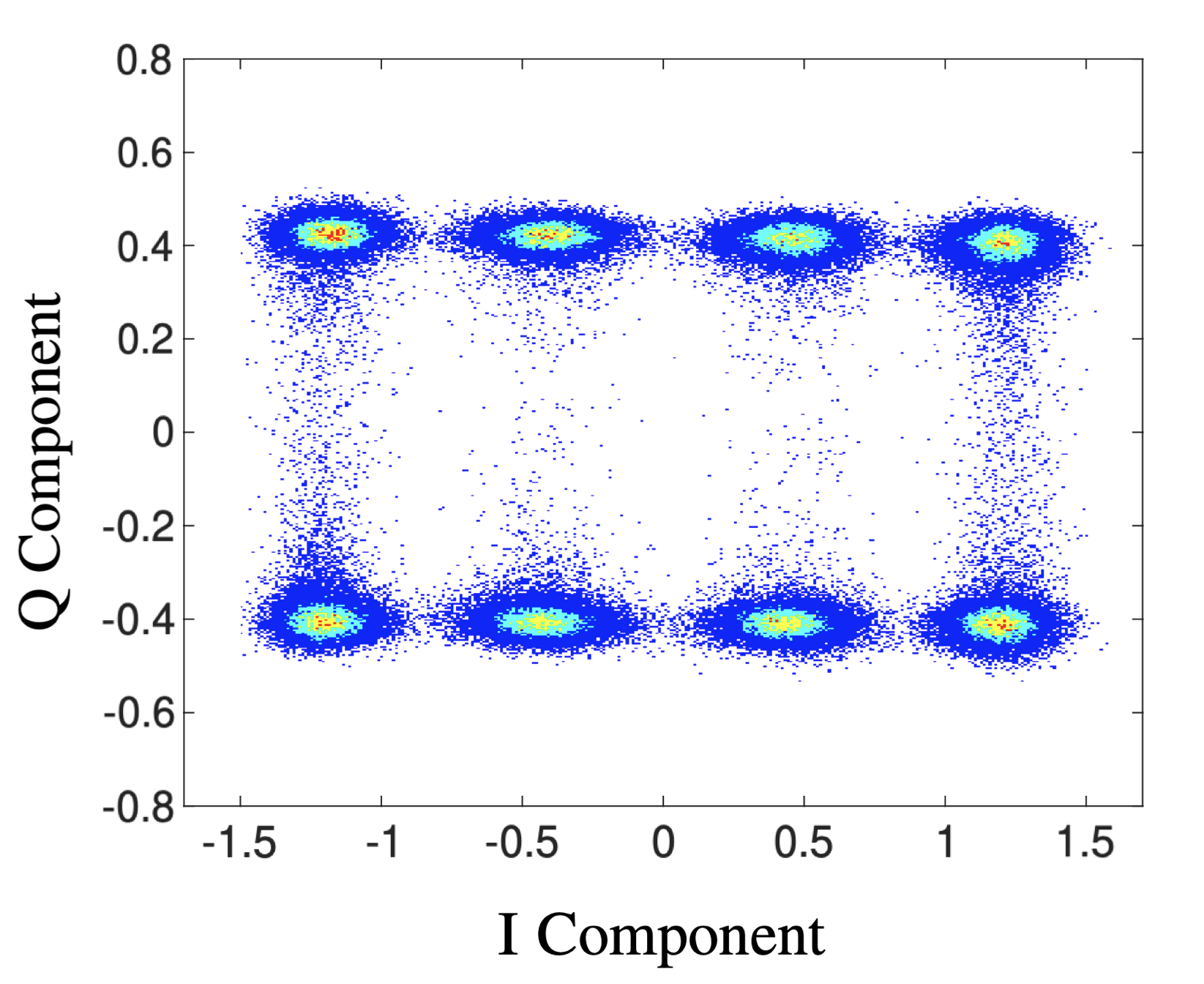}
    \caption{With regularization}
    \label{fig:regularisation_constellations111}
\end{subfigure}
  \caption{8-QAM signal constellations after the NN-based simulated channel equalization when training with mini-batch size equal to 16 without (a) and with (b) $L^2$ regularization.}   \label{fig:compare_with_without_constellations} 
\end{figure}

Fig.~\ref{fig:compare_with_without_constellations} shows the signal constellations of the 8-QAM signal equalized by the NNs trained with the mini-batch size of 16, with and without regularization. We observe that when using the regularization, the degradation of signal constellations can be prevented. It is worth pointing out that although regularization can render better constellations, the equalizing performance is still badly affected by the small mini-batch sizes: even with regularization, we cannot match the performance of the NNs trained with large mini-batches. Thus, while the large weights contribute to the degeneration of constellations, this is not the only detrimental effect caused by a small mini-batch size. This observation suggests that the most important problem with small mini-batch sizes can relate to the insufficient number of unique amplitude levels that are present in each mini-batch, as we suggested above. 

Fig.~\ref{fig:jail_window} and \ref{fig:regularisation_constellations111} demonstrate the signal constellations at the epoch with the highest Q-factor. It was observed that when training the NNs after this level, the performance degeneration took place even with the regularization. Although most of the weights after the regularization turned out to have a relatively low value, there still exist significantly larger outliers. We surmise that these outliers contribute to the distortion of the signal's constellations. Interestingly, using large mini-batch results in a narrower distribution of the weights, which does not feature any significant outliers even when the regularization is not applied. This observation strongly supports our claim that training with an insufficiently large mini-batch brings about the performance degradation of NN-based coherent optical channel equalizers.

\begin{figure}[ht!]
    \centering
    \includegraphics[width=\linewidth]{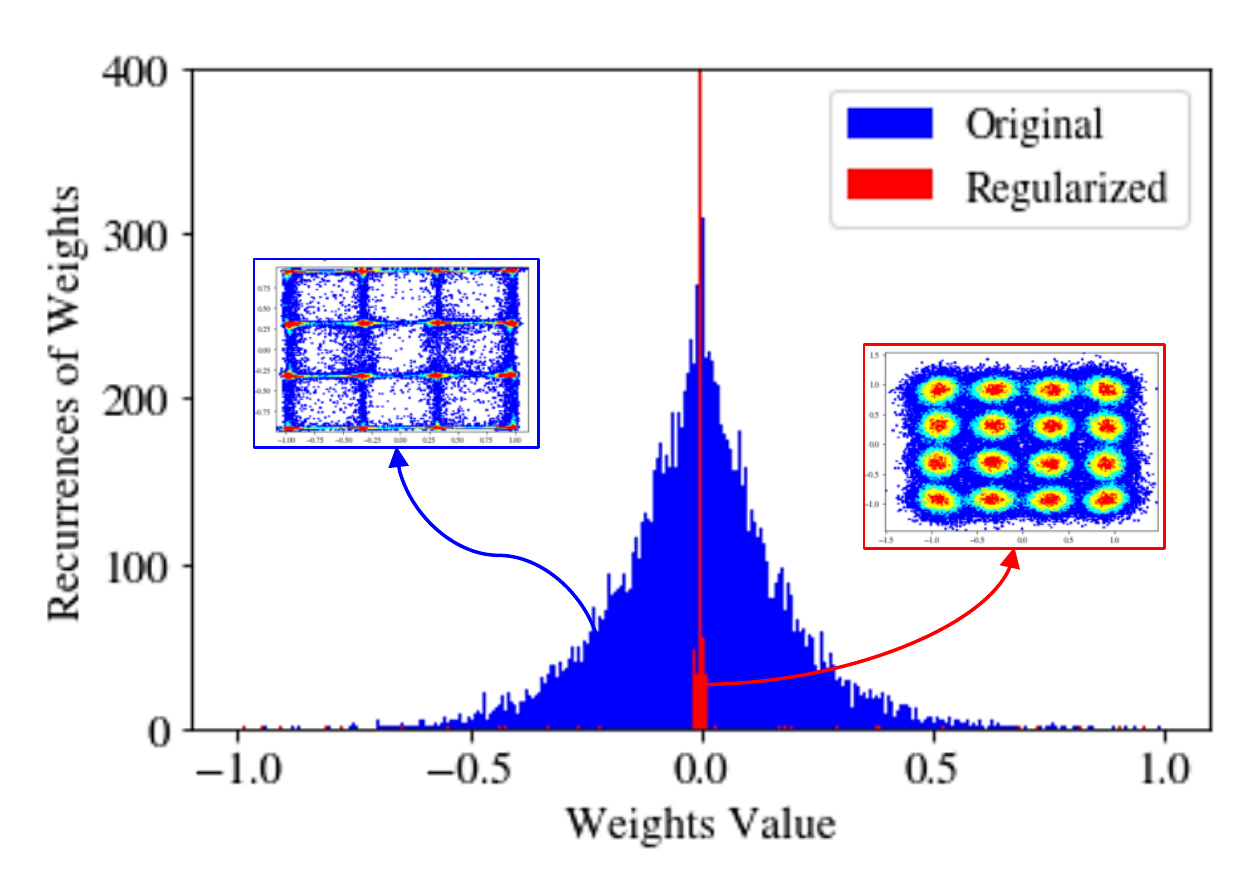}
    \caption{Distribution of MLP weights for the entire model: non-regularized (Original) and regularized cases. The insets show the respective equalized constellations for each case.}
    \label{fig:jail_regularied_MLP}
\end{figure}

Finally, we applied the $L^2$ regularization to all hidden layers of the MLP equalizer in another transmission case where the ``jail window'' was really strong: it was the case (ii) described in Section \ref{Sec:QOT} and given in Fig.~\ref{Fig_constellation_QOT} (b). The weight distribution of the entire model for the MSE (marked as ``Original'') and for the MSE with the $L^2$ regularization (marked as ``Regularized'') is presented in Fig.~\ref{fig:jail_regularied_MLP}. From this result we can see that, this time, for a higher modulation format (16-QAM), the regularization makes the original ``jail window'' constellation be again a set of ''Gaussian-like'' clusters, see the corresponding insets in Fig.~\ref{fig:jail_regularied_MLP}. However, even though the regularization produced a visually different constellation output,  the original and regularized model revealed almost identical performance in terms of the eventual Q-factor: the original one gave 7.89~dB, regularized -- 7.61~dB. This fact shows that regularization is not the solution to make the model learn further\footnote{Using the $L^2$ regularizer can also make the NN model be a single point attractor in the space of its weights, where the attractor is located at the origin. In such a case, any information that was inserted into the NN model dies out exponentially fast.  As described in Ref.\cite{pascanu2013difficulty}, this can hide long-term memory traces that impact the learning process in our task.}, but it rather helped us to identify which element in the NN-model was responsible for the ``jail window'' pattern.

\begin{table*}[ht!]
  \centering
  \caption{Summary of latency values for different hyperparameters sets.}
    \resizebox{\textwidth}{!}{
\begin{tabular}{|c|c|c|c|c|c|c|c|c|}
\hline
                           & \multicolumn{4}{c|}{MLP Equalizer}                                                                                                                                                                                                                       & \multicolumn{4}{c|}{biLSTM  Equalizer}                                                                                                                                                                                                                  \\ \cline{2-9} 
\multirow{-2}{*}{Topology} & Hyperparameters               & \begin{tabular}[c]{@{}c@{}}NN\\  Parameters\end{tabular} & RMpS & \begin{tabular}[c]{@{}c@{}}Latency\\  (s)\end{tabular} & Hyperparameters & \begin{tabular}[c]{@{}c@{}}NN\\  Parameters\end{tabular} & RMpS              & \begin{tabular}[c]{@{}c@{}}Latency \\ (s)\end{tabular} \\ \hline\hline

1  & $N= 35$ ; $n_{1/2}= 600 /518$ & $482,236$                                                                        & $4.82\text{e+}5$            & $7.87\text{e-}5$                                                              & $N= 5$ ; $n_{h}= 73 $                   & $48,180$                                                 & $5.03\text{e+}5$                         & $1.03\text{e-}4$                                                              \\ \hline

2  & $N= 35$ ; $n_{1/2/3}= 400 /460/400$                   & $482,400$                                                                        & $4.82\text{e+}5$            & $7.90\text{e-}5$                                                              & $N= 10$ ; $n_{h}= 52$                   & $27,664$                                                                         & $5.00\text{e+}5$                         & $1.30\text{e-}4$                                                              \\ \hline

3  & $N= 35$ ; $n_{1/2/3/4}= 200 /156/553/555$             & $482,293$                                                                        & $4.82\text{e+}5$            & $7.06\text{e-}5$                                                              & $N= 14$ ; $n_{h}= 44 $                  & $22,000$                                                                         & $5.03\text{e+}5$                         & $1.46\text{e-}4$                                                              \\ \hline

4  & $N= 45$ ; $n_{1/2}= 1250 /458$                        & $1,028,416$                                                                      & $1.02\text{e+}6$            & $1.33\text{e-}4$                                                              & $N= 25$ ; $n_{h}= 48 $                  & $29,760$                                                                         & $1.04\text{e+}6$                         & $2.82\text{e-}4$                                                              \\ \hline

5  & $N= 45$ ; $n_{1/2/3}= 620/564/800$                    & $1,028,160$                                                                      & $1.02\text{e+}6$            & $1.47\text{e-}4$                                                              & $N= 30$ ; $n_{h}= 43 $                  & $26,660$                                                                         & $1.01\text{e+}6$                         & $2.51\text{e-}4$                                                              \\ \hline

6  & $N= 45$ ; $n_{1/2/3/4}= 610 /500/500/501$             & $1,028,542$                                                                      & $1.02\text{e+}6$            & $1.21\text{e-}4$                                                              & $N= 35$ ; $n_{h}= 40$                   & $25,440$                                                                         & $1.03\text{e+}6$                         & $3.14\text{e-}4$                                                              \\ \hline

7  & $N= 60$ ; $n_{1/2/3}= 1000/1000/600$                  & $2,085,200$                                                                      & $2.08\text{e+}6$            & $2.51\text{e-}4$                                                              & $N= 40$ ; $n_{h}= 53$                   & $41,340$                                                                         & $2.00\text{e+}6$                         & $5.34\text{e-}4$                                                              \\ \hline

8  & $N= 50$ ; $n_{1/2/3/4}= 2000 /500/910/200$            & $4,087,000$                                                                      & $4.08\text{e+}6$            & $4.30\text{e-}4$                                                              & $N= 50$ ; $n_{h}= 68$                   & $66,640$                                                                         & $4.02\text{e+}6$ & $9.30\text{e-}4$                                                              \\ \hline
\end{tabular}
}
    \label{Table_Latency}
\end{table*}

\section{Computational complexity analysis: Number of parameters versus number of multiplications} \label{sec:complex}

One of the most critical elements in implementing a deep NN in an end-product is the signal processing latency associated with the signal passing through the NN. The majority of real-world applications demand a really short inference (processing) time\cite{mikami2019field, pokhrel2021towards}, ranging from a few milliseconds to a second.  However, evaluating a NN's inference time, or latency, accurately and efficiently, can be a difficult problem\cite{ponomarev2021latency}. The consequences of latency evaluation errors can lead to poor decisions about the implementation of NN. At the same time, the device's power consumption and the required hardware size for the NN implementation are no less important characteristics; the latter is also related to the NN's CC. Additionally, we note that the latency is actually architecture dependent. For instance, we can implement an MLP in a fully parallel way, which would provide the lowest latency but the highest power consumption. Alternatively, we can implement a single multiplier unit and process the multiplications sequentially. This would result in the lowest power consumption and the highest latency, such that we always ought to seek the desirable balance between the power consumption and latency. In this paper, to better understand the relevance of latency and our CC metrics, we measure the average latency to recover one symbol using a Colab CPU (4vCPU at 2.0~GHz with 26~GB RAM) for sequential NN architectures \footnote{Note that with a different architecture, e.g. on GPU, TPU, different CPU, FPGA, or ASIC, the latency can be drastically different\cite{jouppi2017datacenter}.}. In this section, we will analyze the number of real multiplications per recovered symbol (RMpS) as a CC metric. Technically, if well-defined, this CC metric can help in the assessment of a solution's design appropriateness before going to the hardware level. In the following, we will identify three aspects that should be considered when evaluating the CC of any NN-based equalizer.

First, we address two important points. i) We show that the RMpS metric provides a good estimate of the CC by demonstrating its proportionality to the averaged latency (inference time) for one equalized symbol in a pure sequential architecture. ii) We show that the number of weights of the NN-based equalizer is not a good metric for assessing the NN's complexity. To showcase these points, we have tested eight different topologies for MLP and biLSTM equalizers with different levels of RMpS and measured their one-symbol latency algebraically averaged over 1M recovered symbols. The results of the tests performed in this subsection are summarized in Table~\ref{Table_Latency}. To account for the number of parameters, we used the TensorFlow application for each topology, and for the proper RMpS computation, we used the equations introduced in Ref.~\cite{freire2021performance}. The expressions for the RMpS of the MLP having 2, 3, and 4 layers, and biLSTM equalizers are, respectively:
\begin{equation}
\label{Eq_c1}
  C_{\text{MLP}_2}=
  n_s n_i n_1+
  n_1n_2+ n_2n_o,
\end{equation}
\begin{equation}
\label{Eq_c2}
  C_{\text{MLP}_3}=
  n_s n_i n_1+
  n_1n_2+ n_2n_3+
  n_3n_o,
\end{equation}
\begin{equation}
\label{Eq_c3}
  C_{\text{MLP}_4}=
  n_s n_i n_1+
  n_1n_2+ n_2n_3+
  n_3n_4 + n_4n_o,
\end{equation}
\begin{equation}
\label{Eq_c4}
 C_{\text{biLSTM}}= 2
  n_{s}n_{h}(4n_i+4n_{h}+3+n_o),
\end{equation}
where $n_s$ is the input time sequence size, with the memory size $n_s=2N+1$, with $N$ being the number of neighboring symbols considered, and $n_{i}$ being the number of input features, which, in our case, is equal to $4$ (the number of outputs per symbol); $n_o$ is equal to $2$, $n_{1,2,3,4}$ are the number of neurons in each respective hidden layer, and $n_h$ is the number of hidden units in the LSTM cell.

From Table~\ref{Table_Latency}, we see that for the same equalizer type and similar RMpS, the averaged latency values are also very similar. For example, for topologies $1$, $2$, and $3$, the latency of the MLP equalizer was around $7.8\times10^{-5}$~s, and for the biLSTM equalizer it was around $1.2\times10^{-4}$~s. Also, we can infer that the number of weights in the MLP is exactly equal to the RMpS. However, this equality is true only when the equalizer is solely composed of dense layers, as it is in the MLP case. 

\begin{figure*}[ht!] 
    \centering
  \begin{subfigure}[b]{0.475\linewidth}
    \centering
 \begin{tikzpicture}[scale=0.7]
 \pgfplotsset{every axis/.style={ymin=0}}
\begin{axis}[ scale only axis,  axis y line*=left, xlabel= Averaged Latency ($1\mathrm{e}{-5}$ s), ylabel= \textcolor{blue}{RMpS }]
    \addplot[blue, mark=o, draw] coordinates {
    (7.61, 482236)(13.3,1028416)
    (25.1,2085200)(43,4087000)
    };
\end{axis}
%
\begin{axis}[red, scale only axis, axis y line*=right, axis x line=none, ylabel= Number of Parameters]%
    \addplot[red, mark=x] coordinates {
    (7.61, 482236)(13,1028416)
    (25,2085200)(43,4087000)
    };
\end{axis}

\end{tikzpicture}
    \caption{MLP Equalizer Study}
    \label{fig_latency:a} 
    \vspace{4ex}
  \end{subfigure} 
  \hfill
  \begin{subfigure}[b]{0.475\linewidth}
    \centering
 \begin{tikzpicture}[scale=0.7]
 \pgfplotsset{every axis/.style={ymin=0}}
\begin{axis}[ scale only axis,  axis y line*=left, xlabel= Averaged Latency ($1\mathrm{e}{-4}$ s), ylabel= \textcolor{blue}{RMpS}]
    \addplot[blue, mark=o, draw] coordinates {
    (1.0, 503000)(2.8,1040000)
    (5.34,2005200)(9.3,4021000)
    };
\end{axis}
\begin{axis}[red, scale only axis,  axis y line*=right, axis x line=none, ylabel= Number of Parameters]%
    \addplot[red, mark=x] coordinates {
    (1.0, 48180)(2.8,29760)
    (5.34,41340)(9.3,66640)
    };
\end{axis}

\end{tikzpicture}
    \caption{biLSTM Equalizer Study}
    \label{fig_latency:b} 
    \vspace{4ex}
  \end{subfigure}
  \caption{Processing time to equalize one symbol vs. RMpS comparison, and vs. the number of trainable parameters in the respective NN.}
  \label{fig_latency} 
\end{figure*}
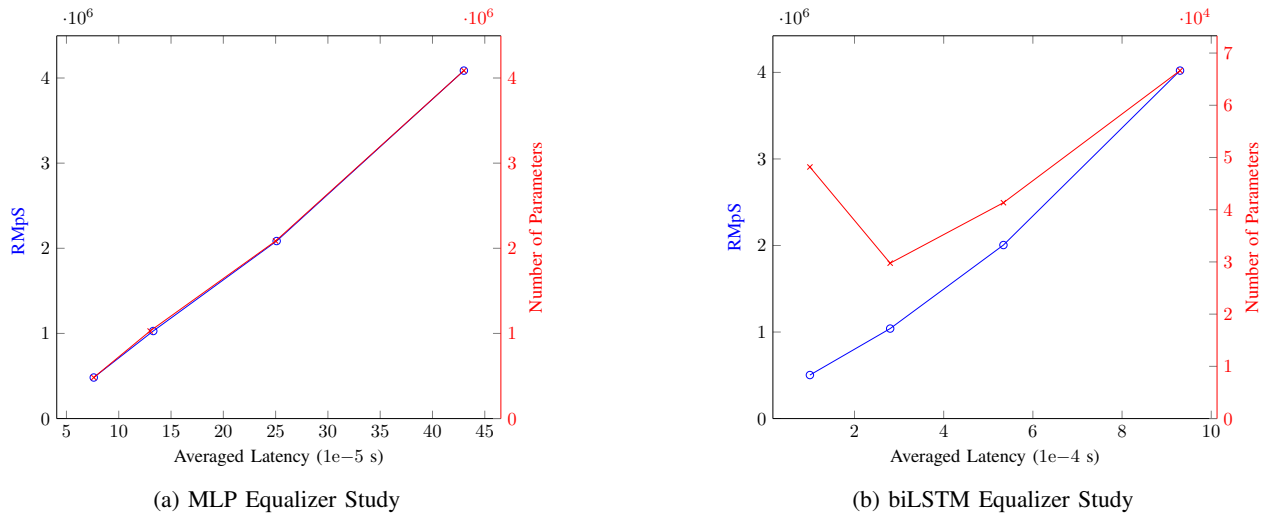

Regarding the number of parameters (trainable weights) in the equalizer, two main conclusions can be drawn from Table~\ref{Table_Latency}. First, for the same level of RMpS, the number of parameters for the MLP is much larger than that number for the biLSTM equalizer. This indicates that comparing equalizers with a different structure in terms of their number of parameters may be misleading. Then, we can also see that, for the biLSTM case, the increase in the number of parameters does not necessarily cause the increase in RMpS (in the CC) and the latency. This is an important observation since, theoretically, for our implemented sequential NN models, the latency should increase with the CC growth. For better visualization, in Fig.~\ref{fig_latency} we show the interrelation between the latency, RMpS, and the number of NN structure parameters. As we can see, the latency grows almost linearly with the RMpS metric, which confirms that our metric is a good estimate for the CC. On the other hand, we claim that the number of parameters does not represent a good estimation for the CC in the case of recurrent and convolutional layers, and it is not a good metric to compare the CC across different NN architectures. Therefore, to address the complexity analyses, we should always consider the RMpS value, while the number of NN parameters is relevant only if we are interested in the memory required by the NN or in the NN's training complexity.   

The second point that we mention, refers to the memory access cost. In the case when we have approximately the same RMpS number, the biLSTM equalizer's latency is almost two times higher than that of the MLP equalizer; see Topology 7 from Table \ref{Table_Latency} with the RMpS equal to $\approx2 \times 10^6$. This effect is the consequence of Keras's implementation of the LSTM layer, in which the input data are subject to additional pre-processing operations to construct the state of each recurrent cell, and this pre-processing requires additional memory usage\footnote{Note that the latency would naturally change when using GPUs since the implementation of LSTM changes for such hardware.}. However, in our expression for the CC using RMpS, we do not account for such a memory usage impact. The cost of communication with the memory is usually much larger than the cost of ``local'' computation\cite{pacheco2011introduction}.
In this context, the term communication refers to the movement of data between the levels of memory or between multiple processors on a network. Therefore, when computing the CC of our equalizers, the proposition of using memory to save/access previous computations to reduce the number of multiplications (RMpS) must be taken with caution, as the cost of moving data (measured in time or energy expenditures) can (and usually does) exceed the cost of an arithmetic operation by orders of magnitude depending on the technology used \cite{demmel2012communication, ballard2014communication}. To illustrate this problem, we notice that by using modern SRAMs~\cite{Imani2016}, the read and write estimated latencies for a 64~Kb SRAM, used for the NN, are $\sim$1~ns; for the most costly operations in NN inference (processing), the multiplications, Ref.\cite{ullah2018area} shows that a 16$\times$16 bits multiplier can have up to $\sim$10~ns estimated latency. Consider the scenario where the multiplication in our NN model is used 100 times. For a parallel implementation, 100 multipliers could be used and the total latency would be just 10~ns. But, turning to the NN implementation using the SRAM to save the multiplications, we need to consider that SRAMs usually has one read port and one write port, so only one parameter can be read at a time. Therefore, the total value would be 10~ns of the multiplication plus 1~ns to write in the SRAM, plus 100 times we read the values of this multiplication, which eventually gives us the estimate: 111~ns. Thus, we see that SRAM usage, in fact, becomes the processing bottleneck that drastically increases overall latency. This is the infamous latency versus complexity trade-off problem in hardware implementation, where we can use fewer resources of our hardware by sacrificing the latency, and vice versa, more resources allow us to reduce the processing time.
Therefore, when going to the level of design for the NN equalizer using some SRAM, it is not completely fair to directly compare the reduced CC (the RMpS number) of such solutions to the CC of some traditional DSP techniques (say, to those of traditional digital back-propagation, the often-used benchmark), since SRAM will also effectively add to the complexity when implementing the particular design in hardware.  Instead, we can say that by using advanced compression techniques and hardware technologies, a model that shares the same multiplications/parameters across its architecture and uses the SRAM to repeatedly access them, can potentially be less complex in terms of the hardware implementation at the expense of worse latency.

Finally, one more direction to account for when we assess the complexity refers to NN pruning and quantization techniques applied to the developed equalizer's structure \cite{fujisawa2022weight,pedro_quantization2021}. The purpose of quantization and pruning is to make the NN implementation in hardware more resource-efficient. In this context, pruning refers to the practice of removing weights from the original trained model. We utilize the number of bit operations (BoPs) metric \cite{hawks2021ps} to measure NN's CC because the BoP metric is especially important when evaluating the performance of mixed-precision arithmetic in hardware implementations on FPGAs or ASICs. As described in Ref.~\cite{hawks2021ps}, the BoPs of a dense layer with $n$ neurons and input with $m$ features can be expressed as:
\begin{equation} \label{bop}
      \mathrm{BoP_{dense}} = nm \, ( b_{w} b_{i}  + b_{w} + b_{i} +\lceil \log_2(m) \rceil),
\end{equation}
%
where, in the context of the NN, $b_w$ is the number of bits used to represent the weights of the NN, $b_i$ is the number of bits used to represent the input/ activation function, and $\lceil x \rceil $ is a ceiling function. From this equation, we observe that  the number of multiplications is multiplied by  $b_w b_i$, and the number of additions is multiplied by  $b_{w} + b_{i} +\lceil \log_2(m)\rceil$.  The latter is the actual bit width of the accumulator needed in MAC operations (the accumulator is a register in which the results of the intermediate arithmetic logic unit are stored). Also, note that this expression considers a dense layer with weight-matrix multiplication and bias addition, otherwise, the number of additions would be $n(m-1)$. Now, considering unstructured pruning (that is, we remove the least important connections/weights rather than entire layers or neurons), we can include the sparsity impact in Eq.~(\ref{bop}) as follows:
\begin{equation} \label{bop2}
\mathrm{BoP_{dense}} = nm(b_{w} b_{i}\times (1- \eta_s)  + b_{w} + b_{i} +\lceil \log_2(m) \rceil).
\end{equation}

According to the equation above, the CC grows quadratically with bit widths and linearly with the pruning/sparsity ratio $\eta_s$, which is the percentage of connections erased from the layer. Now, considering the case with multiple layers, let us take, as an example, the MLP structure with 3 hidden layers (with the number of neurons $n_1$/$n_2$/$n_3$), input features number $m$, and output features number $n_o$. If we assume that the weights have the same quantization characterized by the number of bits $b_w$ throughout the structure, and each input/activation function with $b_i$ bits, as well as each layer is pruned equally with the sparsity ratio $\eta_s$, the BoPs for the MLP can be described in terms of the RMpS metric from Eq.~(\ref{Eq_c2}) as:
\begin{equation} \label{bop3}
\mathrm{BoP_{MLP}} = C_{MLP_3}(b_{w} b_{i}\times (1- \eta_s) + b_{w} + b_{i}) + \mathrm{ACC},
\end{equation}
where ``ACC'' is part of the cost attributed to the accumulator and in this case is equal to $m n_1 \lceil \log_2(m) \rceil+ n_1n_2 \lceil \log_2(n_1) \rceil+n_2n_3 \lceil \log_2(n_2) \rceil+n_3n_o \lceil \log_2(n_3) \rceil$.

However, regarding Eq.~(\ref{bop3}) we, first, stress that it is not applicable when pruning is non-uniform \footnote{Non-uniform pruning means that different layers of the NN model are pruned with different sparsity ratios.}: If another type of pruning is used, the term $C_{MLP_3}\times (1- \eta_s)$ has to be replaced by the total number of multiplications emerging from a particular non-uniform pruning scheme. Second, when dealing with quantization, we should be aware of all quantization levels which are used in the NN equalizer. As mentioned in Ref.~\cite{jacob2018quantization}, different quantization levels can be applied in the activation functions, input, and weights of each layer, and this obviously affects the overall complexity. Therefore, if different quantizations are used in different layers or the input, this must be counted separately, multiplying $b_w b_i$ per layer as in \cite{pedro_quantization2021}. It is a common error to assume that when quantizing refers to just the weights, the complexity would drop quadratically: the CC would drop quadratically only \textit{if we equally reduce the quantization bit width of both the input and activation function}.

\begin{table*}[tbh]
\caption{Overview of Main Issues in Devising Efficient NN-based Equalizers }
    \centering
    \resizebox{\textwidth}{!}{
\begin{tabular}{cll}
\hline\hline
\rowcolor[HTML]{D3D3D3} 
\textbf{\normalsize Topic} & \multicolumn{1}{c}{\cellcolor[HTML]{D3D3D3}\textbf{ \normalsize Problem}} & \multicolumn{1}{c}{\cellcolor[HTML]{D3D3D3}\textbf{\normalsize Solution}} \\ \hline\hline
\textbf{QoT Metrics} & \begin{tabular}[c]{@{}l@{}}The ``jail window'' effect in the NN-based equalizers.  The use of\\ the metrics assuming Gaussian channel statistics, such as \\EVM and SNR can lead to overestimated results in terms of BER.\end{tabular} & \begin{tabular}[c]{@{}l@{}}To avoid overestimation it is recommended to always report the results in terms\\ of BER or Q-factor calculated directly from BER.\end{tabular} \\ \hline
\textbf{PRBS Order} & \begin{tabular}[c]{@{}l@{}}Depending on the PRBS order, the NN can learn the PRBS periodicity\\  or the generation rule of the PRBS sequence that is based on its\\  linear feedback shift register.\end{tabular} & \begin{tabular}[c]{@{}l@{}}The training and testing datasets must be different fragments within \\  the PRBS periodicity range. This periodicity depends on the PRBS \\ order and the modulation format. Also, high PRBS orders should be used, ($\geq$32) \\, and the NNs are to be tested with a B2B system to guarantee the absence  \\ of overestimation and overfitting effects.\end{tabular} \\ \hline
\textbf{DAC Memory} & \begin{tabular}[c]{@{}l@{}}Depending on the DSP implementation, the DAC memory  limits\\  the data variability to a few thousand samples, which makes the NN to \\ overfit quickly into the training dataset statistics.\end{tabular} & \begin{tabular}[c]{@{}l@{}}Increase the variability by, concatenating different traces generated with different\\  random seeds to create  a large training dataset of more than 13M symbols, and \\ train the NN with different random 1M symbols from this training dataset for\\ each epoch.\end{tabular} \\ \hline
\textbf{Regression vs Classification} & \begin{tabular}[c]{@{}l@{}}The regression and classification have some statistical and machine \\ learning drawbacks that can impact the learning in the field of \\ nonlinearity mitigation in coherent optical channels.\end{tabular} & \begin{tabular}[c]{@{}l@{}}In all our tests regression has performed better than classification. However, we\\  mention that this check always needs to be done. For fairness reasons, take into account \\that mini-batch size,  learning rate, and the number of epochs need to be individually \\ tuned for each configuration since the gradients are different when using CEL or MSE.\end{tabular} \\ \hline
\textbf{Batch Size Study} & \begin{tabular}[c]{@{}l@{}}The batch size is an important hyperparameter to guarantee a proper \\ learning process of channel equalization. Using small mini-batch sizes,\\ in the regression can make the NN move the learning in a direction \\ that is not the best.\end{tabular} & \begin{tabular}[c]{@{}l@{}}We observed that using a large mini-batch size ($\geq$1000) is good for generalization \\ when using regression for NN-based equalizers. We see that when using small \\ mini-batches, the  NN was not generalizing for the different levels of nonlinearity \\ for points in the QAM constellation\end{tabular} \\ \hline
\textbf{Computational Complexity} & \begin{tabular}[c]{@{}l@{}}The computational complexity can be confused  with the number of NN \\ parameters. Also, it can be mistakenly underestimated if one uses memory \\ to save some common multiplications.  The underestimation can take place\\ as well if quantization and pruning are not accounted for accurately.\end{tabular} & \begin{tabular}[c]{@{}l@{}}We recommend reporting the computational complexity  in terms of  real\\  multiplications per recovered symbol. In the same way, when using the quantization\\and pruning, the CC should be reported in terms of BoPs, Eq~(\ref{bop}). Also,\\   when reporting the CC values considering that some computations can\\ be saved to be used later, it needs to be clearly stated that latency is not a constraint.\end{tabular} \\ \hline\hline
\end{tabular}
    }
    \label{tab:overview}
\end{table*}

Notice that, theoretically, the signal/activation function quantization has a floor depending on the modulation format used in the transmission. Recall that each M-QAM symbol denotes a single number from a set of $M$ symbols in the QAM alphabet. That is, the real and imaginary components of each M-QAM symbol are represented by a number in the range $\sqrt M$. As a result, both real and imaginary parts' resolution should be significantly higher than $ \log_2(\sqrt M)$. For example, in a 64-QAM system, the real/imaginary parts of each symbol must be represented by at least $\log_2(\sqrt 64) = 3$ quantization levels (bits in a quantized quantity's representation), and the minimal number of levels in the quantized complex symbol's representation should be no less than 3+3 bits in total. However, in the exemplary case of 64-QAM, using a value close to 6 bits to represent the NN's complex-valued input typically results in a substantial performance degradation because the system would virtually always make hard decisions and lose crucial equalization features. Regarding the quantization of the weights, the bit-precision for the weights is more flexible: it can be as low as allowed by the performance level of the model, and such advanced techniques as, e.g., the quantization-aware training\cite{jacob2018quantization} can be used to achieve the low-bit quantization of the weights.

\section{Conclusion}

In this paper, we revealed and studied important hidden caveats and pitfalls that we have observed in recent publications and our own research dealing with the applications of machine learning methods and, particularly, the NNs for nonlinear impairments' compensation in coherent optical communications. We have grouped our original results here in such a way that this paper can also serve as a guidance and tutorial in this rapidly-growing field, drawing out the lessons learned and aiming at somewhat navigating the researchers in the area. We note that this work is not purely a traditional review insofar as we presented a number of completely new results regarding the difficulties in designing efficient NN equalizers. We believe that our results can foster new concepts and artificial intelligence techniques, allowing researchers to avoid known design process pitfalls and misinterpretation of results or findings. We underline again the challenges and common misconception errors that often occur in the development of NN-based equalizers applied in high-speed coherent optical communication systems: a poor model generalization, dependent dataset characteristics (periodicity), overfitting behavior and indications, performance overestimation, and inaccuracy in estimating the computational complexity. For convenience, our findings and recommendations are summoned up in Table.~\ref{tab:overview}. Although this exploration of pitfalls is, without a doubt, far from complete, we believe that we have covered the most common problems that pose a particularly high risk for the design of efficient NN-based equalizers and other machine learning structures used in optical transmission lines. 

\bibliographystyle{IEEEtran}
\bibliography{references}

\begin{thebibliography}{10}
\providecommand{\url}[1]{#1}
\csname url@samestyle\endcsname
\providecommand{\newblock}{\relax}
\providecommand{\bibinfo}[2]{#2}
\providecommand{\BIBentrySTDinterwordspacing}{\spaceskip=0pt\relax}
\providecommand{\BIBentryALTinterwordstretchfactor}{4}
\providecommand{\BIBentryALTinterwordspacing}{\spaceskip=\fontdimen2\font plus
\BIBentryALTinterwordstretchfactor\fontdimen3\font minus
  \fontdimen4\font\relax}
\providecommand{\BIBforeignlanguage}[2]{{%
\expandafter\ifx\csname l@#1\endcsname\relax
\typeout{** WARNING: IEEEtran.bst: No hyphenation pattern has been}%
\typeout{** loaded for the language `#1'. Using the pattern for}%
\typeout{** the default language instead.}%
\else
\language=\csname l@#1\endcsname
\fi
#2}}
\providecommand{\BIBdecl}{\relax}
\BIBdecl

\bibitem{jarajreh2014artificial}
M.~A. Jarajreh, E.~Giacoumidis, I.~Aldaya, S.~T. Le, A.~Tsokanos,
  Z.~Ghassemlooy, and N.~J. Doran, ``Artificial neural network nonlinear
  equalizer for coherent optical ofdm,'' \emph{IEEE Photonics Technology
  Letters}, vol.~27, no.~4, pp. 387--390, 2014.

\bibitem{Darko2015}
D.~Zibar, M.~Piels, R.~Jones, and C.~G. Schäeffer, ``Machine learning
  techniques in optical communication,'' \emph{Journal of Lightwave
  Technology}, vol.~34, no.~6, pp. 1442--1452, 2016.

\bibitem{Hoydis}
T.~O’Shea and J.~Hoydis, ``An introduction to deep learning for the physical
  layer,'' \emph{IEEE Transactions on Cognitive Communications and Networking},
  vol.~3, no.~4, pp. 563--575, 2017.

\bibitem{Hunt_2009}
S.~Hunt, Y.~Sun, A.~Shafarenko, R.~Adams, N.~Davey, B.~Slater, R.~Bhamber,
  S.~Boscolo, and S.~K. Turitsyn, ``Adaptive electrical signal post-processing
  with varying representations in optical communication systems,'' in
  \emph{Engineering Applications of Neural Networks}.\hskip 1em plus 0.5em
  minus 0.4em\relax Springer Berlin Heidelberg, 2009, pp. 235--245.

\bibitem{eriksson2017applying}
T.~A. Eriksson, H.~B{\"u}low, and A.~Leven, ``Applying neural networks in
  optical communication systems: possible pitfalls,'' \emph{IEEE Photonics
  Technology Letters}, vol.~29, no.~23, pp. 2091--2094, 2017.

\bibitem{WANG20171}
\BIBentryALTinterwordspacing
D.~Wang, M.~Zhang, Z.~Li, C.~Song, M.~Fu, J.~Li, and X.~Chen, ``System
  impairment compensation in coherent optical communications by using a
  bio-inspired detector based on artificial neural network and genetic
  algorithm,'' \emph{Optics Communications}, vol. 399, pp. 1--12, 2017.
  [Online]. Available:
  \url{https://www.sciencedirect.com/science/article/pii/S0030401817303346}
\BIBentrySTDinterwordspacing

\bibitem{hager2018nonlinear}
C.~H{\"a}ger and H.~D. Pfister, ``Nonlinear interference mitigation via deep
  neural networks,'' in \emph{2018 Optical Fiber Communications Conference and
  Exposition (OFC)}.\hskip 1em plus 0.5em minus 0.4em\relax IEEE, 2018, pp.
  1--3.

\bibitem{zhang2019field}
S.~Zhang, F.~Yaman, K.~Nakamura, T.~Inoue, V.~Kamalov, L.~Jovanovski,
  V.~Vusirikala, E.~Mateo, Y.~Inada, and T.~Wang, ``Field and lab experimental
  demonstration of nonlinear impairment compensation using neural networks,''
  \emph{Nature communications}, vol.~10, no.~1, pp. 1--8, 2019.

\bibitem{Zibar19}
F.~{Musumeci} \emph{et~al.}, ``An overview on application of machine learning
  techniques in optical networks,'' \emph{IEEE Commun. Surveys Tuts.}, vol.~21,
  no.~2, pp. 1383--1408, Jun. 2019.

\bibitem{Khan19}
F.~N. Khan, Q.~Fan, C.~Lu, and A.~P.~T. Lau, ``An optical communication's
  perspective on machine learning and its applications,'' \emph{Journal of
  Lightwawe Technology}, vol.~37, no.~2, pp. 493--516, Jan. 2019.

\bibitem{freire2020complex}
P.~J. Freire, V.~Neskornuik, A.~Napoli, B.~Spinnler, N.~Costa, G.~Khanna,
  E.~Riccardi, J.~E. Prilepsky, and S.~K. Turitsyn, ``Complex-valued neural
  network design for mitigation of signal distortions in optical links,''
  \emph{Journal of Lightwave Technology}, vol.~39, no.~6, pp. 1696--1705, 2021.

\bibitem{schadler2021soft}
M.~Sch{\"a}dler, G.~B{\"o}cherer, and S.~Pachnicke, ``Soft-demapping for short
  reach optical communication: A comparison of deep neural networks and
  volterra series,'' \emph{Journal of Lightwave Technology}, vol.~39, no.~10,
  pp. 3095--3105, 2021.

\bibitem{9645206}
X.~Lin, S.~Luo, S.~K.~O. Soman, O.~A. Dobre, L.~Lampe, D.~Chang, and C.~Li,
  ``Perturbation theory-aided learned digital back-propagation scheme for
  optical fiber nonlinearity compensation,'' \emph{Journal of Lightwave
  Technology}, vol.~40, no.~7, pp. 1981--1988, 2022.

\bibitem{freire2021performance}
P.~J. Freire, Y.~Osadchuk, B.~Spinnler, A.~Napoli, W.~Schairer, N.~Costa, J.~E.
  Prilepsky, and S.~K. Turitsyn, ``Performance versus complexity study of
  neural network equalizers in coherent optical systems,'' \emph{Journal of
  Lightwave Technology}, vol.~39, no.~19, pp. 6085--6096, 2021.

\bibitem{freire2021transfer}
P.~J. Freire, D.~Abode, J.~E. Prilepsky, N.~Costa, B.~Spinnler, A.~Napoli, and
  S.~K. Turitsyn, ``Transfer learning for neural networks-based equalizers in
  coherent optical systems,'' \emph{Journal of Lightwave Technology}, vol.~39,
  no.~21, pp. 6733--6745, 2021.

\bibitem{hawkins2004problem}
D.~M. Hawkins, ``The problem of overfitting,'' \emph{Journal of chemical
  information and computer sciences}, vol.~44, no.~1, pp. 1--12, 2004.

\bibitem{kuschnerov2010data}
M.~Kuschnerov, M.~Chouayakh, K.~Piyawanno, B.~Spinnler, E.~De~Man,
  P.~Kainzmaier, M.~S. Alfiad, A.~Napoli, and B.~Lankl, ``Data-aided versus
  blind single-carrier coherent receivers,'' \emph{IEEE Photonics Journal},
  vol.~2, no.~3, pp. 387--403, 2010.

\bibitem{agrawal2013nonlinear}
\BIBentryALTinterwordspacing
G.~P. Agrawal, \emph{Nonlinear Fiber Optics}, 5th~ed.\hskip 1em plus 0.5em
  minus 0.4em\relax Boston: Academic Press, 2013. [Online]. Available:
  \url{http://www.sciencedirect.com/science/article/pii/B9780123970237000024}
\BIBentrySTDinterwordspacing

\bibitem{gulli2017deep}
A.~Gulli and S.~Pal, \emph{Deep learning with Keras}.\hskip 1em plus 0.5em
  minus 0.4em\relax Packt Publishing Ltd, 2017.

\bibitem{matsumoto1998mersenne}
M.~Matsumoto and T.~Nishimura, ``Mersenne twister: a 623-dimensionally
  equidistributed uniform pseudo-random number generator,'' \emph{ACM
  Transactions on Modeling and Computer Simulation (TOMACS)}, vol.~8, no.~1,
  pp. 3--30, 1998.

\bibitem{klein2017fast}
A.~Klein, S.~Falkner, S.~Bartels, P.~Hennig, and F.~Hutter, ``Fast bayesian
  optimization of machine learning hyperparameters on large datasets,'' in
  \emph{Artificial intelligence and statistics}.\hskip 1em plus 0.5em minus
  0.4em\relax PMLR, 2017, pp. 528--536.

\bibitem{Alex01}
A.~Alvarado, E.~Agrell, D.~Lavery, R.~Maher, and P.~Bayvel, ``Replacing the
  soft-decision fec limit paradigm in the design of optical communication
  systems,'' \emph{Journal of Lightwave Technology}, vol.~33, no.~20, pp.
  4338--4352, 2015.

\bibitem{mckinley2004evm}
M.~D. McKinley, K.~A. Remley, M.~Myslinski, J.~S. Kenney, D.~Schreurs, and
  B.~Nauwelaers, ``Evm calculation for broadband modulated signals,'' in
  \emph{64th ARFTG Conf. Dig}.\hskip 1em plus 0.5em minus 0.4em\relax Orlando,
  2004, pp. 45--52.

\bibitem{schmogrow2011error}
R.~Schmogrow, B.~Nebendahl, M.~Winter, A.~Josten, D.~Hillerkuss, S.~Koenig,
  J.~Meyer, M.~Dreschmann, M.~Huebner, C.~Koos \emph{et~al.}, ``Error vector
  magnitude as a performance measure for advanced modulation formats,''
  \emph{IEEE Photonics Technology Letters}, vol.~24, no.~1, pp. 61--63, 2011.

\bibitem{freude2012quality}
W.~Freude, R.~Schmogrow, A.~Josten, D.~Hillerkuss, S.~Koenig, J.~Meyer,
  M.~Dreschmann, C.~Koos, J.~Leuthold, B.~Nebendahl \emph{et~al.}, ``Quality
  metrics for optical transmission,'' in \emph{2012 International Conference on
  Photonics in Switching (PS)}.\hskip 1em plus 0.5em minus 0.4em\relax IEEE,
  2012, pp. 1--4.

\bibitem{shafik2006extended}
R.~A. Shafik, M.~S. Rahman, and A.~R. Islam, ``On the extended relationships
  among evm, ber and snr as performance metrics,'' in \emph{2006 International
  Conference on Electrical and Computer Engineering}.\hskip 1em plus 0.5em
  minus 0.4em\relax IEEE, 2006, pp. 408--411.

\bibitem{eriksson2017characterization}
T.~A. Eriksson, T.~Fehenberger, and W.~Idler, ``Characterization of nonlinear
  fiber interactions using multidimensional mutual information over time and
  polarization,'' \emph{Journal of Lightwave Technology}, vol.~35, no.~6, pp.
  1204--1210, 2017.

\bibitem{eriksson2015four}
T.~A. Eriksson, T.~Fehenberger, N.~Hanik, P.~A. Andrekson, M.~Karlsson, and
  E.~Agrell, ``Four-dimensional estimates of mutual information in coherent
  optical communication experiments,'' in \emph{2015 European Conference on
  Optical Communication (ECOC)}.\hskip 1em plus 0.5em minus 0.4em\relax IEEE,
  2015, pp. 1--3.

\bibitem{catuogno2018non}
T.~Catuogno, M.~R. Camara, and M.~Secondini, ``Non-parametric estimation of
  mutual information with application to nonlinear optical fibers,'' in
  \emph{2018 IEEE International Symposium on Information Theory (ISIT)}.\hskip
  1em plus 0.5em minus 0.4em\relax IEEE, 2018, pp. 736--740.

\bibitem{rousseeuw1999fast}
P.~J. Rousseeuw and K.~V. Driessen, ``A fast algorithm for the minimum
  covariance determinant estimator,'' \emph{Technometrics}, vol.~41, no.~3, pp.
  212--223, 1999.

\bibitem{kramer2016scikit}
O.~Kramer, ``Scikit-learn,'' in \emph{Machine learning for evolution
  strategies}.\hskip 1em plus 0.5em minus 0.4em\relax Springer, 2016, pp.
  45--53.

\bibitem{schaedler2019deep}
M.~Schaedler, C.~Bluemm, M.~Kuschnerov, F.~Pittal{\`a}, S.~Calabr{\`o}, and
  S.~Pachnicke, ``Deep neural network equalization for optical short reach
  communication,'' \emph{Applied Sciences}, vol.~9, no.~21, p. 4675, 2019.

\bibitem{zhang2019functional}
J.~Zhang, P.~Lei, S.~Hu, M.~Zhu, Z.~Yu, B.~Xu, and K.~Qiu, ``Functional-link
  neural network for nonlinear equalizer in coherent optical fiber
  communications,'' \emph{IEEE Access}, vol.~7, pp. 149\,900--149\,907, 2019.

\bibitem{liu2018ols}
L.~Liu, M.~Bi, S.~Xiao, J.~Fang, T.~Huang, and W.~Hu, ``Ols-based rbf neural
  network for nonlinear and linear impairments compensation in the co-ofdm
  system,'' \emph{IEEE Photonics Journal}, vol.~10, no.~2, pp. 1--8, 2018.

\bibitem{liu2018effective}
S.~Liu, P.-C. Peng, C.-W. Hsu, S.~Chen, H.~Tian, and G.-K. Chang, ``An
  effective artificial neural network equalizer with s-shape activation
  function for high-speed 16-qam transmissions using low-cost directly
  modulated laser,'' in \emph{2018 International Conference on Network
  Infrastructure and Digital Content (IC-NIDC)}.\hskip 1em plus 0.5em minus
  0.4em\relax IEEE, 2018, pp. 269--273.

\bibitem{Jerart}
J.~L. Julus, D.~Manimegalai, A.~C. Beaula, J.~J. Athanesious, and A.~A.
  Roobert, ``Advanced nonlinear equalizer for filter bank multicarrier-based
  long reach-passive optical network system,'' \emph{International Journal of
  Communication Systems}, vol.~34, no.~14, 2021.

\bibitem{Kotlyar:21}
\BIBentryALTinterwordspacing
O.~Kotlyar, M.~Kamalian-Kopae, M.~Pankratova, A.~Vasylchenkova, J.~E.
  Prilepsky, and S.~K. Turitsyn, ``Convolutional long short-term memory neural
  network equalizer for nonlinear fourier transform-based optical transmission
  systems,'' \emph{Opt. Express}, vol.~29, no.~7, pp. 11\,254--11\,267, Mar
  2021. [Online]. Available:
  \url{http://www.opticsexpress.org/abstract.cfm?URI=oe-29-7-11254}
\BIBentrySTDinterwordspacing

\bibitem{ming2021ultralow}
H.~Ming, X.~Chen, X.~Fang, L.~Zhang, C.~Li, and F.~Zhang, ``Ultralow complexity
  long short-term memory network for fiber nonlinearity mitigation in coherent
  optical communication systems,'' \emph{Journal of Lightwave Technology}, pp.
  1--1, 2022.

\bibitem{fatadin2016estimation}
I.~Fatadin, ``Estimation of ber from error vector magnitude for optical
  coherent systems,'' in \emph{Photonics}, vol.~3, no.~2.\hskip 1em plus 0.5em
  minus 0.4em\relax Multidisciplinary Digital Publishing Institute, 2016,
  p.~21.

\bibitem{zhang2021understanding}
C.~Zhang, S.~Bengio, M.~Hardt, B.~Recht, and O.~Vinyals, ``Understanding deep
  learning (still) requires rethinking generalization,'' \emph{Communications
  of the ACM}, vol.~64, no.~3, pp. 107--115, 2021.

\bibitem{yang2020overfitting}
Z.~Yang, F.~Gao, S.~Fu, M.~Tang, and D.~Liu, ``Overfitting effect of artificial
  neural network based nonlinear equalizer: from mathematical origin to
  transmission evolution,'' \emph{Science China Information Sciences}, vol.~63,
  pp. 1--13, 2020.

\bibitem{bishop1995neural}
C.~M. Bishop \emph{et~al.}, \emph{Neural networks for pattern
  recognition}.\hskip 1em plus 0.5em minus 0.4em\relax Oxford university press,
  1995.

\bibitem{an1996effects}
G.~An, ``The effects of adding noise during backpropagation training on a
  generalization performance,'' \emph{Neural computation}, vol.~8, no.~3, pp.
  643--674, 1996.

\bibitem{neelakantan2015adding}
A.~Neelakantan, L.~Vilnis, Q.~V. Le, I.~Sutskever, L.~Kaiser, K.~Kurach, and
  J.~Martens, ``Adding gradient noise improves learning for very deep
  networks,'' \emph{arXiv preprint arXiv:1511.06807}, 2015.

\bibitem{809097}
C.~Wang and J.~Principe, ``Training neural networks with additive noise in the
  desired signal,'' \emph{IEEE Transactions on Neural Networks}, vol.~10,
  no.~6, pp. 1511--1517, 1999.

\bibitem{6796498}
Y.~Grandvalet, S.~Canu, and S.~Boucheron, ``Noise injection: Theoretical
  prospects,'' \emph{Neural Computation}, vol.~9, no.~5, pp. 1093--1108, 1997.

\bibitem{yin2015noisy}
S.~Yin, C.~Liu, Z.~Zhang, Y.~Lin, D.~Wang, J.~Tejedor, T.~F. Zheng, and Y.~Li,
  ``Noisy training for deep neural networks in speech recognition,''
  \emph{EURASIP Journal on Audio, Speech, and Music Processing}, vol. 2015,
  no.~1, pp. 1--14, 2015.

\bibitem{rifai2011adding}
S.~Rifai, X.~Glorot, Y.~Bengio, and P.~Vincent, ``Adding noise to the input of
  a model trained with a regularized objective,'' \emph{arXiv preprint
  arXiv:1104.3250}, 2011.

\bibitem{brown2003use}
W.~M. Brown, T.~D. Gedeon, and D.~I. Groves, ``Use of noise to augment training
  data: a neural network method of mineral--potential mapping in regions of
  limited known deposit examples,'' \emph{Natural Resources Research}, vol.~12,
  no.~2, pp. 141--152, 2003.

\bibitem{freire2021neural}
P.~J. Freire, J.~E. Prilepsky, Y.~Osadchuk, S.~K. Turitsyn, and V.~Aref, ``Deep
  neural network-aided soft-demapping in optical coherent systems: Regression
  versus classification,'' \emph{arXiv preprint arXiv:2109.13843}, 2021.

\bibitem{rady2011shannon}
H.~A.~K. Rady, ``Shannon entropy and mean square errors for speeding the
  convergence of multilayer neural networks: A comparative approach,''
  \emph{Egyptian Informatics Journal}, vol.~12, no.~3, pp. 197--209, 2011.

\bibitem{Goodfellowbook2016}
I.~Goodfellow, Y.~Bengio, and A.~Courville, \emph{Deep Learning}.\hskip 1em
  plus 0.5em minus 0.4em\relax MIT Press, 2016,
  \url{http://www.deeplearningbook.org}.

\bibitem{binh2019noises}
L.~N. Binh, \emph{Noises in optical communications and photonic systems}.\hskip
  1em plus 0.5em minus 0.4em\relax CRC Press, 2019.

\bibitem{zhang2018generalized}
Z.~Zhang and M.~R. Sabuncu, ``Generalized cross entropy loss for training deep
  neural networks with noisy labels,'' in \emph{32nd Conference on Neural
  Information Processing Systems (NeurIPS)}, 2018.

\bibitem{Georg2021book}
\BIBentryALTinterwordspacing
G.~Bocherer. (2021) Lecture notes on machine learning for communications.
  [Online]. Available: \url{http://georg-boecherer.de/mlcomm-v0.pdf}
\BIBentrySTDinterwordspacing

\bibitem{pazzani1994reducing}
M.~Pazzani, C.~Merz, P.~Murphy, K.~Ali, T.~Hume, and C.~Brunk, ``Reducing
  misclassification costs,'' in \emph{Machine Learning Proceedings 1994}.\hskip
  1em plus 0.5em minus 0.4em\relax Elsevier, 1994, pp. 217--225.

\bibitem{gutierrez2015ordinal}
P.~A. Guti{\'e}rrez, M.~Perez-Ortiz, J.~Sanchez-Monedero, F.~Fernandez-Navarro,
  and C.~Hervas-Martinez, ``Ordinal regression methods: survey and experimental
  study,'' \emph{IEEE Transactions on Knowledge and Data Engineering}, vol.~28,
  no.~1, pp. 127--146, 2015.

\bibitem{lathuiliere2019comprehensive}
S.~Lathuili{\`e}re, P.~Mesejo, X.~Alameda-Pineda, and R.~Horaud, ``A
  comprehensive analysis of deep regression,'' \emph{IEEE transactions on
  pattern analysis and machine intelligence}, vol.~42, no.~9, pp. 2065--2081,
  2019.

\bibitem{bosman2020visualising}
A.~S. Bosman, A.~Engelbrecht, and M.~Helbig, ``Visualising basins of attraction
  for the cross-entropy and the squared error neural network loss functions,''
  \emph{Neurocomputing}, vol. 400, pp. 113--136, 2020.

\bibitem{9662308}
S.~Wang, F.~Liu, and B.~Liu, ``Escaping the gradient vanishing: Periodic
  alternatives of softmax in attention mechanism,'' \emph{IEEE Access}, vol.~9,
  pp. 168\,749--168\,759, 2021.

\bibitem{end_to_end_MSE}
Z.~Niu, H.~Yang, H.~Zhao, C.~Dai, W.~Hu, and L.~Yi, ``End-to-end deep learning
  for long-haul fiber transmission using differentiable surrogate channel,''
  \emph{Journal of Lightwave Technology}, pp. 1--1, 2022.

\bibitem{li2016online}
Y.~Li, C.~Lan, J.~Xing, W.~Zeng, C.~Yuan, and J.~Liu, ``Online human action
  detection using joint classification-regression recurrent neural networks,''
  in \emph{European conference on computer vision}.\hskip 1em plus 0.5em minus
  0.4em\relax Springer, 2016, pp. 203--220.

\bibitem{mukherjee2021joint}
L.~Mukherjee, M.~A.~K. Sagar, J.~N. Ouellette, J.~J. Watters, and K.~W.
  Eliceiri, ``Joint regression-classification deep learning framework for
  analyzing fluorescence lifetime images using nadh and fad,'' \emph{Biomedical
  Optics Express}, vol.~12, no.~5, pp. 2703--2719, 2021.

\bibitem{liu2018joint}
M.~Liu, J.~Zhang, E.~Adeli, and D.~Shen, ``Joint classification and regression
  via deep multi-task multi-channel learning for alzheimer's disease
  diagnosis,'' \emph{IEEE Transactions on Biomedical Engineering}, vol.~66,
  no.~5, pp. 1195--1206, 2018.

\bibitem{liu2017deep}
M.~Liu, J.~Zhang, and et. al., ``Deep multi-task multi-channel learning for
  joint classification and regression of brain status,'' in \emph{International
  conference on medical image computing and computer-assisted
  intervention}.\hskip 1em plus 0.5em minus 0.4em\relax Springer, 2017, pp.
  3--11.

\bibitem{wu2021joint}
J.-Y. Wu, M.~Wu, Z.~Chen, X.~Li, and R.~Yan, ``A joint
  classification-regression method for multi-stage remaining useful life
  prediction,'' \emph{Journal of Manufacturing Systems}, vol.~58, pp. 109--119,
  2021.

\bibitem{chen2019joint}
J.~Chen, L.~Cheng, X.~Yang, J.~Liang, B.~Quan, and S.~Li, ``Joint learning with
  both classification and regression models for age prediction,'' in
  \emph{Journal of Physics: Conference Series}, vol. 1168, no.~3.\hskip 1em
  plus 0.5em minus 0.4em\relax IOP Publishing, 2019, p. 032016.

\bibitem{lin2017focal}
T.-Y. Lin, P.~Goyal, R.~Girshick, K.~He, and P.~Doll{\'a}r, ``Focal loss for
  dense object detection,'' in \emph{Proceedings of the IEEE international
  conference on computer vision}, 2017, pp. 2980--2988.

\bibitem{Kingma2015}
D.~P. Kingma and J.~L. Ba, ``{Adam: A method for stochastic optimization},''
  \emph{3rd International Conference on Learning Representations, ICLR 2015 -
  Conference Track Proceedings}, pp. 1--15, 2015.

\bibitem{Wilson2003}
D.~R. Wilson and T.~R. Martinez, ``{The general inefficiency of batch training
  for gradient descent learning},'' \emph{Neural Networks}, vol.~16, no.~10,
  pp. 1429--1451, 2003.

\bibitem{Smith2017}
\BIBentryALTinterwordspacing
S.~L. Smith and Q.~V. Le, ``A bayesian perspective on generalization and
  stochastic gradient descent,'' 2017. [Online]. Available:
  \url{https://arxiv.org/abs/1710.06451}
\BIBentrySTDinterwordspacing

\bibitem{ShirishKeskar2017}
\BIBentryALTinterwordspacing
N.~S. Keskar, D.~Mudigere, J.~Nocedal, M.~Smelyanskiy, and P.~T.~P. Tang, ``On
  large-batch training for deep learning: Generalization gap and sharp
  minima,'' 2016. [Online]. Available: \url{https://arxiv.org/abs/1609.04836}
\BIBentrySTDinterwordspacing

\bibitem{1036395}
M.~Wandel, P.~Kristensen, T.~Veng, Y.~Qian, Q.~Le, and L.~Gruner-Nielsen,
  ``Dispersion compensating fibers for non-zero dispersion fibers,'' in
  \emph{Optical Fiber Communication Conference and Exhibit}, 2002, pp.
  327--329.

\bibitem{peng2013bit}
L.~Peng, M.~H{\'e}lard, and S.~Haese, ``On bit-loading for discrete multi-tone
  transmission over short range pof systems,'' \emph{Journal of lightwave
  technology}, vol.~31, no.~24, pp. 4155--4165, 2013.

\bibitem{kim2020modulation}
J.-W. Kim and C.-H. Lee, ``Modulation format identification of square and
  non-square m-qam signals based on amplitude variance and osnr,'' \emph{Optics
  Communications}, vol. 474, p. 126084, 2020.

\bibitem{nadal2015experimental}
L.~Nadal, J.~M. F{\`a}brega, J.~V{\'\i}lchez, and M.~S. Moreolo, ``Experimental
  analysis of 8-qam constellations for adaptive optical ofdm systems,''
  \emph{IEEE Photonics Technology Letters}, vol.~28, no.~4, pp. 445--448, 2015.

\bibitem{wang2014capacity}
J.~Wang, N.~Li, W.~Shi, Y.~Ma, X.~Liang, and X.~Dong, ``Capacity of 60 ghz
  wireless communications based on qam,'' \emph{Journal of Applied
  Mathematics}, vol. 2014, 2014.

\bibitem{brownlee2018better}
J.~Brownlee, ``{Better Deep Learning. Train Faster, Reduce Overfitting, and
  Make Better Predictions},'' \emph{Machine Learning Mastery With Python},
  vol.~1, no.~2, 2018.

\bibitem{650112}
S.~Ong, C.~You, S.~Choi, and D.~Hong, ``A decision feedback recurrent neural
  equalizer as an infinite impulse response filter,'' \emph{IEEE Transactions
  on Signal Processing}, vol.~45, no.~11, pp. 2851--2858, 1997.

\bibitem{pascanu2013difficulty}
R.~Pascanu, T.~Mikolov, and Y.~Bengio, ``On the difficulty of training
  recurrent neural networks,'' in \emph{International conference on machine
  learning}.\hskip 1em plus 0.5em minus 0.4em\relax PMLR, 2013, pp. 1310--1318.

\bibitem{mikami2019field}
M.~Mikami and H.~Yoshino, ``Field trial on 5g low latency radio communication
  system towards application to truck platooning,'' \emph{IEICE Transactions on
  Communications}, p. 2018TTP0021, 2019.

\bibitem{pokhrel2021towards}
S.~R. Pokhrel, N.~Kumar, and A.~Walid, ``Towards ultra reliable low latency
  multipath tcp for connected autonomous vehicles,'' \emph{IEEE Transactions on
  Vehicular Technology}, vol.~70, no.~8, pp. 8175--8185, 2021.

\bibitem{ponomarev2021latency}
E.~Ponomarev, S.~Matveev, I.~Oseledets, and V.~Glukhov, ``Latency estimation
  tool and investigation of neural networks inference on mobile gpu,''
  \emph{Computers}, vol.~10, no.~8, p. 104, 2021.

\bibitem{jouppi2017datacenter}
N.~P. Jouppi, C.~Young, N.~Patil, D.~Patterson, G.~Agrawal, R.~Bajwa, S.~Bates,
  S.~Bhatia, N.~Boden, A.~Borchers \emph{et~al.}, ``In-datacenter performance
  analysis of a tensor processing unit,'' in \emph{Proceedings of the 44th
  annual international symposium on computer architecture}, 2017, pp. 1--12.

\bibitem{pacheco2011introduction}
P.~Pacheco, \emph{An introduction to parallel programming}.\hskip 1em plus
  0.5em minus 0.4em\relax Elsevier, 2011.

\bibitem{demmel2012communication}
J.~Demmel, ``Communication avoiding algorithms,'' in \emph{2012 SC Companion:
  High Performance Computing, Networking Storage and Analysis}.\hskip 1em plus
  0.5em minus 0.4em\relax IEEE, 2012, pp. 1942--2000.

\bibitem{ballard2014communication}
G.~Ballard, E.~Carson, J.~Demmel, M.~Hoemmen, N.~Knight, and O.~Schwartz,
  ``Communication lower bounds and optimal algorithms for numerical linear
  algebra,'' \emph{Acta Numerica}, vol.~23, pp. 1--155, 2014.

\bibitem{Imani2016}
M.~Imani, S.~Patil, and T.~Rosing, ``Low power data-aware stt-ram based hybrid
  cache architecture,'' in \emph{2016 17th International Symposium on Quality
  Electronic Design (ISQED)}, 2016, pp. 88--94.

\bibitem{ullah2018area}
S.~Ullah, S.~Rehman, B.~S. Prabakaran, F.~Kriebel, M.~A. Hanif, M.~Shafique,
  and A.~Kumar, ``Area-optimized low-latency approximate multipliers for
  fpga-based hardware accelerators,'' in \emph{Proceedings of the 55th annual
  design automation conference}, 2018, pp. 1--6.

\bibitem{fujisawa2022weight}
S.~Fujisawa, F.~Yaman, H.~G. Batshon, M.~Tanio, N.~Ishii, C.~Huang, T.~F.
  de~Lima, Y.~Inada, P.~R. Prucnal, N.~Kamiya \emph{et~al.}, ``Weight pruning
  techniques towards photonic implementation of nonlinear impairment
  compensation using neural networks,'' \emph{Journal of Lightwave Technology},
  vol.~40, no.~5, pp. 1273--1282, 2022.

\bibitem{pedro_quantization2021}
\BIBentryALTinterwordspacing
D.~R. Argüello, P.~J. Freire, J.~E. Prilepsky, M.~Kamalian-Kopae, A.~Napoli,
  and S.~K. Turitsyn, ``Experimental implementation of a neural network optical
  channel equalizer in restricted hardware using pruning and quantization,''
  2022. [Online]. Available: \url{https://doi.org/10.21203/rs.3.rs-1236203/v1}
\BIBentrySTDinterwordspacing

\bibitem{hawks2021ps}
B.~Hawks, J.~Duarte, N.~J. Fraser, A.~Pappalardo, N.~Tran, and Y.~Umuroglu,
  ``Ps and qs: Quantization-aware pruning for efficient low latency neural
  network inference,'' \emph{arXiv preprint arXiv:2102.11289}, 2021.

\bibitem{jacob2018quantization}
B.~Jacob, S.~Kligys, B.~Chen, M.~Zhu, M.~Tang, A.~Howard, H.~Adam, and
  D.~Kalenichenko, ``Quantization and training of neural networks for efficient
  integer-arithmetic-only inference,'' in \emph{Proceedings of the IEEE
  conference on computer vision and pattern recognition}, 2018, pp. 2704--2713.

\end{thebibliography}

  \IEEEbiographynophoto{Pedro J. Freire} received his Bachelor's and Master's degrees in Electronic Engineering from the Federal University of Pernambuco, with a one-year and a-half period at the State University of New York and the State University of San Francisco. His Master’s thesis in the photonics area was dealt with a new metaheuristic using evolutionary algorithms to solve the RMLSA problem. Currently, he holds a Marie-Curie (MSCA) doctoral fellowship and works as an Early Stage Researcher at Aston University and Infinera.  His interests focus on advanced digital signal processing and coding, network monitoring and planning, artificial intelligence, machine learning, and hardware DSP realization.
  \endIEEEbiographynophoto
  
  \IEEEbiographynophoto{Antonio Napoli} has a Ph.D. degree at PoliTO, 2006. He joined Infinera in 2018 and working on XR Optics. He led three special issues on JOCN/JLT. He served as an OFC TPC member and co-organized 4 OFC workshops. He is the technical coordinator of three H2020 projects. Interests: DSP, wide-band systems, modeling, ML/AI. Dr. Napoli represents Infinera at OIF. He is (co)-author of 7 patents, 193 articles, and a book chapter. 
  \endIEEEbiographynophoto

  \IEEEbiographynophoto{Bernhard Spinnler} received the Dipl.-Ing. Degree in Communications Engineering and the Dr.-Ing. Degree from the University of Erlangen-Nürnberg, Germany, in 1994 and 1997, respectively. Since 1997, he has worked on low-complexity modem design of wireless radio relay systems at Siemens AG, Information and Communication Networks. In 2002, he joined the optical networks group of Siemens Corporate Technology which became Nokia Siemens Networks and later Coriant. He is now working for Infinera on the design and modeling of robust, tolerant, and automated optical communications systems. His interests focus on advanced digital signal processing and coding, network automation, artificial intelligence, and machine learning.
  \endIEEEbiographynophoto
  
    \IEEEbiographynophoto{Nelson Costa}
    was born in Tomar, Portugal, in 1983. He received the Licenciatura and Ph.D. degrees in electrical and computer engineering from the Instituto Superior Técnico (IST), Lisbon Technical University, Lisbon, Portugal, in 2006 and 2012, respectively. He is currently with Infinera Portugal, Carnaxide, Portugal. He has authored or coauthored more than 50 publications in international conferences and journals. His current research interests include advanced coherent optical modulation formats, nonlinear fiber transmission effects, and network optimization.
  \endIEEEbiographynophoto
  
    \IEEEbiographynophoto{Professor Sergei K. Turitsyn}
     graduated from the Department of Physics of Novosibirsk University in 1982 and received his Ph.D. degree in Theoretical and Mathematical Physics from the Budker Institute of Nuclear Physics, Novosibirsk, Russia, in 1986. In 1992 he moved to Germany, first, as a Humboldt Fellow and then working in the collaborative projects with Deutsche Telekom. Since 2012 he serves as a director of the Aston Institute of Photonic Technologies, which is a world-known photonics research center, with a strong track record of academic achievements, a range of developed technologies, and industrial collaborations. Sergei Turitsyn is the originator of several key concepts in the fields of nonlinear science, optical fiber communications, and fiber lasers. He was/is a principal investigator in 67 national and international, research and industrial projects. Turitsyn serves as a member of the Editorial Board (Electronics, Photonics and Device Physics) of the Scientific Reports, Nature Publishing Group, and an Associate Editor of JLT. Turitsyn was the recipient of a Royal Society Wolfson Research Merit Award in 2005. In 2011 he was awarded the European Research Council, Advanced Grant. He received the Lebedev medal from the Rozhdestvensky Optical Society in 2014, Aston 50th Anniversary Chair medal in 2016, and Chancellor’s Medal in 2018. He is a Fellow of the Optical Society of America and the Institute of Physics.
  \endIEEEbiographynophoto
  
    \IEEEbiographynophoto{Jaroslaw E. Prilepsky} received the M.E. degree in theoretical physic (Hons.)  from the  National  University of  Kharkiv,  Ukraine,  in  1999,  and the  Ph.D. degree  in  theoretical  physics  from  the  B. Verkin Institute for Low-Temperature Physics and Engineering, Kharkiv, Ukraine, in 2003, with a focus on nonlinear excitation   in  low-dimensional   systems.   From   2003   to  2010,   he   was   a Research  Fellow  at  the  B. Verkin Institute for Low-Temperature Physics and Engineering. From  2010  to  2012,  he  was  a Research  Associate  with  the  Nonlinearity  and  Complexity  Research  Group, Aston  University,  U.K.,  and  since  2012,  he  has  been  a  Research  Fellow with  the  Aston  Institute  of  Photonics  Technologies,  Aston  University.  He has  authored  over  80  journal  papers  and  conference  contributions  in  the fields of nonlinear physics and mathematics, solitons, nonlinear signal-noise interaction, optical transmission, and signal processing. His current research interests include (but are  not  limited  to)  optical  transmission  systems  and  networks,  nonlinearity mitigation  methods,  neural networks and machine learning, nonlinear  Fourier-based  optical  transmission  methods, optical  signal processing.
  \endIEEEbiographynophoto

\clearpage

\end{document}